\documentclass[nonacm]{acmart}

\usepackage[table]{colortbl}
\usepackage{multirow}

\AtBeginDocument{%
  \providecommand\BibTeX{{%
    \normalfont B\kern-0.5em{\scshape i\kern-0.25em b}\kern-0.8em\TeX}}}

\begin{document}

\title[All the Way There \& Back: Wayfinding Apps for Blind Travelers]{All the Way There and Back: Inertial-Based, Phone-in-Pocket Indoor Wayfinding and Backtracking Apps for Blind Travelers}


\author{Chia Hsuan Tsai}
\authornote{All three authors contributed equally to this research.}
\email{ctsai24@ucsc.edu}
\author{Fatemeh Elyasi}
\authornotemark[1]
\email{felyasi@ucsc.edu}
\author{Peng Ren}
\authornotemark[1]
\email{pren1@ucsc.edu}
\affiliation{%
  \institution{University of California, Santa Cruz}
  \streetaddress{1156 High Street}
  \city{Santa Cruz}
  \state{California}
  \country{USA}
}
\author{Roberto Manduchi}
\affiliation{%
  \institution{University of California, Santa Cruz}
  \streetaddress{1156 High Street}
  \city{Santa Cruz}
  \state{California}
  \country{USA}
}
\email{manduchi@ucsc.edu}
\begin{abstract}
We introduce two iOS apps that have been designed to support wayfinding and backtracking for blind travelers navigating in indoor building environments. Wayfinding involves determining and following a route through the building's corridors to reach a destination, and assumes that the app has access to the floor plan of the building. Backtracking one's route, on the other hand, requires no map knowledge. Our apps only use the inertial and magnetic sensors of the smartphone, and thus require no infrastructure modification (e.g., installation and support of BLE beacons). Unlike systems that use the phone's camera, users of our apps can conveniently keep their phone tucked inside a pocket, while interacting with the apps using a smartwatch. Routing directions are given via speech. Both apps were tested in a user study with seven blind participants, who used them while navigating a campus building.
\end{abstract}




\maketitle

\section{Introduction}
Navigating a building can be confusing and disorienting for anyone who is visiting the building for the first time. For those who are blind, this experience can be even more challenging, as these travelers cannot rely on visual feedback. While multiple research efforts have focused on accessible technology to support independent travel in indoor space, at the time of this writing there are no commercially available systems ready for widespread use. Indoor spaces are challenging because GPS cannot be relied upon. This doesn't mean that indoor localization is not feasible: a variety of techniques have been demonstrated, some of which have been adapted for navigation by blind individuals. One example is NavCog in its various versions~\cite{ahmetovic2016navcog,murata2018smartphone,sato2019navcog3,sato2017navcog3}, which uses Bluetooth Low Energy (BLE) for accurate localization.  BLE localization is a mature technology. Unfortunately, BLE-based navigation is only possible if a specific infrastructure (a possibly large set of BLE beacons) has been installed, and a laborious calibration procedure (fingerprinting) has been conducted. This raises questions of scalability, as the success of this technology hinges on the good will of the agency that manages the building. It would thus be desirable that the technology powering the navigational aid would not depend on external dedicated infrastructure.

In recent years, and with the advent of powerful AI, there has been intense interest in systems that use visual sensors (e.g., the camera embedded in a smartphone) to extract positional data, and to provide users with information about the world around them. For example, navigation apps built on Apple's ARKit have been developed specifically for use by blind travelers~\cite{crabb2023lightweight,yoon2019leveraging}. Visual-inertial odometry technology, coupled with Particle Filtering, can produce extremely accurate localization. Its core strength, the use of visual data, is unfortunately also its main drawback. Users of these apps must hold the smartphone (in their hand, or perhaps attached to a lanyard~\cite{williams2015not}) in such a way that the camera has a good view of the environment. This is not always possible, convenient, or desirable. Blind travelers normally use a long cane or a dog guide, and thus have one hand already occupied maneuvering the cane or holding the dog. Indeed,  it has been often observed that navigational aid for blind walkers should be hands-free~\cite{NYTimes21,sato2017navcog3}. 

In this article, we describe an experiment we conducted with 7 blind travelers, who tested two apps created in our lab that were designed to assist with navigation in buildings characterized by a network of corridors. Both apps utilize data from the inertial sensors of a smartphone, and thus require no external infrastructure. Since the apps don't use data from the smartphone camera, users can conveniently keep the phone tucked in a pocket (indeed, this is how the apps have been tested in our experiment). The first one is a {\em Wayfinding} app: it has knowledge of the floor plan of the building to be navigated, computes the shortest length route to destination, and provides navigational support in an accessible way. The second one is a {\em Backtracking} app. Its sole purpose is to help an individual, who has previously walked on a certain route (e.g., from the front entrance of a building to a certain office room), to trace back their steps. While arguably less useful than a full blown wayfinding application, backtracking support can help a blind traveler feel more confident in some situations. Importantly, the Backtracking app requires no prior knowledge of the building layout: it only uses data recorded during the first route traversal ({\em way-in}) to generate support for the user when walking back ({\em return}). Both apps have provision for error recovery (i.e., can provide corrective directions if a participant missed a turn, or took the wrong turn). 

These are the main contributions of this work:
\begin{itemize}
    \item We demonstrate an accessible iPhone app (\textit{Wayfinding}) that uses two different localization technologies, complemented by Particle Filtering, for inertial-based localization in a building with known floor plan.
    \item We propose a new hybrid approach for backtracking that uses magnetic and inertial data, with no knowledge of the building layout (and thus usable in any venue even without access to a floor plan). This algorithm was implemented and tested in the \textit{Backtracking} app.
    \item We demonstrate the effectiveness of a simple speech-based user interface mechanism, consistent for both apps,  designed to mitigate the unavoidable inaccuracy associated with the dead-reckoning nature of inertial-based localization.
    \item We describe a user study with 7 blind travelers. The participants first traversed three routes (292 meters and 13 turns in total) using our Wayfinding app; then, they traversed the same routes, in the opposite direction, using the Backtracking app. 
\end{itemize}

This article is structured as follows. After reviewing the related work in Sec.~\ref{sec:RW}, we describe our Wayfinding and Backtracking apps (including their shared user interface mechanisms) in Sec.~\ref{sec:Apparatus}. Our experiment with seven blind participant is described in Sec.~\ref{sec:Experiment}, and results, along with limitations of our approach, are discussed in Sec.~\ref{sec:Discussion}. Sec.~\ref{sec:Conclusions} has the conclusions.


\section{Related Work}\label{sec:RW}
Multiple research considered smartphone app implementations to enhance the mobility of individuals with visual impairments, especially in GPS-denied indoor environments. 
These apps leverage a variety of technology, including BLE beaconing, Wi-Fi, IMUs, video cameras, or sensor fusion, to offer navigational support~\cite{cheng2005accuracy, fusco2020indoor, ishihara2017beacon, luca2016towards, manduchi2010blind, sato2019navcog3}.
As an example, NavCog~\cite{ahmetovic2016navcog} incorporates a precise localization algorithm based on BLE beacons, and offers customizable voice and non-vocal sound instructions for user interaction. Additionally, LuzDeploy~\cite{gleason2018crowdsourcing}, built upon NavCog, has adopted crowdsourcing technology to maintain the localization infrastructure.
 Subsequent versions~\cite{ahmetovic2017achieving} integrated  BLE and inertial data positioning, and Particle Filtering (NavCog3~\cite{murata2018smartphone,sato2019navcog3,sato2017navcog3}). A virtual navigation app (VirtualNav~\cite{guerreiro2020virtual}) was built upon the foundation of NavCog3, to explore the transfer of virtually acquired route knowledge to practical navigation. Other projects using  BLE-based localization include ASSIST)~\cite{nair2022assist} and GuideBeacon~\cite{cheraghi2017guidebeacon}.  Alternative technologies considered for localization include RFID tags (e.g.,  PERCEPT~\cite{ganz2012percept}), color tags (e.g., NaviLens~\cite{martin2020accessible}), Wi-Fi beaconing~\cite{abu2017isab}, and magnetic navigation~\cite{riehle2012indoor,giudice2019evaluation}. 

 In recent work, visual-inertial odometry (VIO) has been shown to provide accurate localization without relying on dedicated  infrastructure~\cite{crabb2023lightweight}. Backtracking apps using similar localization technologies have also been developed~\cite{yoon2019leveraging,flores2018easy,Shu2019}. As mentioned earlier, these system require use of a camera with unoccluded visual field.

 Research on navigation using inertial sensing (the technology considered in this article) includes~\cite{riehle2013indoor,fallah2012user,apostolopoulos2014integrated,ren2023experiments}. Riehle et al.~\cite{riehle2013indoor} described a classic pedestrian dead-reckoning system (PDR) with explicit turn detection, which was tested with 8 blind or low vision participants in a simple indoor route. Participants could request assistance from the experimenter if they felt they needed it. Compared to a condition in which the route was described at the beginning, but no guidance was provided during the trial, it was shown that use of the PDR system resulted in fewer requests for assistance made and  in more instances of successful localization of the route's destination. Fallah et al.~\cite{fallah2012user} and Apostolopoulos et al~\cite{apostolopoulos2014integrated} proposed an inertial-based navigation system designed to overcome inherent localization inaccuracy through the ``user as a sensor modality''. Routes were represented via a sequence of perceivable landmarks. Participants were asked to follow directions to the next landmark; once arrived at the landmark, they were asked to confirm it on the app. This ingenious strategy allowed the localization system to reset itself at the correct location at each detected landmark. The inertial system was based on step counting (through an accelerometer) and orientation sensing (through a compass). A Particle Filter was used to track the state (location and orientation) of the user. In addition, different methods were considered to estimate the user's step length using the Particle Filter. An experiment with 6 blind and low vision participants was conducted on two floors of a building with 11  routes, most of which contained one or two turns. One route included climbing a staircase, while another route included taking an elevator. These routes contained multiple landmarks that needed to be discovered by the participants, including doors that needed to be counted, hallways, stairs, ramps. Peng et al.~\cite{ren2023experiments} demonstrated an indoor/outdoor wayfinding system, designed for navigation in a transit hub, that tracked the user's location through fusion of inertial and GPS data.
 
  Along with localization, research has focused on the design of accessible user interface modalities for navigational systems~\cite{krainz2016accessible,hossain2020sightless,shahini2022friendly,zahabi2022design,kuriakose2023turn}, using modalities such as speech~\cite{ahmetovic2016navcog,giudice2019evaluation,fallah2012user}, sound~\cite{fiannaca2014headlock,ross2000wearable,manduchi2014last}, and vibration~\cite{azenkot2011smartphone,sato2017navcog3,flores2015vibrotactile}.

\section{Apparatus}\label{sec:Apparatus}
We developed two iOS apps for this study: {\em Wayfinding} and {\em Backtracking}. The Wayfinding app has access to a floor plan of the building to be navigated. It is designed to find a route from the current user's location to the desired destination, update the route as necessary during traversal, and provide directions to the user as they are following the route. The Backtracking app has no prior information about the building layout. Similarly to \cite{flores2018easy}, Backtracking operates in two phases. In the first phase ({\em way-in}), it simply tracks the route taken by the user to reach a destination. Then, in the {\em return} phase, it produces directions to help the user re-trace the same way-in route. To do so, the app progressively matches the partial return route with the way-in route (in reverse). 

As will be clear in the following, the Backtracking app needs to use different localization strategies than the Wayfinding app, as a consequence of its lack of building layout information. The user interface, however, was designed to be almost identical for the two apps, shielding the user from the different underlying mechanisms.

\subsection{Wayfinding}

\subsubsection{Localization}\label{sec:localization}
We implemented two different pedestrian dead-reckoning (PDR) algorithms for user localization and tracking using data from the phone's inertial sensors. These two techniques are termed {\em Azimuth/Steps} and {\em RoNIN}, respectively~\cite{ren2021smartphone}. A Particle Filter is applied to the output of either system. During the tests, we ran both systems in parallel, though only one of them was used to provide guidance to the user. The reason for running both algorithms in parallel was twofold. First, we wanted to have a ``fail safe'' mechanism: if an algorithm failed to correctly track the participant, we could resort to switching to the other algorithm. Note that this only happened once in the whole experiment. Second, it allowed us to comparatively assess localization data from both algorithms in a variety of situations. 

Both algorithms compute the user's locations in terms of an arbitrary (but fixed) {\em world} reference frame, with the Z axis pointing downwards (in the direction of gravity). We used a simple initial calibration procedure (described below)  to find the angle of the rotation around the Z axis that brings the world reference frame to the frame used to define the floor plan (the {\em floor plan} frame).

{\bf\em Azimuth/Steps (A/S)}. This algorithm produces, at each detected step (heel strike), a 2-D vector $\Delta_p$ with a length equal to the estimated step length, and direction given by the phone's {\em azimuth} (or {\em heading}) angle. This vector is then fed as input to the Particle Filter (discussed later in this section). Heel strikes (steps) are computed by an LSTM-based algorithm that processes data from the phone's inertial sensors ~\cite{ren2021smartphone,edel2015advanced}.
Each participant's step length was estimated by the following calibration procedure. The participant was asked to walk along a straight corridor path (38.25 meters in length). By dividing the length of this path by the number of steps (from the step counter), we obtained the user's approximate step length. A similar step length calibration was used in~\cite{riehle2013indoor, apostolopoulos2014integrated}.
Note that after calibration, the step length is further updated by the Particle Filter (in a similar fashion to~\cite{apostolopoulos2014integrated}). 

In spite of its simplicity, A/S was shown to produce good path reconstruction results when coupled with a Particle Filter, which can mitigate the drift effect inherent in the dead-reckoning nature of the algorithm~\cite{ren2021smartphone}. One drawback of this approach is that the orientation of the vector $\Delta_p$ (azimuth) is that of the phone, rather than that of the walker. This means that moving the phone to a different location on one's body (e.g., pulling it out of the pocket to take a call) may incorrectly be interpreted as a change in walking direction. In our experiment, participants kept the phone tucked inside their back pocket throughout the trials, hence the orientation of the phone with respect to the user's body could be considered to be constant.

{\bf\em RoNIN}. RoNIN~\cite{herath2020ronin} is a machine learning algorithm for dead-reckoning from inertial data. Its feasibility for the reconstruction of paths taken by blind walkers, using a long cane or a dog guide, was demonstrated in~\cite{ren2021smartphone}. RoNIN produces velocity vectors at a rate of 25 Hz. We used the authors’  open-source ResNet18 implementation\footnote{https://github.com/Sachini/ronin}, and integrated the velocity vectors over individual step periods to obtain displacement vectors $\Delta_p$. 

Unlike the A/S algorithm, RoNIN produces displacement vectors that are defined with respect to the world frame (rather than the phone's frame). This means that the heading direction estimated by RoNIN is independent of the phone's orientation with respect to one's body~\cite{herath2020ronin}. Hence, the user is not constrained to hold the phone in a fixed location. However, RoNIN is as liable as A/S to orientation drift due to the integration of noisy sensor data.

Experiments reported in~\cite{ren2021smartphone} showed that while RoNIN worked remarkably well for some individuals, it produced less satisfactory results for others. In particular, the magnitude of the velocity vector returned in output for certain users was found to be either smaller or larger than their actual speed. To account for this user-dependent error, we employed the same calibration procedure described above to find an adjustment coefficient ({\em RoNIN multiplier}) for each participant. In practice, we divided the length of the path traversed during calibration by the estimation of the same path length produced by RoNIN. Then, during the trials with the same participants, we multiplied the velocity vectors produced by the RoNIN multiplier thus computed.  

{\bf\em Particle Filtering (PF)}. Particle Filtering is a form of Bayesian filtering that is commonly used for spatial tracking in the presence of prior information or constraints~\cite{fox2003bayesian}. 
In our case, knowledge of the floor plan allows us to define ``impenetrable walls" that are unlikely to be traversed by a traveler~\cite{yu2019shoesloc}. Using a Particle Filter, the posterior distribution of the walker's location is expressed by means of a set of samples ({\em particles}). In our experiments, we used 500 particles, which was found to be a reasonable trade-off between computational speed and accuracy of tracking.  

Each $i$-th particle is characterized by a X-Y location $p_i$, a drift angle value $\Delta_{\theta,i}$, and, for A/S localization, a step length value $s_i$. 
At the beginning of a trial, all particles are located at the (known) initial user location. The drift angles associated with the particles are sampled from a zero-mean Gaussian distribution with $\sigma=30^{\circ}$, while the step lengths are sampled from a Gaussian distribution with $\sigma=6$~cm centered at the step length found during the initial calibration. Each particle has a (positive) weight $w_i$. Weights are all initialized to $1/500$.

At each time period (in our case, at each detected step), each particle is spatially propagated by a vector that is a function of the displacement vector $\Delta_p$ produced by A/S or by RoNIN. Specifically, $\Delta_p$ is first rotated by the particle's drift angle $\Delta_{\theta,i}$. For A/S localization, its length is changed to $s_i$. The particle's drift angle $\Delta_{\theta,i}$ is  updated by addition of Gaussian noise  ($\sigma=1^{\circ}$). Then, Gaussian noise is added to its modulus ($\sigma=0.5\cdot s_i$) and phase ($\sigma=0.5^{\circ}$). The particle's location is then updated by adding the resulting vector to $p_i$.  Additionally, the particle weights can be modified during a positional update as described later in this section (after re-weighting, the  weights are normalized to sum to 1.) 

At each time step, a new set of 500 particles is resampled with replacement from the current set of particles. This means that, for 500 times, a particle $j$ in the current set is picked at random, with the probability of a particle being sampled equal to its weight. A new particle is generated with the same characteristics (location $p_j$, angular drift $\Delta_{\theta,j}$, and, for A/S, step length $s_j$). Gaussian noise ($\sigma$=10~cm) is added to the X and Y components of its location. The old particles are then discarded, and the new ones are retained until the next update. The user's location is taken to be the weighted sum of the particles' locations: $p=\sum_i w_i p_i$. 

Particles are re-weighted at each time period according to the following rules:

\begin{enumerate}
    \item If a particle is found to be crossing a wall, its weight is set to 0.
    \item If a particle is at a distance from the weighted average $p$ larger than a threshold value $D$, its weight is set to 0.
    \item  If at any time a particle is found to be inside a room, its weight is multiplied by 0.9. 
\end{enumerate} 

\begin{figure}[h]
  \centering
  \begin{tabular}{cc}
  \includegraphics[width=0.5\linewidth]{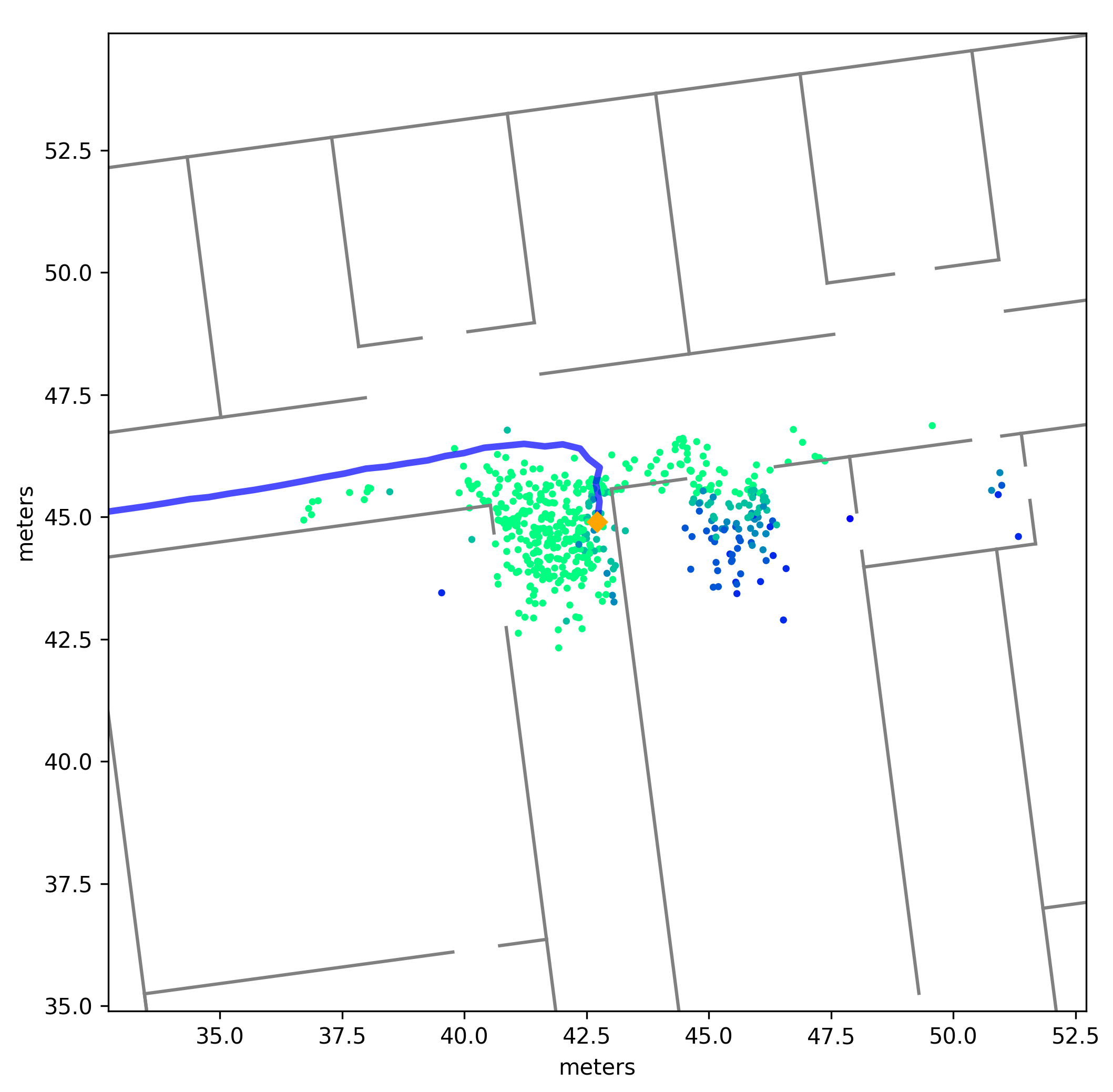} &
    \includegraphics[width=0.25\linewidth]{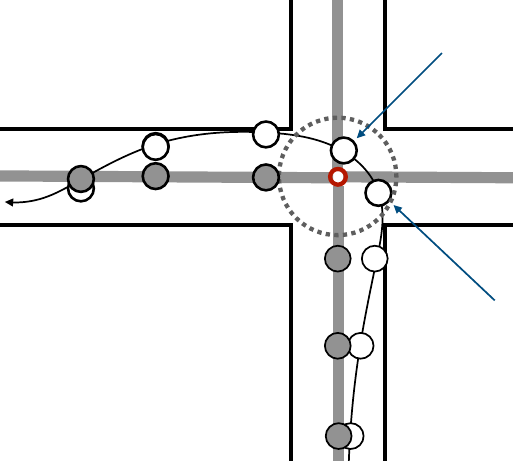} 
\\
  (a)&(b)
  \end{tabular}
  \caption{(a): Blue line: Path from participant P5 measured with A/S. The particle cloud is shown with colors ranging from green (high weight) to blue (low weight). Note that while the bulk of the particles follow the actual path, turning into a corridor, some particles enter nearby rooms through their openings. 
 (b): An illustration of route segment assignment. Four route segments meet at a junction (red circle). The walker's path is shown with white circles, while its projection to the assigned route path is shown with grey circles. When the walker is within a circle with radius $T$ around the junction, no segment assignment is made. }
  \Description{This figure has two panels, marked as (a) and (b). To the left, we see a building's floor plan, with axes marked in meters. There is a blue line ending in a small orange diamond. There are also a multitude of dots, some close the the orange diamond, others spread around. 
 To the right, we see a diagram representing an intersection of corridors. There are 4 grey segments, centered at the intersecting corridors, meeting at a red circle. Three is a dotted circle, with diameter approximately equal to that of the corridors, centered at the junction. There is a thin black line with an arrow at the end, which starts from the bottom through one corridor, then turns left. On this curving line there are 8 white circles. Two of these circles are inside the dotted circle. For the remaining circles, the orthogonal projection of these circles onto the closest grey segment is shown as a grey circle.}
  \label{fig:PF}
\end{figure}

Provision (1) reduces the risk of trajectories going through impenetrable walls. Compactness of the particle set is enforced by (2). This is particularly useful for large open spaces, including long corridors, where the particle set could otherwise expand boundlessly. 
The threshold values $D$ for provision (2) were set (based on trial-and-error experiments) to 5.5~m for A/S localization and to 3.5~m for RoNIN.  Provision (3) reduces the likelihood that a user walking in a corridor be mistakenly localized inside a room. It is important to note that all rooms in the building are considered ``open'' (see Fig.~\ref{fig:PF}~(a)). Even though the selected routes are defined on corridors, walkers could indeed enter a room through an open door (as it happened with participant P7; see Fig.~\ref{fig:R3W}~(d)), and should be tracked in those spaces as well. However, we noted that without provision (3), sizeable amounts of particles would often enter through the open doors (and get temporarily stuck in) one or more rooms when a participant was walking on a  corridor nearby (Fig.~\ref{fig:PF}~(a)). By reducing their weights, we decrease the likelihood that particles in these isolated clusters be selected for resampling.

Resampling also ensures that only angular drift values (and, for A/S, step lengths) that lead to ``legitimate" trajectories are preserved. Explicitly modeling the drift angle was shown in ~\cite{ren2021smartphone} to effectively mitigate the effect of accumulating angular drift, a well-known issue associated with dead-reckoning. For A/S localization, we also found that, through resampling, the particles' step length values tended to coalesce towards a value that was often different from that measured during initial calibration, as discussed later in Sec.~\ref{sec:population}. 

In our iPhone implementation, Particle Filtering updates are performed each time a footstep is detected. An update normally requires between 0.01 to 0.02 seconds to complete.

{\bf\em Initial Orientation Calibration}. For each trial in our experiment, we assumed that participants would start from a known location, and that they initially walked along a known direction. After 6 steps, we found the angle between the reconstructed trajectory (defined, as mentioned earlier, with respect to an arbitrary world reference frame) and the known walking direction, as defined in the floor plan frame.

\subsubsection{Routes and Waypoints}
The walkable area in a building characterized by a network of corridors can be represented by a set of {\em waypoints} (located at the corridors' junctions) and {\em route segments}, which join any two waypoints if there is a traversable straight path between the two. A {\em route} joining any two waypoints is a sequence of interconnected route segments.

At each time $t$, the localization algorithm (whether A/S or RoNIN, followed by Particle Filtering) produces an estimated location $p(t)$ of the user. Based on this location, the app issues notifications to the user as appropriate. However, rather than directly using the 2-D locations $p$, we consider the {\em projected} locations $\bar p(t)$ onto its associated route segments. This approach is justified by the nature of typical buildings with networks of corridors, and simplifies the logic used to produce notifications, which is described in Sec.~\ref{sec:notifications}.

When users walk on a long corridor, far from junctions with other corridors, projecting their location onto the route segment associated with that corridor is a safe operation. More care must be taken when in the proximity of a junction, due to potential localization errors leading to incorrect assignments.
For example, as shown in Fig.~\ref{fig:PF}~(b) (see locations marked by blue arrows), associating the user's location with the closest route segment may lead to incorrect results when the distance of the user to the junction is comparable with the radius of localization uncertainty. 
To deal with such situations, we adopted the following mechanism. A walker whose position is currently associated with a certain route segment, maintains this association until the projected location $\bar p$ is closer than a threshold distance $T$ to a junction with other route segments ($T$ was set to 1.5~m in our experiment). At that point, route segment association becomes ambiguous, and $p$ is no longer associated with any segment. Segment association is resumed when the projection of $p$ onto any of the route segments ending at that junction is at a distance of $T$ (or larger) from the junction. From that point on, the walker is associated with this new segment (see Fig.~\ref{fig:PF}~(b)).

Our wayfinding app determines the shortest  route to a destination using the iOS GameplayKit toolkit\footnote{https://developer.apple.com/documentation/gameplaykit}. Routes are constantly updated as the user moves, and re-routing is computed as necessary. In addition to waypoints, we defined a number of ``landmarks'', whose presence was communicated to the walker when nearby. These landmarks, whose location is shown in Figs.~\ref{fig:R1W}--~\ref{fig:R3W}~(a), are listed in Tab.~\ref{tab:landmarks}.

\subsection{Backtracking}
As mentioned earlier, our Backtracking app is structured in two distinct phases. In the {\em way-in} phase, the app is in charge of recording the route taken by the user (who may be accompanied by a sighted guide, or perhaps may be exploring the environment on their own). During the {\em return} phase, the app is in charge of tracking the path of the walker and matching it with the way-in path. In particular, the system needs to determine when the walker is getting close to a location where a turn was taken during the way-in, in which case it can produce an appropriate notification. 
\subsubsection{Path Reconstruction: Way-in}
Unlike the Wayfinding app, impenetrable wall constraints cannot be used to correct for orientation drift, because the Backtracking app has no knowledge of the floor map of the building. In buildings characterized by networks of corridors, however, it is conceivable that walkers would proceed along relatively straight paths until they turn at a corridor junction. The angle made by two intersecting corridors, for typical buildings, is often equal to $\pm 90^{\circ}$ or to a multiple of $45^{\circ}$. This geometric structural constraint can be leveraged to sidestep orientation drift. Following~\cite{flores2018easy,tsai2021finding}, we represent both way-in and return paths as a sequence of straight segments interleaved with discrete angle turns.

We use the robust turn detector described in~\cite{ren2021smartphone}, which processes azimuth information using a Mixture Kalman Filter~\cite{chen2000mixture}. Although this algorithm was shown to work well even for multiples of $45^{\circ}$ turns, we constrained detection to multiples of $90^{\circ}$ for the purpose of this experiment (this reflects the type of junctions found in the building considered for our tests, which were all of $\pm 90^{\circ}$). The way-in path can be depicted as a 2-D polygonal chain (polyline), where the length of each segment is equal to the number of steps taken in that segment, and consecutive segments have an angle as measured by the turn detector (see e.g. Fig.~\ref{fig:projection}). Steps are detected using the same LSTM network used for A/S localization.

\subsubsection{Path Matching: Return}
 Our strategy for matching the return path with the way-in path is based on the coordination of two different algorithms: {\em projected return sequence} and {\em sequence alignment}.

\begin{figure}
  \centering  
  \begin{tabular}{cc}
    \includegraphics[width=0.35\linewidth]{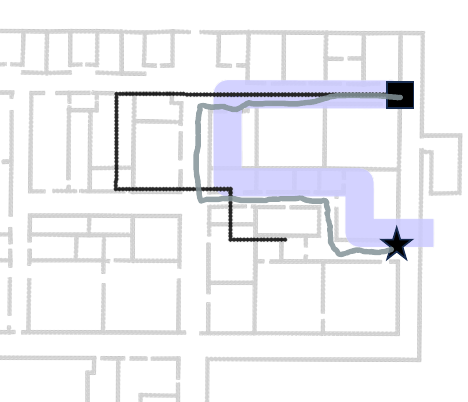} &
  \includegraphics[width=0.35\linewidth]{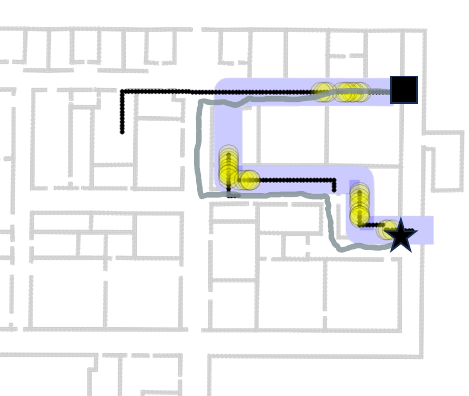} \\
    (a)&(b)
    \end{tabular}
    \caption{Examples of return path matching using projected sequence (a) and hybrid matching (b). The way-in path is shown with a thick purple line, ending at the black square. The length of each segment is given by the number of steps recorded, multiplied by the step length measured during calibration.  The actual path of the participant during the return phase is shown by a gray line.  Projected sequences are shown with black lines. In (b), reliable matches are shown as yellow circles.  Note that in (a), the length of the initial segment appears to be longer than during the way-in, possibly because the walker took shorter steps, or took additional steps while looking for a place where to turn. In (b), the trajectory is corrected as soon as a new reliable match is found.} 
 \Description{This figure has two panels, marked as (a) and (b). Each panel shows a partial floor plan of a building. There is a thick purple polyline, made by a chain of segments connecting at 90 degrees, starting from a black square and ending at a black star. There is a grey thin line, also starting at the black square, and approximately following the purple thick line. In the panel to the left, there is a black poly line, starting from the black square, also made by a chain of segments connecting at 90 degrees. In the panel to the right, on the purple thick line, there are multiple yellow circles. In addition, there are short segments made by black dots, that are either straight or turn by 90 degrees. Each such short segment starts at a yellow circle. } \label{fig:projection}
\end{figure}

{\em Projected Return Sequence}. This algorithm builds a 2-D trajectory as a polyline, as described above for the way-in path. 
This simple algorithm works reasonably well in ``ideal'' conditions but fails if (1) the walker's step length is different between way-in and return (leading to a different number of steps for the same segment; see Fig.~\ref{fig:projection}~(a)); (2) the turn detector fails to recognize a turn (e.g., because taken too slowly) or produces a false positive (due to an irregular trajectory); or (3) the walker does not follow exactly the same route (in reverse) as during way-in, e.g. because they missed a turn then turned around. All of these situations may be expected, especially when someone walks without visual feedback. 

 {\em Sequence Alignment}. As described in~\cite{riehle2012indoor,tsai2021finding}, one may cast the path matching problem as one of (sub)sequence alignment. Assume that the sequence of way-in measurements has been reversed, which is convenient since the route is being backtracked.  At each time during return, we determine the initial way-in subsequence of measurements that best matches the current sequence of return measurements. In symbols: given the (reversed) way-in sequence of measurements (observations) $W = (o_{w}(1),\dots,o_w(N))$,  and the current sequence of return measurements $R = (o_{r}(1),\dots,o_r(J))$, the goal is to find a sequence of indices $i_1,\dots,i_J$ such that $(o_{w}(i_1),\dots,o_w(i_J))$ best matches $R$ under an appropriate criterion. 
 For real-time guidance, we are interested in the last matching point $i_J$:  we will assume that the walker at the current return time index $J$ is in the same location they were at time index $i_J$ during way-in. 
 Standard dynamic programming approaches (e.g. Dynamic Time Warping~\cite{listgarten2004multiple,riehle2012indoor}) can then be used to find an optimal match.

 The measurements we consider are magnetic field vectors and turns detected. Step detection is also considered implicitly: for both way-in and return, the sequences of time indices are defined such that there are three regularly spaced time intervals between two consecutive detected steps. We found that this choice gives enough spatial granularity for magnetic field matching, while ensuring parsimonious sampling (e.g., no samples are recorded when the user is stationary).

 It is well known that the magnetic field recorded in different positions within a building is not uniform, due to reasons such as the presence of large metallic objects and magnetic field generating appliances. ``Magnetic signatures'' can thus be attached to specific locations, enabling sophisticated localization mechanisms \cite{Shu2019,Kuang2018,Storms2010,chen2020meshmap}(e.g., IndoorAtlas' magnetic positioning\footnote{https://www.indooratlas.com} \cite{vstancel2021indoor}.)  The magnetic field at a certain location can be measured by a 3-axis magnetometer, of the type embedded in any modern smartphone. (Prior calibration of the magnetometers is necessary for good results.) 
 From the measured 3-D magnetic field vector we derive a 2-D vector that is invariant to the orientation of the phone as described in~\cite{li2012feasible,fan2017accurate}.
 
 Given the measurements, one can create a directed graph with nodes indexed as $(i,j)$, where $i$ is a way-in time index, and $j$ is a return time index. (Note that the graph is constantly expanded as the walker progresses along the return path.) A node $(i,j)$ in the graph has only three edges, to $(i+1,j)$, $(i+1,j+1)$, and $(i,j+1)$, respectively. This is consistent with the assumption that the walker is normally moving in the same direction as in the (reversed) way-in path, but possibly with a different step length (resulting in steps detected in either phase that cannot be matched in the other phase, which are accounted for by the edges to $(i+1,j)$ and $(i,j+1)$ ). The edges to $(i+1,j)$ and $(i,j+1)$ carry a non-null cost, while node costs are defined as a function of the  magnetic field vector discrepancy (measured as the Euclidean difference between the recorded vectors)  between the measurements $o_w(i)$ and $o_r(j)$. 
 Sequence alignment is obtained by finding the minimum cost path starting from $(0,0)$ and arriving at a node $(i,J)$. 
 
 Unfortunately, this simple approach, originally proposed in~\cite{riehle2012indoor}, did not produce satisfactory results in our preliminary tests, especially if the return path is not perfectly identical to the way-in path. Indeed, we found that, in some cases,  the magnetic field can vary rather dramatically when moving across the width of a building corridor\cite{subbu2013locateme}.


 An improvement to pure magnetic-based alignment can be obtained by considering the turns taken by the walker, which, in an ideal case, should be matched between way-in and return. We use the mechanism described in~\cite{tsai2021finding}, which considers the difference in walking direction between way-in and return by structuring the graph as a sequence of layered planar graphs, where each layer represents a possible orientation discrepancy between way-in and return.

The minimum cost path in the graph is recomputed at each new return sample. We use the incremental Dynamic Time Warping (iDTW) proposed by Riehle et al.~\cite{riehle2012indoor}, which uses a sliding window defined around the endpoint of the previously found optimal path (we set the window size equal to 200 samples). Although this algorithm produces a suboptimal solution, it represents a good compromise between precision and computational cost.

 {\em Hybrid Matching Strategy}.  The sequence alignment method described above was shown to work well even for moderately complex situations. However, since the optimal path is recomputed at each iteration, the estimated user location at the current time may become ``jittery''.
This may cause instability when producing guidance notifications, complicating the task of deciding when exactly to notify the walker of an upcoming turn.

 We propose a simple solution to this problem,  based on the notion of {\em last reliable match }. We use local properties of the current minimum cost graph path to decide, at the current time $J$, whether the match $(i_J,J)$ can be considered ``reliable'', meaning that it is likely to be preserved even after later observations are recorded. 
 In practice, we look at the terminal part (the last 21 samples) of the minimum cost path in the graph ending at $(i_J,J)$, and judge this last match to be reliable if this path segment is well approximated by a line with unitary slope. This simple algorithm was shown to work well in most cases.
 
 When a reliable match is detected, a projected return sequence is initialized and continuously updated, as discussed earlier, until a new reliable match is found. This projected sequence starts from the last reliable match, with an initial direction equal to the current walker orientation (as defined by the sequence of  turns taken during return). Guidance notifications are produced based on the walker's location identified using this projected sequence. As soon as a new reliable match is detected, the projected sequence is re-initialized at that match (see Fig.~\ref{fig:projection}~(b)). Examples of reliable matches  and projected paths are shown later in Sec.~\ref{sec:TGPB}.

\subsubsection{Way-in Path Simplification}
In some cases, the way-in path contains redundant turns or loops. This may happen when, for example, if the walker took a short detour (perhaps because they were unclear about the route to take), or if they followed a zig-zag course instead of walking straight. 
It would be desirable to remove these redundant features before backtracking. In our experiment, we implemented a very simple ``path simplification'' algorithm for the way-in path. In short, we define a conservative radius of uncertainty $D$ equal to 7 steps. We then merge together nearby parallel edges that are at a distance of $D$ or less from each other. We do the same with turn locations (vertices). Then, we re-create a simplified path as a shortest length polyline that goes through the remaining edges and vertices. 

\begin{figure}
  \centering  
  \begin{tabular}{c}
    \includegraphics[width=0.8\linewidth]{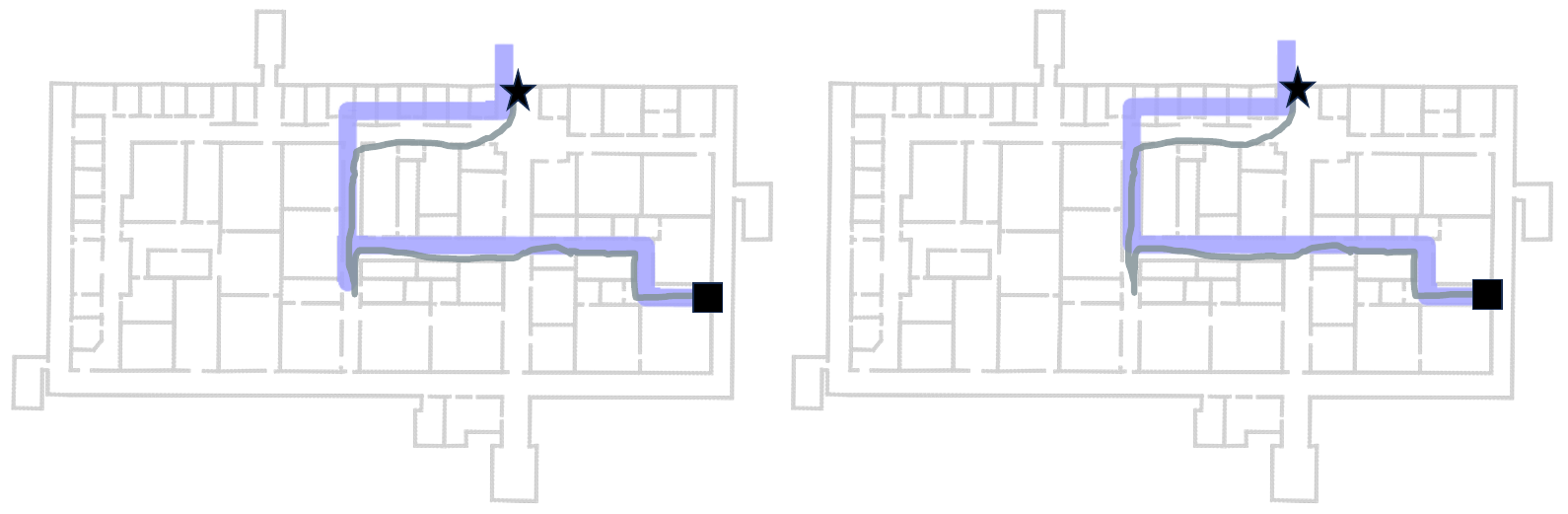} \\
    (a)\\
  \includegraphics[width=0.8\linewidth]{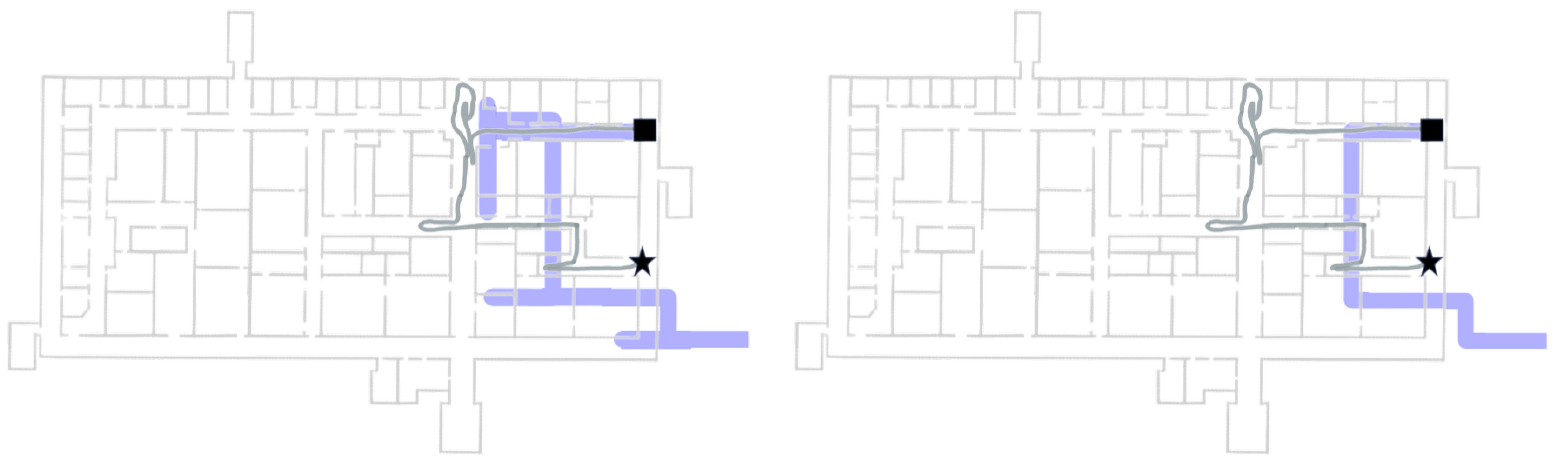} \\
    (b)
    \end{tabular}
  \caption{Examples of successful (a) and unsuccessful (b) way-in path optimization. The original reconstructed way-in path is shown in the left panels with a thick purple line, ending at the black square, along with the approximate actual path taken by the walker (as measured from the video), shown with a gray line. The optimized way-in paths are shown in the right panels. }
   \Description{This figure has four panels, organized in two rows of two (marked as (a) and (b)). Each panel shows a partial floor plan of a building. There is a thick purple polyline, made by a chain of segments connecting at 90 degrees, starting from a black square and ending at a black star. There is a grey thin line, also starting at the black square, and ending at a gray star.}\label{fig:optimization}
\end{figure}

An example of successful way-in path simplification is shown in Fig.~\ref{fig:optimization}~(a). Note that the original path (shown to the left) contains a short detour with a $180^{\circ}$ turn, which was due to the walker originally missing a turn. This spurious piece was correctly removed by the simplification algorithm.  Fig.~\ref{fig:optimization}~(b) shows a case with a complex way-in path, containing multiple loops. Although our algorithm was able to remove these loops, and the resulting simplified path maintained the correct path geometry, its first segment turned out to be substantially shorter than in the original path, which led to an unsuccessful backtracking trial.

\subsection{User Interface}\label{sec:UI}

We list in the following the general principles that guided our interface design for the two apps.

{\bf \em 1. Consistency.} Even though the two apps utilize different localization technologies, the user interface is almost identical between the two. Both apps ultimately provide directions towards a destination in similar environments (networks of corridors), thus it is only natural that they should afford similar user experiences. The small differences (highlighted in Sec.~\ref{sec:notifications}) are a consequence of the different prior knowledge the apps can rely on. 

\begin{figure}
  \centering  
  \begin{tabular}{cc}
    \includegraphics[width=0.2\linewidth]{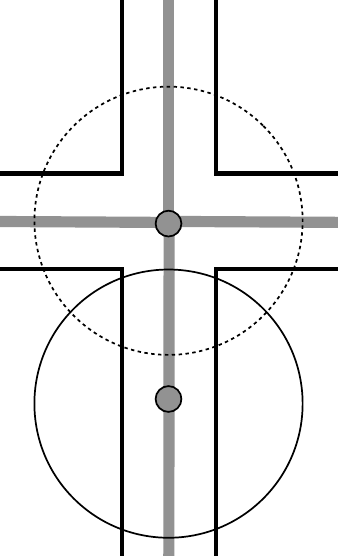} &
  \includegraphics[width=0.75\linewidth]{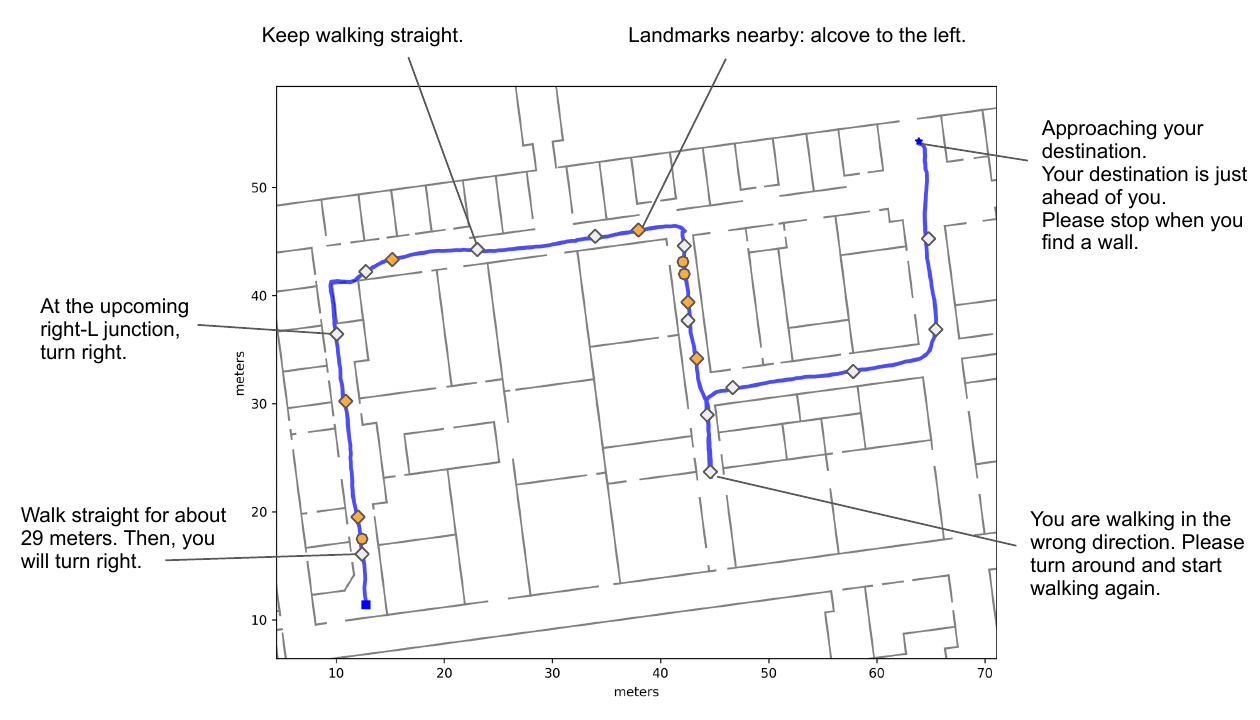} \\
    (a)&(b)
    \end{tabular}
  \caption{(a): An hypothetical junction, with the walker's location (as estimated by the app) shown by either gray-filled circle. The larger circles represented the expected radius of location uncertainty. If the walker is located by the app at the junction center (top circle), their actual location can be anywhere in the dashed circle of uncertainty (including before or after the junction). If the app locates the walker at the lower circle,  the actual location of the walker is certainly before the junction. (b): An example of notifications produced during route traversal (participant P5, route R1W). Directional notifications are shown as symbols filled in gray, while landmark notifications are shown as filled in orange. For the symbols shown as diamonds, an actual notification was produced. For the symbols shown as circles, a notification was not produced as it would have interrupted an ongoing notification. For a few selected notifications, the content of the speech produced is also shown.}
   \Description{This figure has two panels. In the left panel, marked as (a), we see a floor plan with two corridors intersecting. There is a thick gray line in the middle of each corridor. At the intersection of the two gray lines is a gray-filled small circle with a black edge, which is encircled by a co-centered circle with a much larger radius. This larger circle is shown with a dashed black line. Below the grey-filled circle, on one of the grey lines, is another gray-filled circle, also encircled by a co-centered circle with a much larger radius (shown with a continuous black line). This large circle touches the beginning of the junction in the topmost part. In the right panel, marked as (b), we see a floor plan of a building, with a blue line, and a few symbols on this line. Some of these symbols are circles, and some are diamonds. Some are filled in grey, some in orange. For 6 of these symbols, there is a callout with some short text. This is the text content of these callouts: "Walk straight for about 29 meters. Then, you will turn right."; "At the upcoming 
right-L junction, turn right."; "Keep walking straight."; "Landmarks nearby: alcove to the left."; "Approaching your destination. Your destination is just ahead of you. Please stop when you find a wall."; "You are walking in the wrong direction. Please turn around and start walking again.".}\label{fig:inaccuracies}
\end{figure}

{\bf \em 2. Robustness to localization inaccuracy.}
The localization system of both apps relies on dead reckoning from a smartphone's inertial sensors. In spite of mitigation techniques (Particle Filtering for Wayfinding, sequence alignment for Backtracking), localization errors of up to 2-3 meters should be expected. Neglecting these potential errors when issuing notifications may lead to catastrophic results. For example, suppose that the app issues a notification shortly before it  locates the walker at a junction. 
As shown in Fig.~\ref{fig:inaccuracies}~(a), the walker's real position may happen to be at some distance before or after the junction. In either case, if the walker were to turn, they would face a wall. What's worse, they would have no way to  know whether  to search for the junction  to their right or to their left. 

To account for the expected localization inaccuracy, we made the following design choice: the presence of a junction where a turn should be taken is announced with substantial advance notice, enough to ensure that, with high likelihood, the walker has not yet reached the junction.  Although conceptually very simple, it was not obvious {\em a priori} that this approach would be viable and acceptable. The notification is given at a variable distance from the junction (depending on the current localization error). Users are thus in charge of searching for the next available opening upon hearing the notification, which involves active exploration with the long cane or with the dog guide. An important goal of our experiments was to verify whether our participants would be successful in this task, and,  importantly, whether they would find this strategy acceptable.


Based on preliminary tests, we decided that a notification should be issued when the walker is located by the app at 7 meters from the next waypoint (for the Wayfinding app), or 14 steps from the next turn point in the reversed way-in path (for the Backtracking app). While this  distance may seem large, it is important to observe that our participants often kept walking while listening to the notification; by the time the notification is completed and they have processed it, they may have already advanced by 2-3 meters.

{\bf \em 3.  Minimal disruption.} Walking without visual feedback in an unknown space requires a good deal of concentration. 
It is important that the notifications generated during walking be designed to minimize the required cognitive load for processing, and that they are not distracting. Our apps issue notifications in the form of short synthetic speech sentences (Sec.~\ref{sec:notifications}). For the most time-critical notifications (e.g., announcing an upcoming turn), the notification is preceded by a short chime, and accompanied by a short vibration on the Watch. The user can, at any time, have the last notification repeated, or hear a description of the remaining route. 
In addition, walkers heard a short sound at each footstep. Originally implemented for debugging purposes, we decided to leave this feature in the apps, giving each participant the option to turn it off. 

{\bf \em 4.  Watch-based control.} In our experiment, we asked our participants to keep the phone in a pocket throughout the trials. We also asked our participants to wear an Apple Watch, that they used to control the app. 
These are the gestures considered for the Watch:
\begin{itemize}
\item{\em Before the beginning of a trial:} Participants were asked to select a specific route (e.g., ``Path number 2'') from a list. The list could be traversed in both directions through left and right swipes on the Watch's face, with the name of the current item in the list read by VoiceOver (see Fig.~\ref{fig:pictures}~(b)). 
\item{\em To start the app:} Participants were asked to start the either app by rotating the Watch's crown, and keep rotating (in either direction) until they heard a ``ding'' sound from the Watch, signifying that the app had started. At that time, participants heard the notification ``Please start walking.''
\item{\em To hear the last notification again:} At any time during a trial, participants could do a right swipe on the Watch's face to hear the last notification issued by the app. 
\item{\em To hear a route description:}  At any time during a trial, participants could do a left swipe on the Watch's face to hear a description (in terms of route segments and turns) of the remaining route, from their current location till destination. 
\item{\em To stop the app:} Upon arrival at destination, or if the trial was aborted, participants were asked to again rotate the Watch's crown to stop the app. 
\end{itemize}

\subsubsection{Notifications}\label{sec:notifications}
Notifications are produced based on the current route and the location of the user. More specifically, we consider the distance of the user's location
(projected on the associated route segment for the Wayfinding app, or to the associated way-in path segment for the Backtraking app) to the next waypoint in the route or to nearby landmarks (Wayfinding), or to the next turn in the reversed way-in  (Backtracking).

Our apps support up to four distinct types of notifications. Note that, in case of conflict (a notification triggered while a prior notification was being delivered), the ongoing notification is never interrupted, except in case the new notification is of type (1) as listed below (alerting the user of an upcoming turn). This type of notification is considered time-critical, and therefore takes priority. Also note that the same notification is never repeated, except for notifications of type (3).
\begin{enumerate}
\item{\em When the next waypoint (Wayfinding) or turn (Backtracking) is at less than 7 meters.} If a turn should be taken at that waypoint, the notification is:  ``At the upcoming [X,left/right L,T] junction, turn [left, right].'' Note that the junction morphology (X, L, T) requires access to a floor plan, and for this reason, this detail is not announced by the Backtracking app. 

If the walker is not supposed to turn at the next waypoint, the message is simply: ``Keep walking straight.'' This last notification is meant to reduce the risk that the walker, having found a junction, may mistakenly turn there. Note that this information is not available for the Backtracking app, which is only aware of the turns taken during the way-in. 

If the next waypoint is the destination (final) waypoint, or the end of the way-in route, the notification is: ``Approaching your destination, Your destination is just ahead of you''. If the route ends at a wall or at a closed door, this notification is followed by: ``Please stop when you find a wall''. 

\item{\em Upon entering a new route segment (Wayfinding) or when the last reliable match is updated to a new segment(Backtracking).} Notification: ``Walk straight for about XX [meters/feet/steps]. Then, you will turn [left, right].'' The last sentence was removed for the Backtracking app after participant P4, for reasons that will be made clear later. Some route segments may contain protruding obstacles (e.g., a steel cabinet) on either side. In this case, the notification is preceded by ``Please keep to the [left/right]''.

\item{\em When the user has been walking in the wrong direction on a route segment for at least 4.5 meters (Wayfinding), or is at a distance of more than 16 steps from the way-in path (Backtracking).} ``You are walking in the wrong direction. Please turn around and start walking again.'' This length (4.5~m) was chosen through trial-and-error in initial experiments, and is consistent with the expected radius of uncertainty of localization. This notification is repeated if they continue walking in the wrong direction for another 4.5 meters.


\item {\em When the next landmark is at less than 2 meters (Wayfinding only).} ``Landmark nearby. [Landmark name] to the [left/right].'' Note that a shorter threshold distance than for approaching junctions is used here. This is justified by the fact that, while a junction needs to be announced {\em before} the walker has reached it, advance notice is less critical for announcing nearby ``landmarks''.

\end{enumerate}

\section{Experiment}\label{sec:Experiment}

\begin{table}[ht]
\caption{Characteristics of the participants in our study. For blindness onset, `B' indicates `since birth', while `L' indicates `later in life'. Step length (final) represents the average step length value over all particles at the end of the trials (averaged over the three Wayfinding trials). Note that the Particle Filters included step length as a state only beginning from P4. }\label{tab:participants}
\centering
\begin{tabular}{c|c|c|c|c|c|c|c|c|}
&\multirow{ 2}{*}{Gender}&\multirow{ 2}{*}{Age}&Blindness&\multirow{ 2}{*}{Mobility aid}&Step length (cm)&Step length (cm)&RoNIN&Preferred \\
&&&onset&&(calibration)&(final)&multiplier&units\\
\hline
P1&F&73&L&Dog&48&--&0.96&Steps\\
\hline
P2&M&69&B&Cane&51&--&1.08&Feet\\
\hline
P3&M&53&B&Cane&54&--&1.14&Feet\\
\hline
P4&F&69&B&Cane&51&44&1.0&Feet\\
\hline
P5&M&75&L&Cane&44&41&1.21&Meters\\
\hline
P6&F&76&L&Cane&40&40&1.11&Steps\\
\hline
P7&F&72&L&Dog&63&58&1.08&Feet\\
\hline
\end{tabular}

\end{table}

\subsection{Population}\label{sec:population}
We recruited 7  participants for this experiment. The participants' characteristics are summarized in Tab.~\ref{tab:participants}. All participants were blind, with at most some residual light perception.
All participants were experienced independent walkers, although P6 had recently switched to using a long cane after many years of walking with a dog guide, and was still re-learning to use the cane. P5  had a hearing impairment and used hearing aids. All participants were iPhone users, except for P7, who used a cell phone with a physical keypad. Only P1 regularly wore a smartwatch (Apple Watch). Tab.~\ref{tab:participants} also shows the average step length values measured during the initial calibration, as well as the average step length from the particles at the end of the trials (remember that step length was added as a state in the Particle Filter only beginning with P4). Note that the step length measured in the initial calibration is always larger than or equal to that found by the Particle Filter. This suggests that participants may have walked with a different step length during calibration (which was conducted in a ``safe'', straight corridor stretch) and during the actual test, where they might have felt less confident and therefore walked with shorter steps. 

\begin{figure}
  \centering  
  \begin{tabular}{cc}
    \includegraphics[height=5cm]{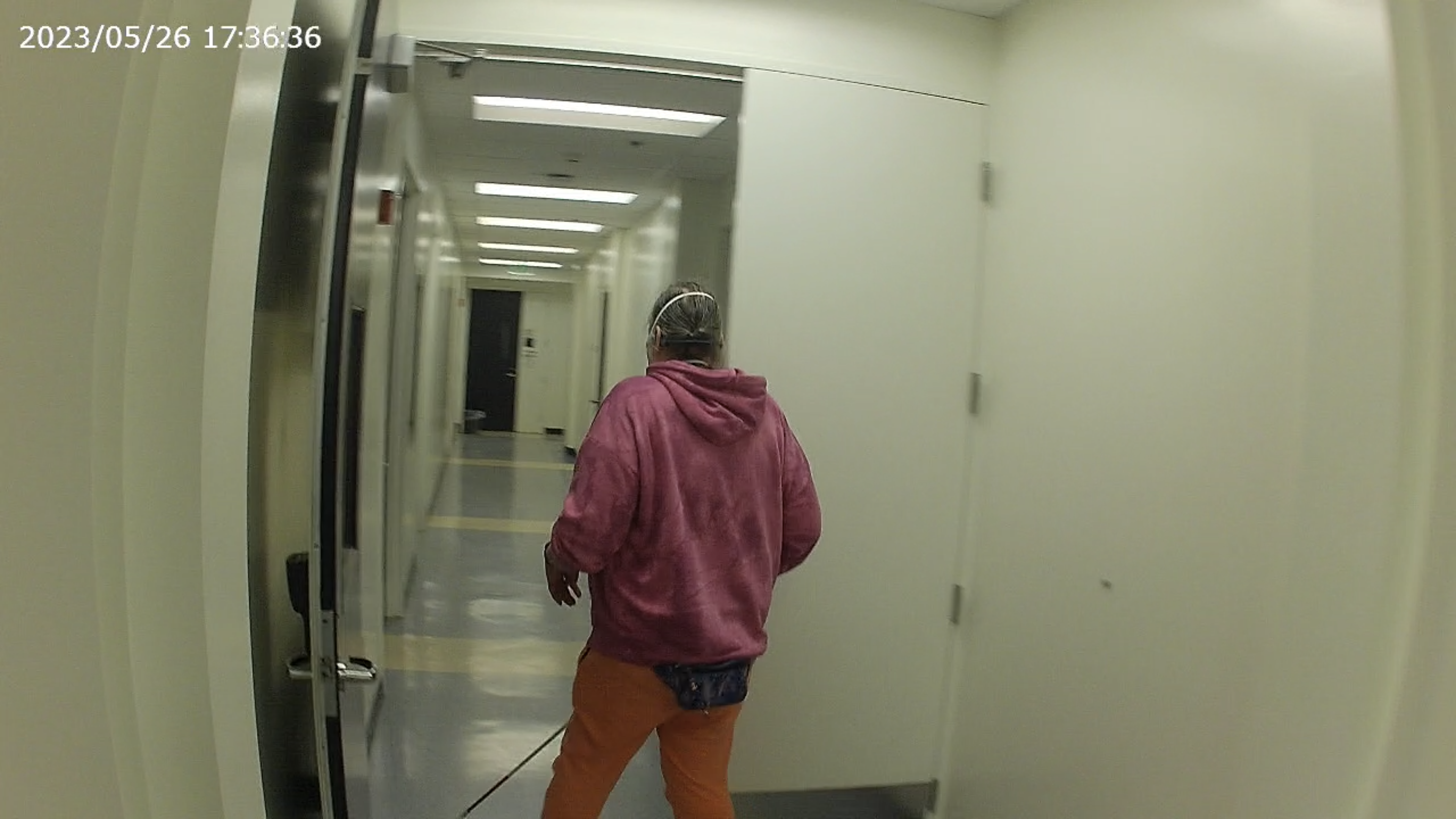} &
  \includegraphics[height=5cm]{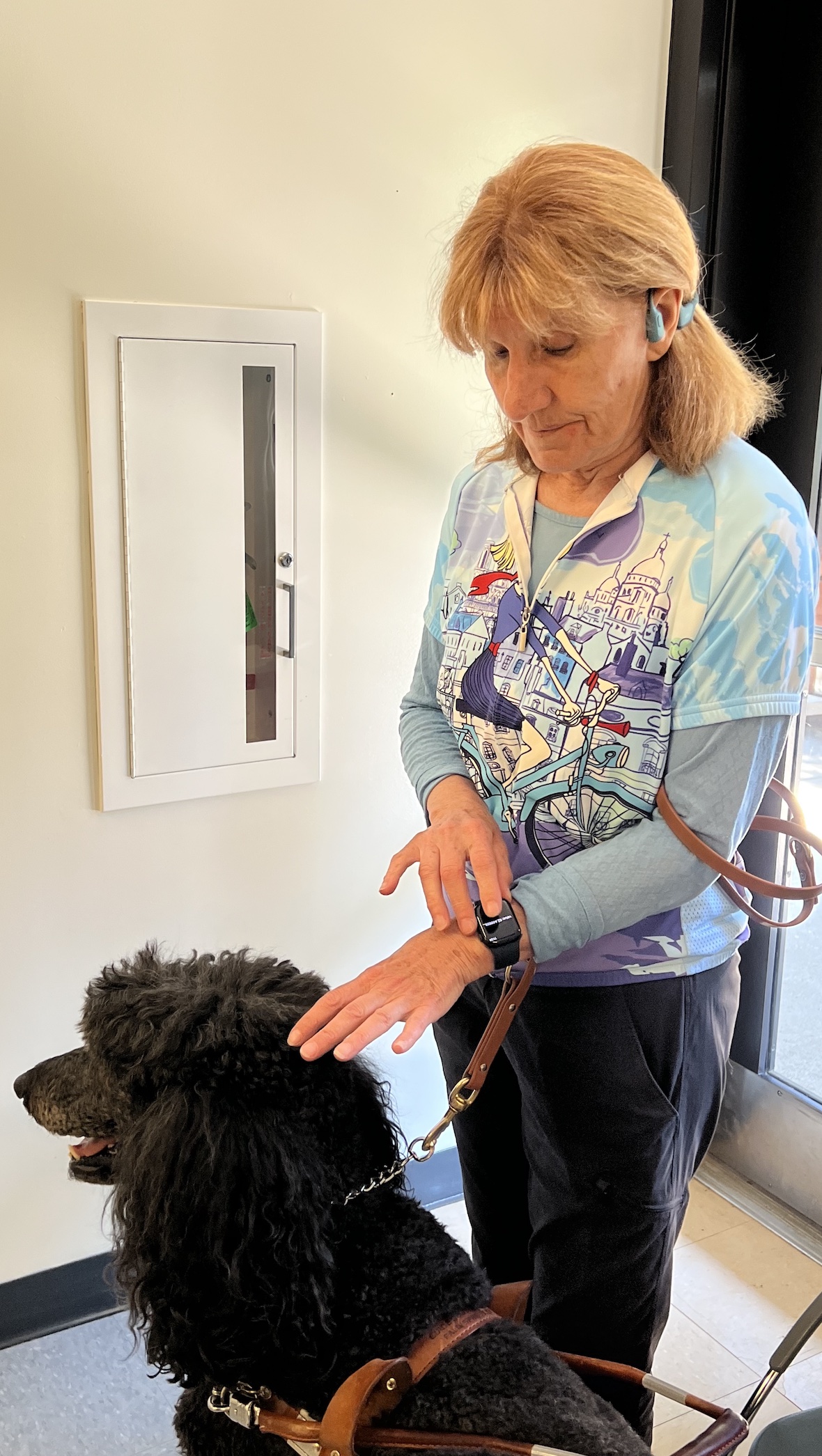} \\   
  (a)&(b)\\
  \includegraphics[height=5cm]{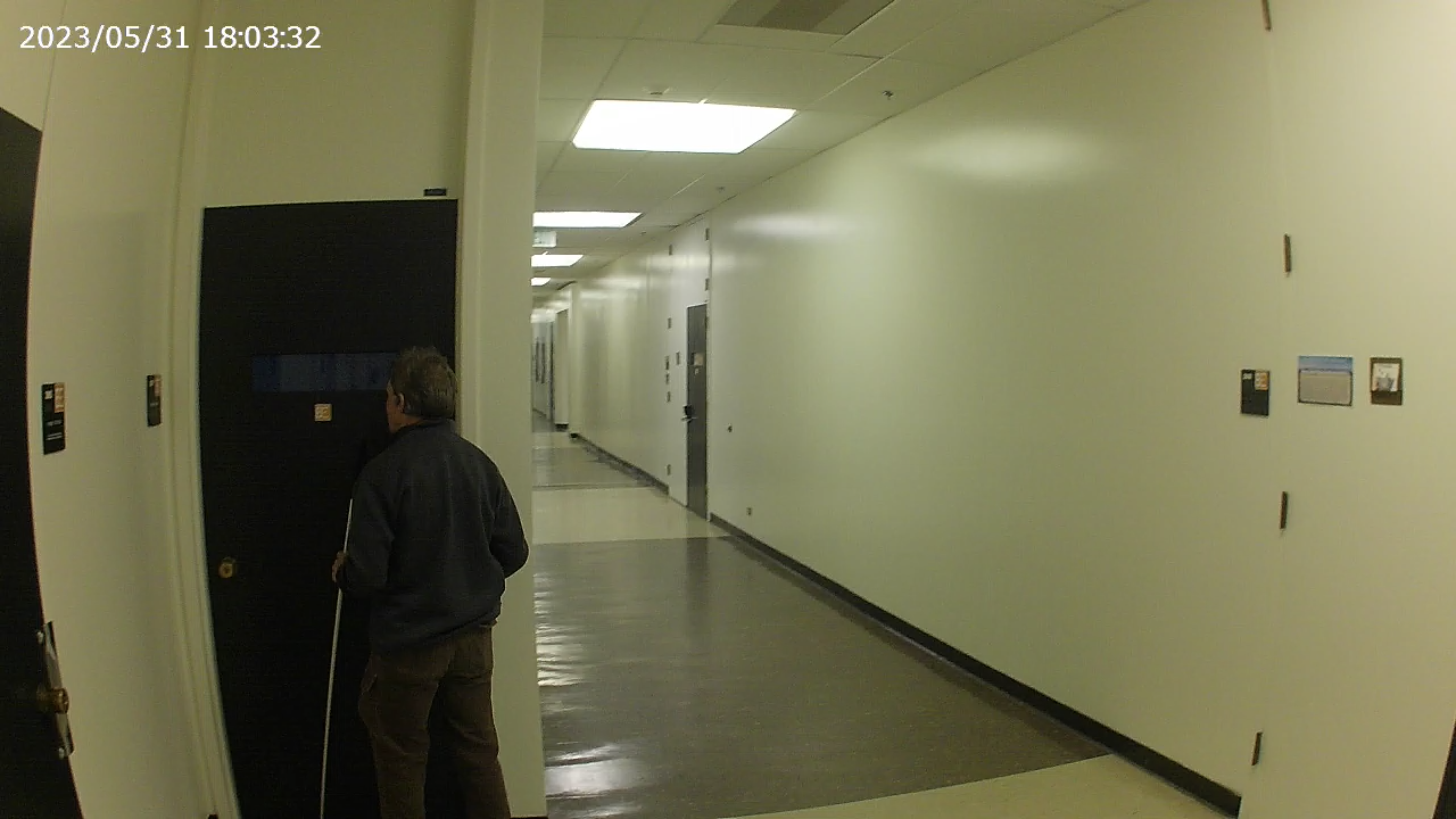} &
  \includegraphics[height=5cm]{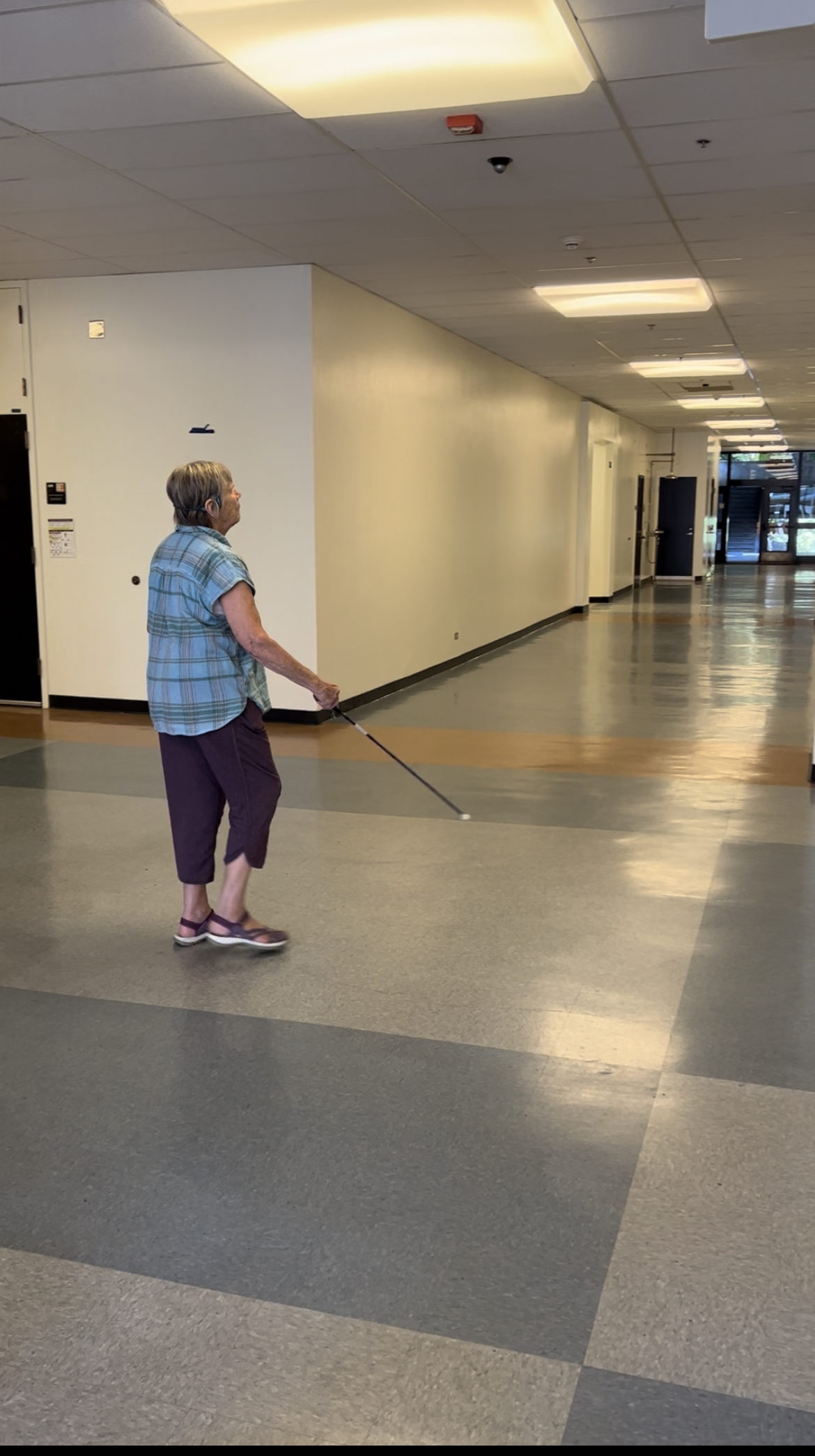} \\   (c)&(d)\\
  \includegraphics[height=5cm]{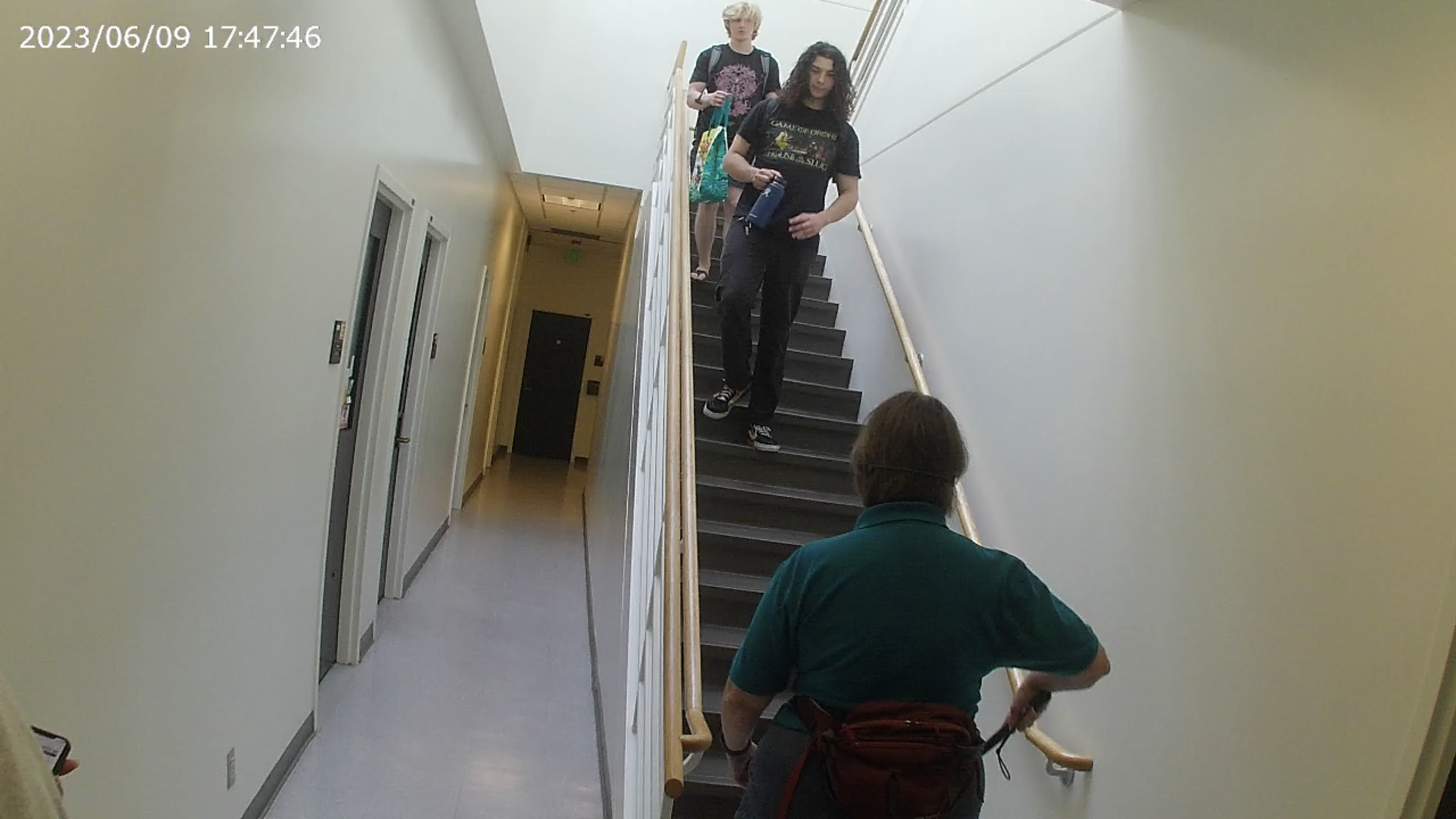} &
  \includegraphics[height=5cm]{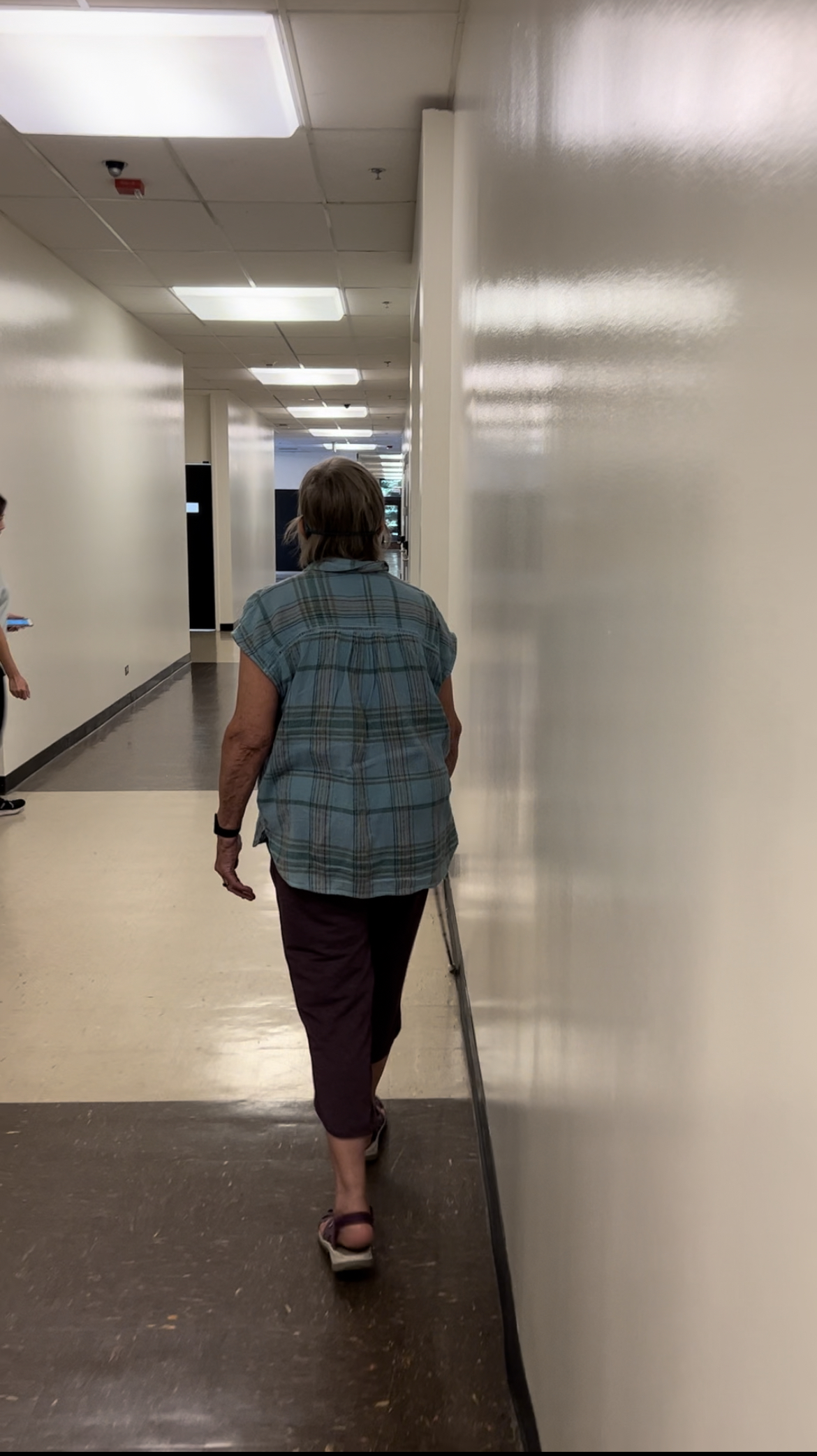}\\
  (e)&(f)
    \end{tabular}
    \caption{Pictures of our participants during the trials. (a): P2 in R2W. (b) P7 at the beginning of R3W. (c) P3 in R1W. (d) P6 in R2W. (e) P4 in R1W. (f) P6 in R2W.} \Description{This figure has 6 pictures. Each picture shows a participant during the trials. All participants use a long cane, except for one picture (marked (b)) in which the participant uses a dog guide.}\label{fig:pictures}
\end{figure}

\subsection{Procedure}
\subsubsection{Environment}\label{sec:env}
\begin{figure}[h]
  \centering
  \begin{tabular}{cc}
  \includegraphics[width=0.55\linewidth]{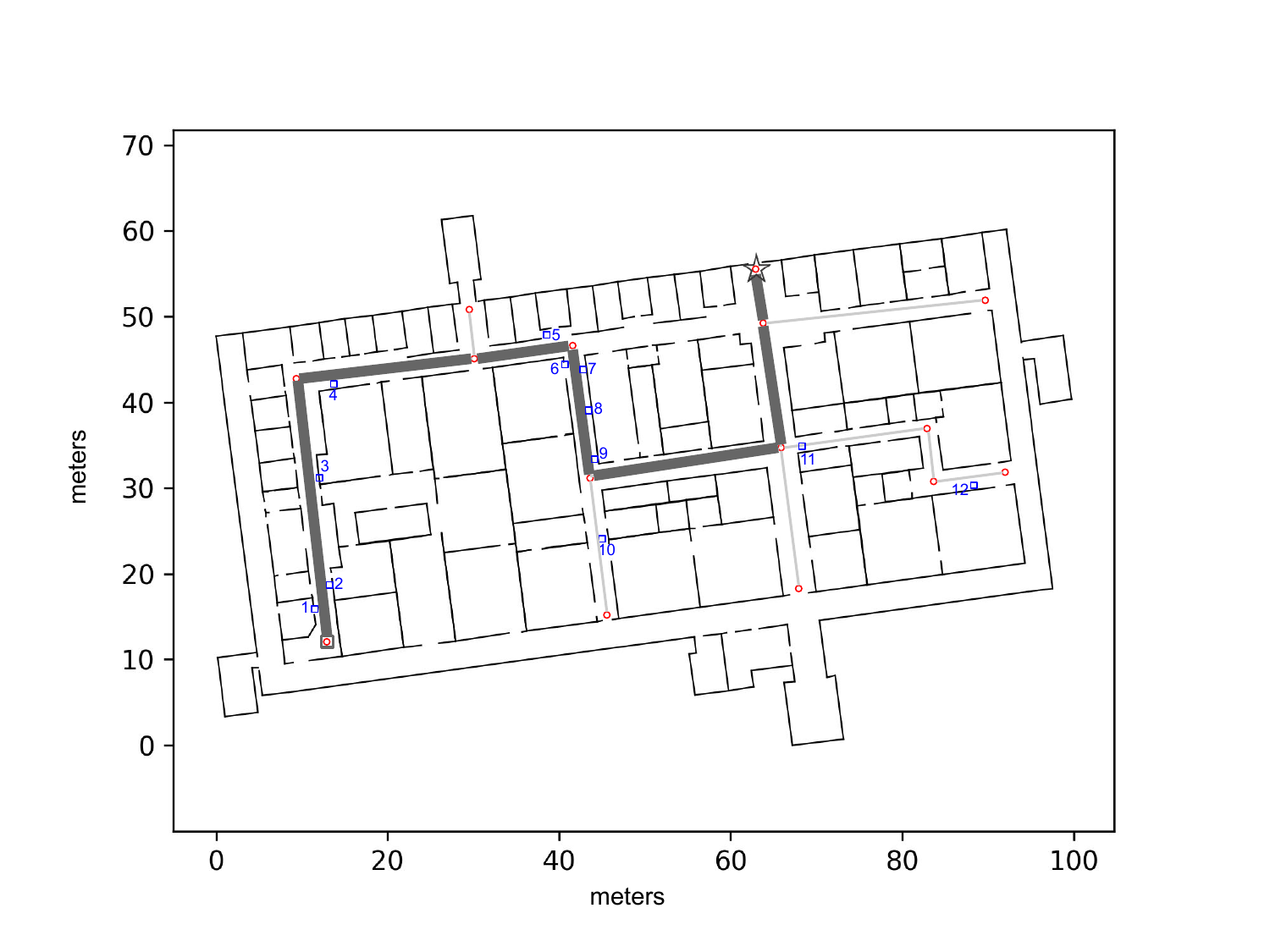} &
  \includegraphics[width=0.45\linewidth]{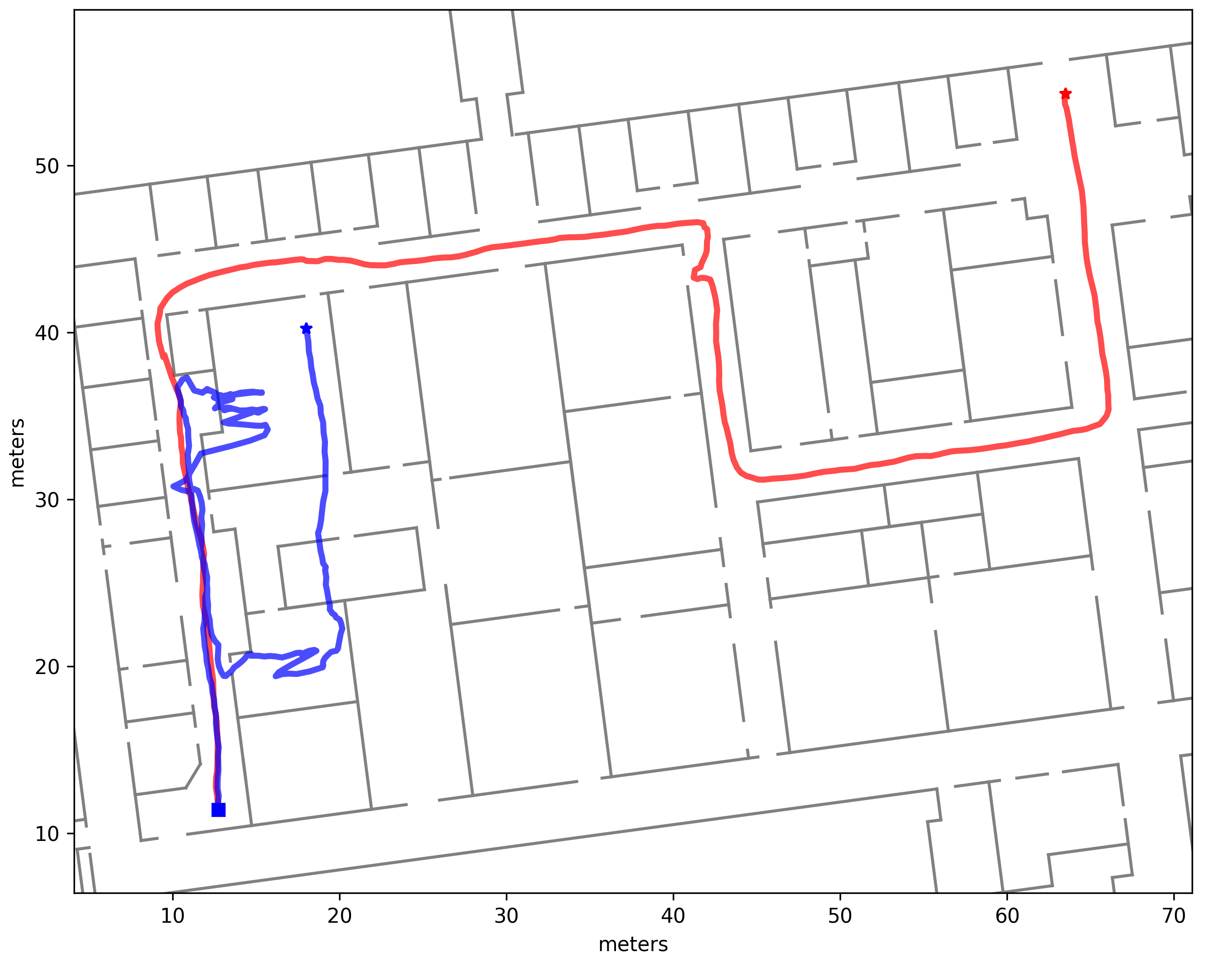}\\
  (a)&(b)\\
  \includegraphics[width=0.45\linewidth]{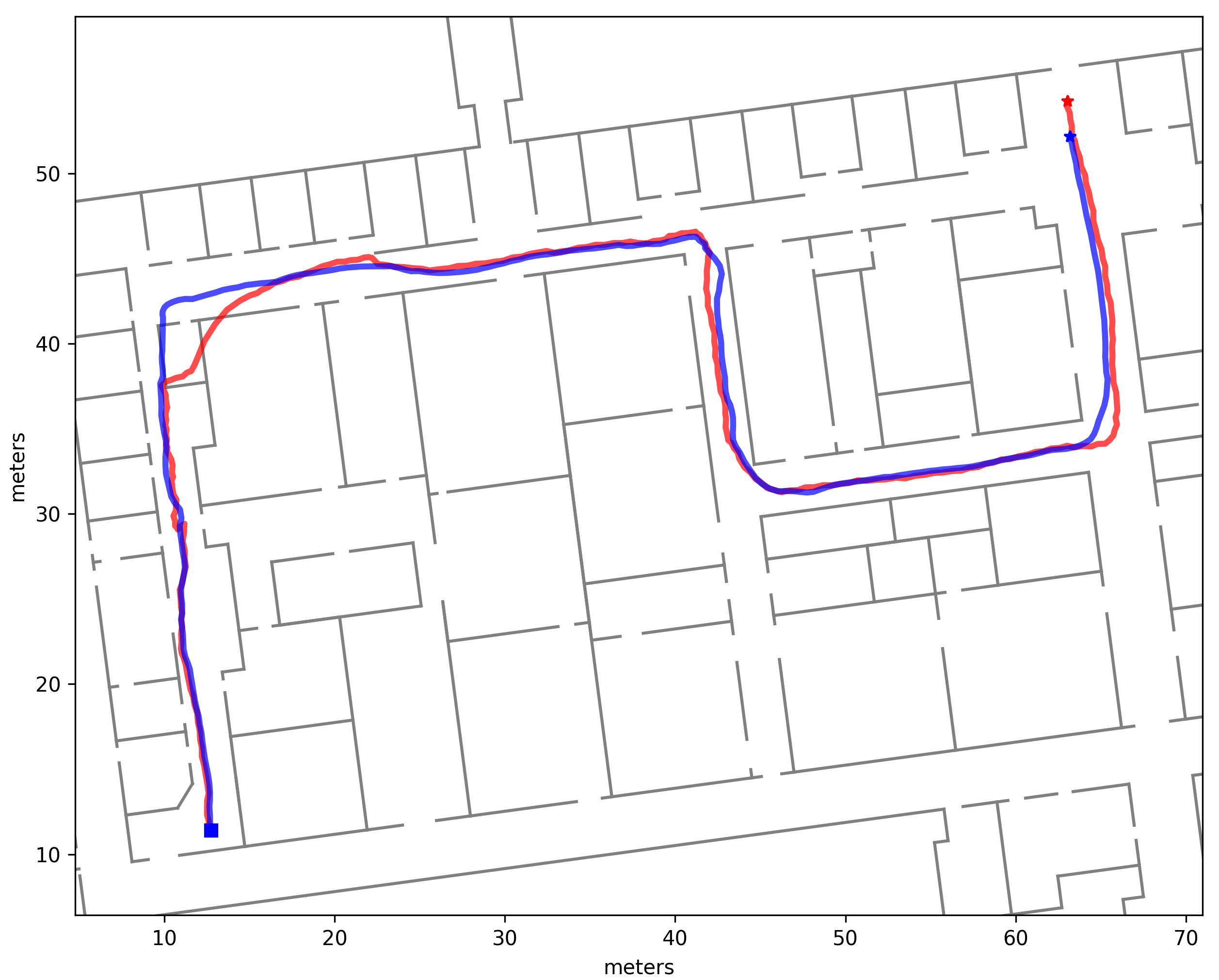}&
  \includegraphics[width=0.45\linewidth]{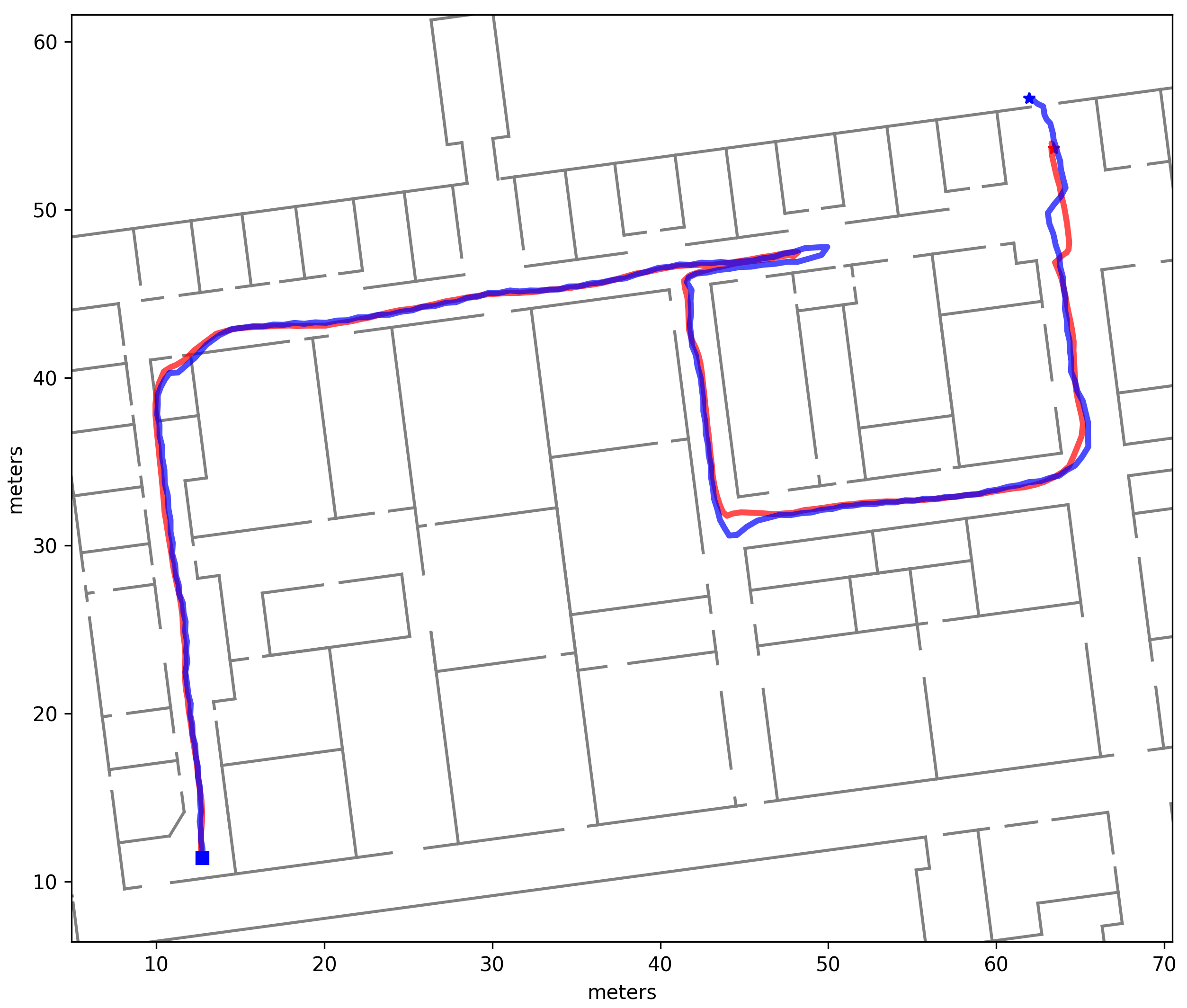}\\
  (c)&(d)
  \end{tabular}
  \caption{Route R1W. (a): Floor plan of the considered building. Waypoints are shown in red, traversability graph edges are shown in gray. The start and end waypoints are marked with a square and a star, respectively. The shortest path is shown with a thick dark gray line. Landmarks are shown in blue and enumerated (see Tab.~\ref{tab:landmarks} for landmark listing.) (b)--(d): recorded paths using A/S (blue line) and RoNIN (red line). (b): P1. (c): P4. (d): P7.}
  \Description{This figure has four panels. The top left panel shows a floor plan, with a number of small red circles at corridor junctions. Several pairs of these are joined by grey edges. Along the corridors, there are some numbered blue dots. A dashed grey line, overlapping a sequence of grey edges, joins a red circle framed by a square, with a red circle framed by a star.}
  \label{fig:R1W}
\end{figure}

\begin{figure}
  \centering
  \begin{tabular}{cc}
  \includegraphics[width=0.55\linewidth]{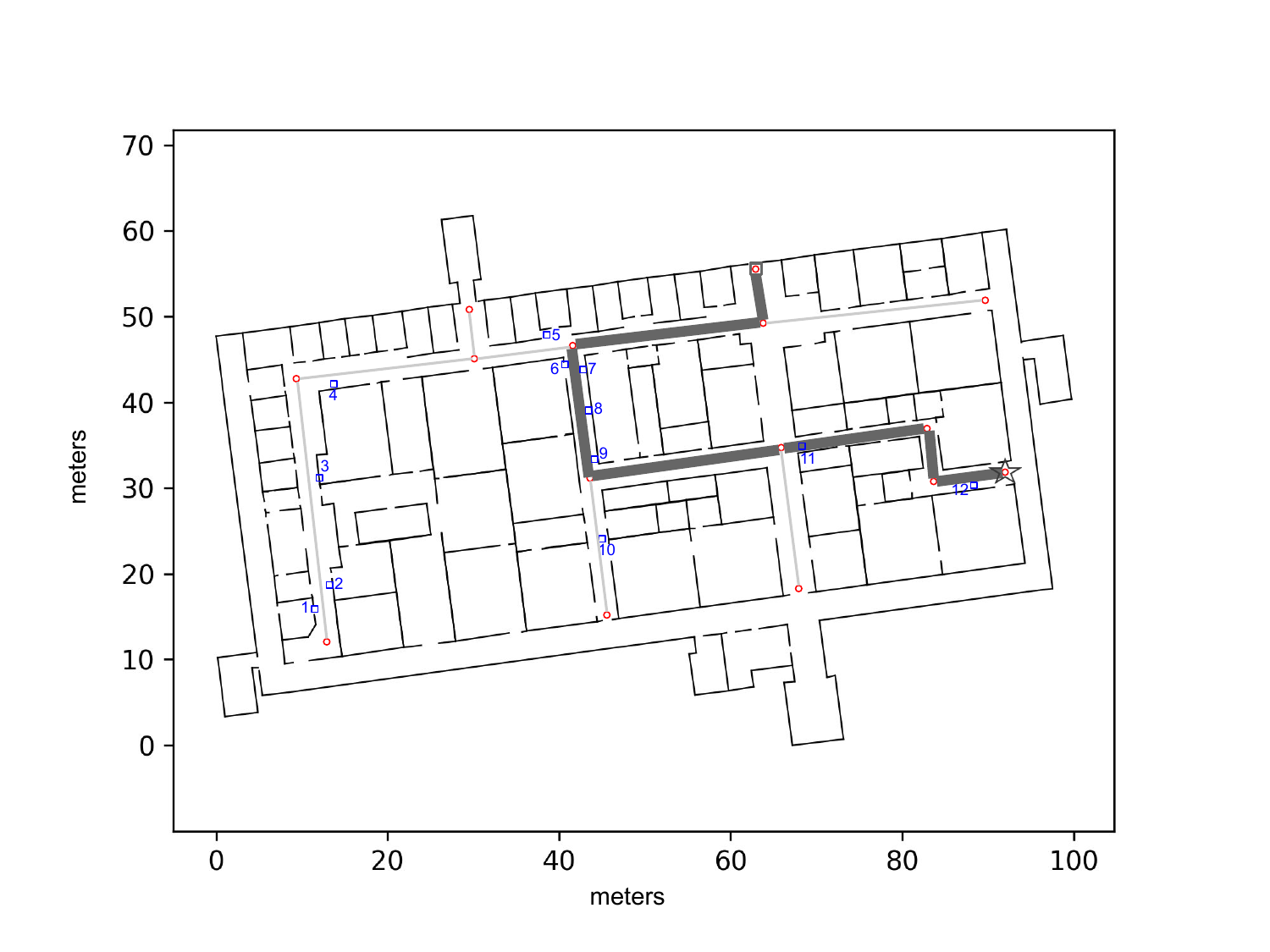} &
  \includegraphics[width=0.45\linewidth]{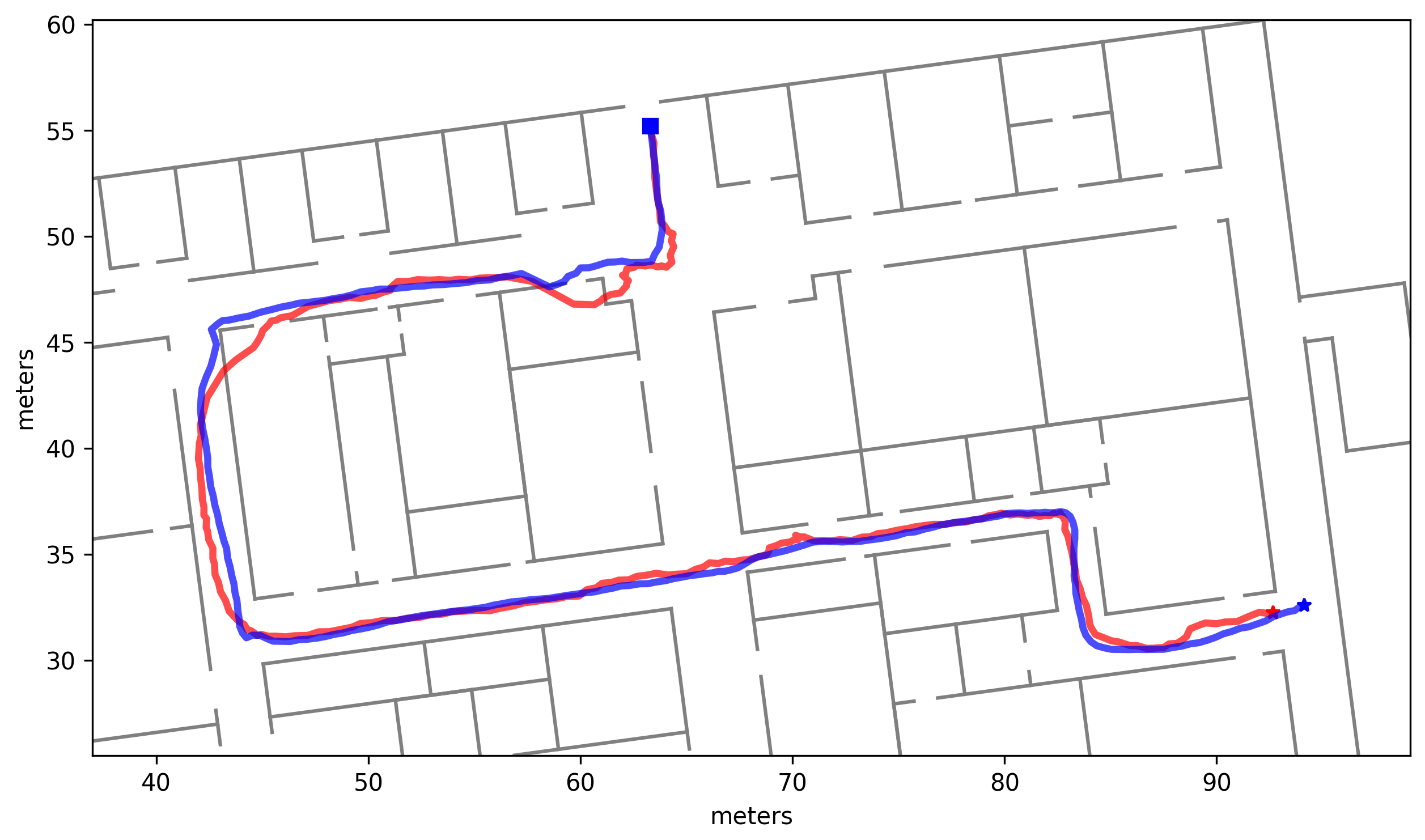}\\
  (a)&(b)\\
  \includegraphics[width=0.45\linewidth]{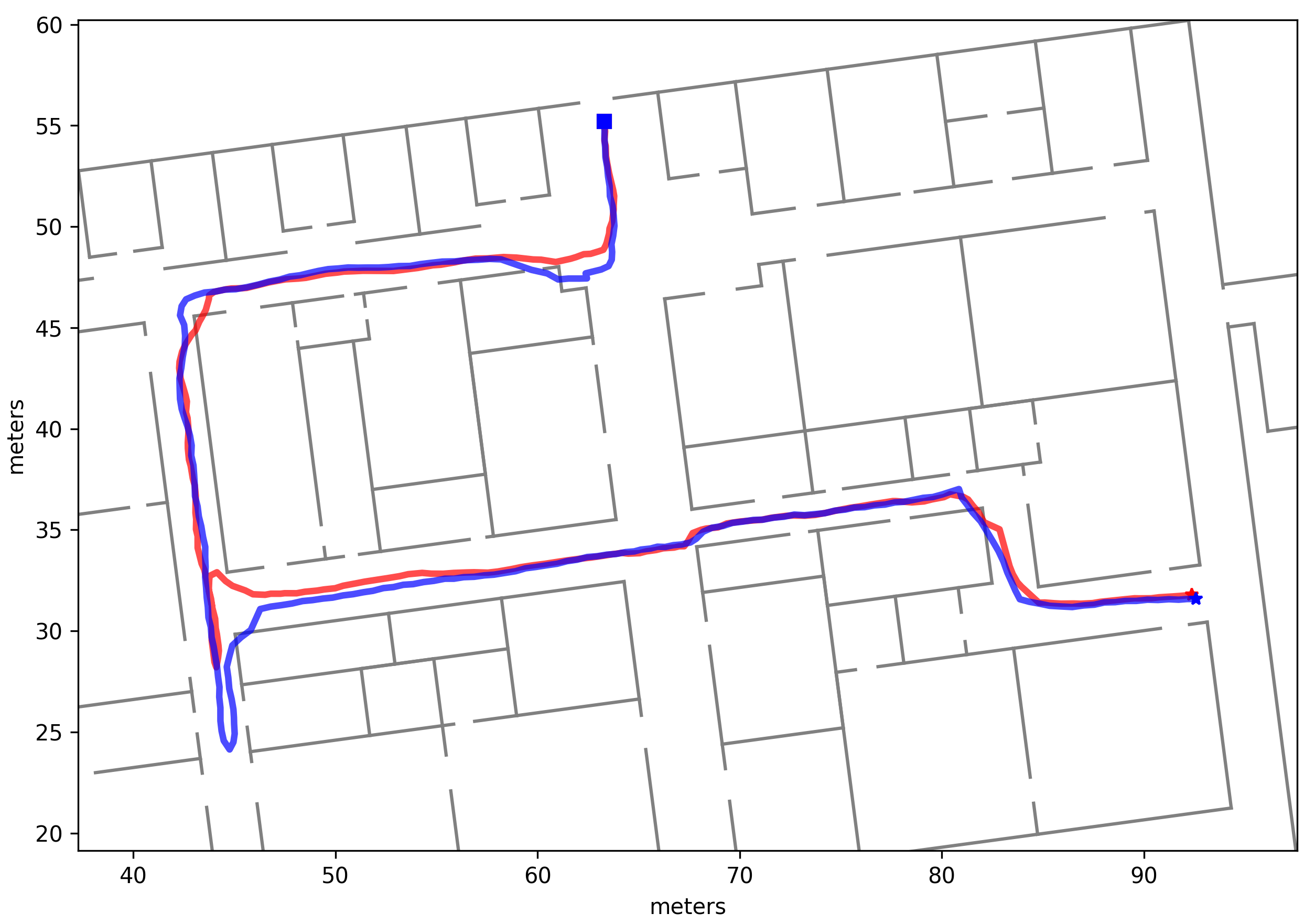}&
  \includegraphics[width=0.45\linewidth]{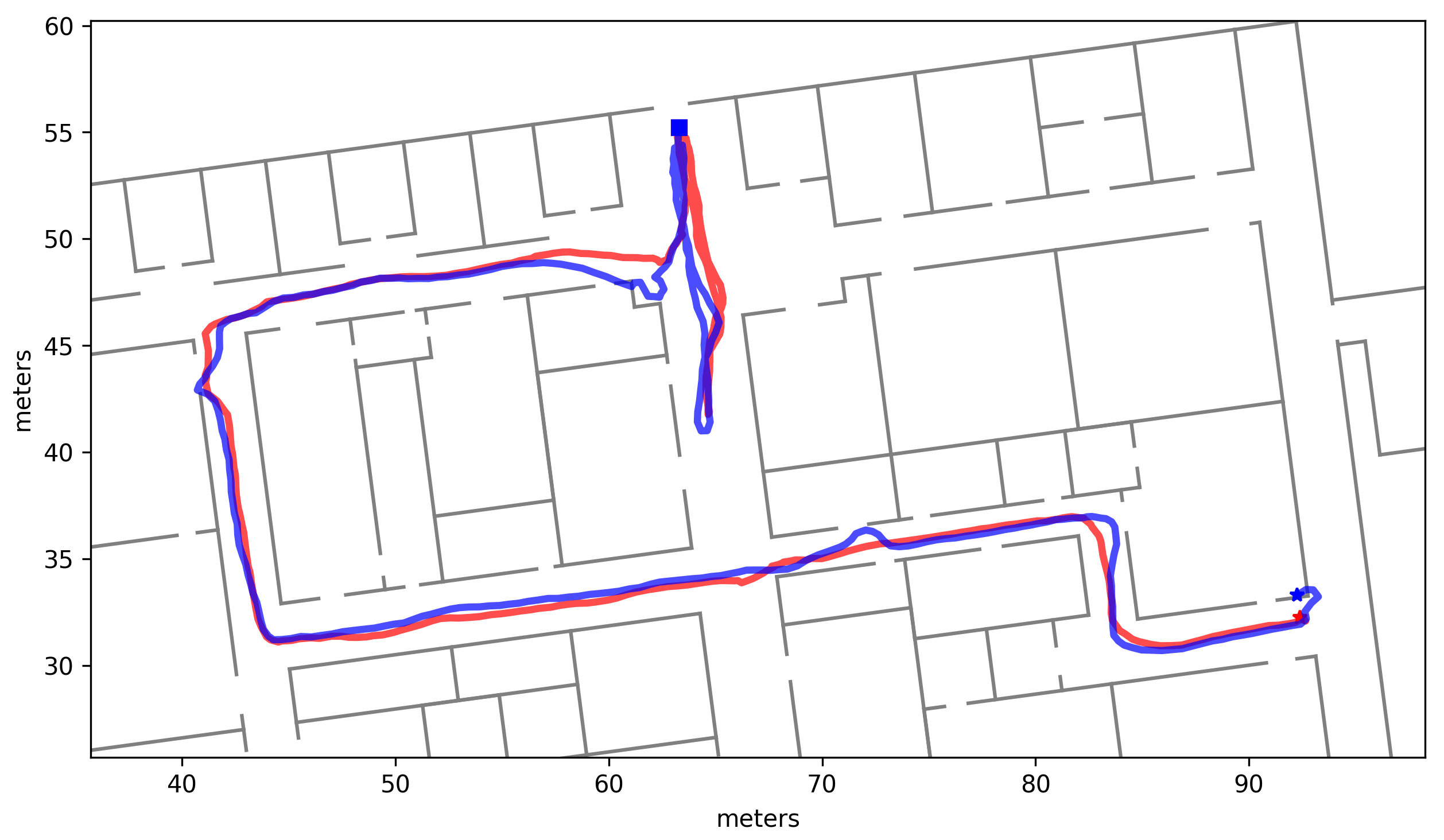}\\
  (c)&(d)
  \end{tabular}
  \caption{Route R2W. (a): Floor plan of the considered building. Waypoints are shown in red, traversability graph edges are shown in gray. The start and end waypoints are marked with a square and a star, respectively. The shortest path is shown with a thick dark gray line. Landmarks are shown in blue and enumerated (see Tab.~\ref{tab:landmarks} for landmark listing.) (b)--(d): recorded paths using A/S (blue line) and RoNIN (red line). (b): P5. (c): P2. (d): P1.}
  \Description{This figure has four panels. The top left panel shows a floor plan, with a number of small red circles at corridor junctions. Several pairs of these are joined by grey edges. Along the corridors, there are some numbered blue dots. A dashed grey line, overlapping a sequence of grey edges, joins a red circle framed by a square, with a red circle framed by a star.}  
  \label{fig:R2W}
  
\end{figure}

\begin{figure}
  \centering
  \begin{tabular}{cc}
  \includegraphics[width=0.45\linewidth]{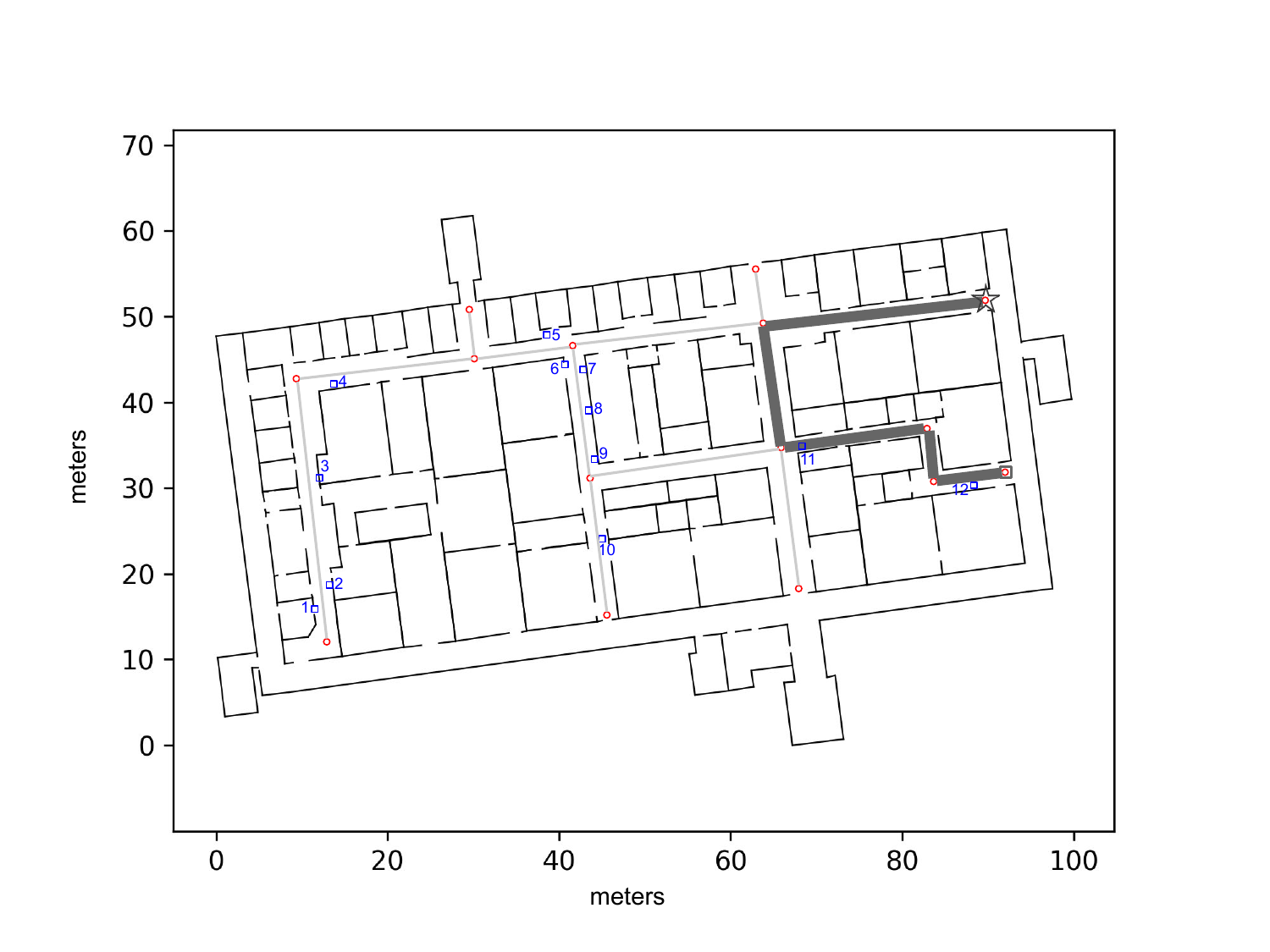} &
  \includegraphics[width=0.45\linewidth]{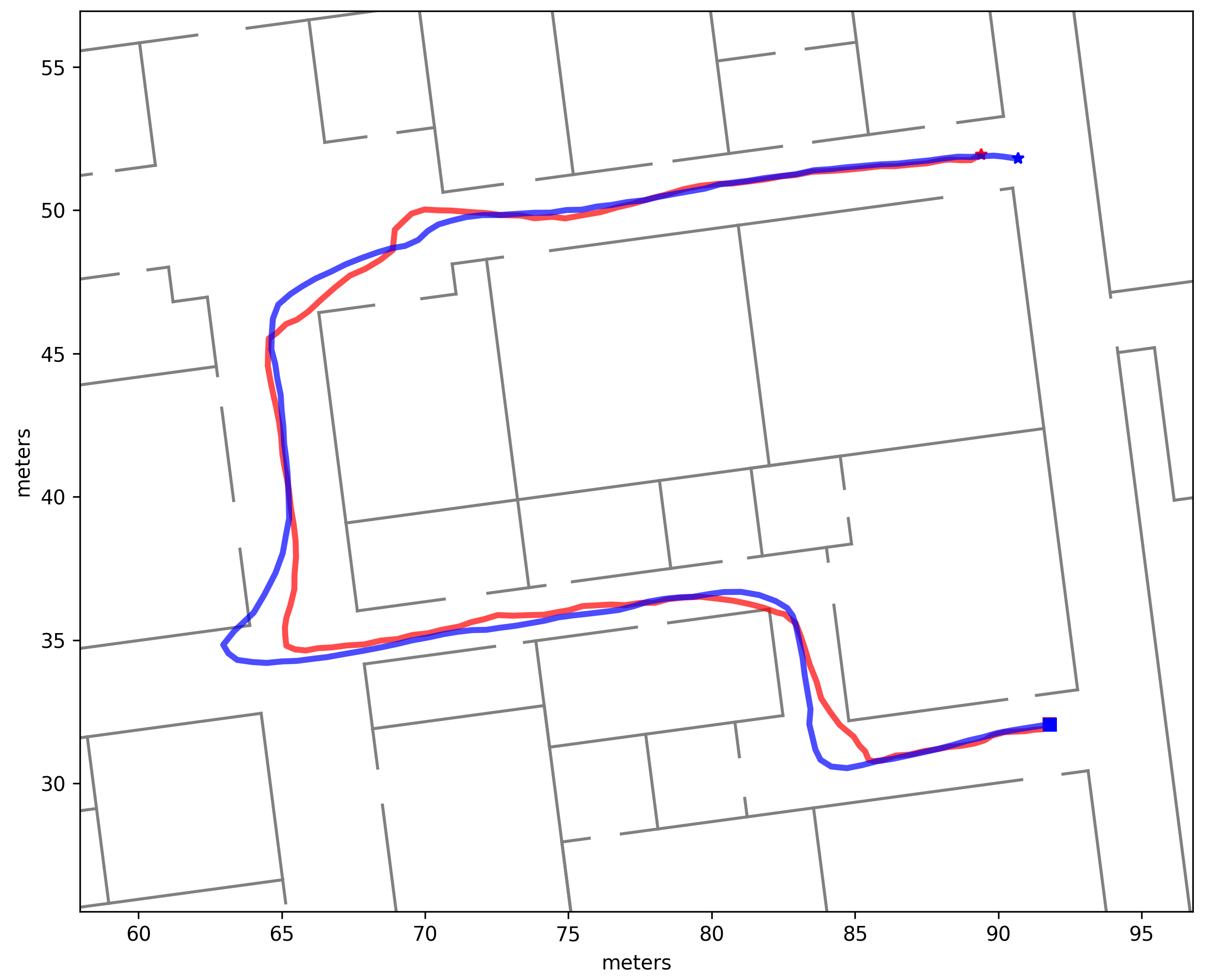}\\
  (a)&(b)\\
  \includegraphics[width=0.45\linewidth]{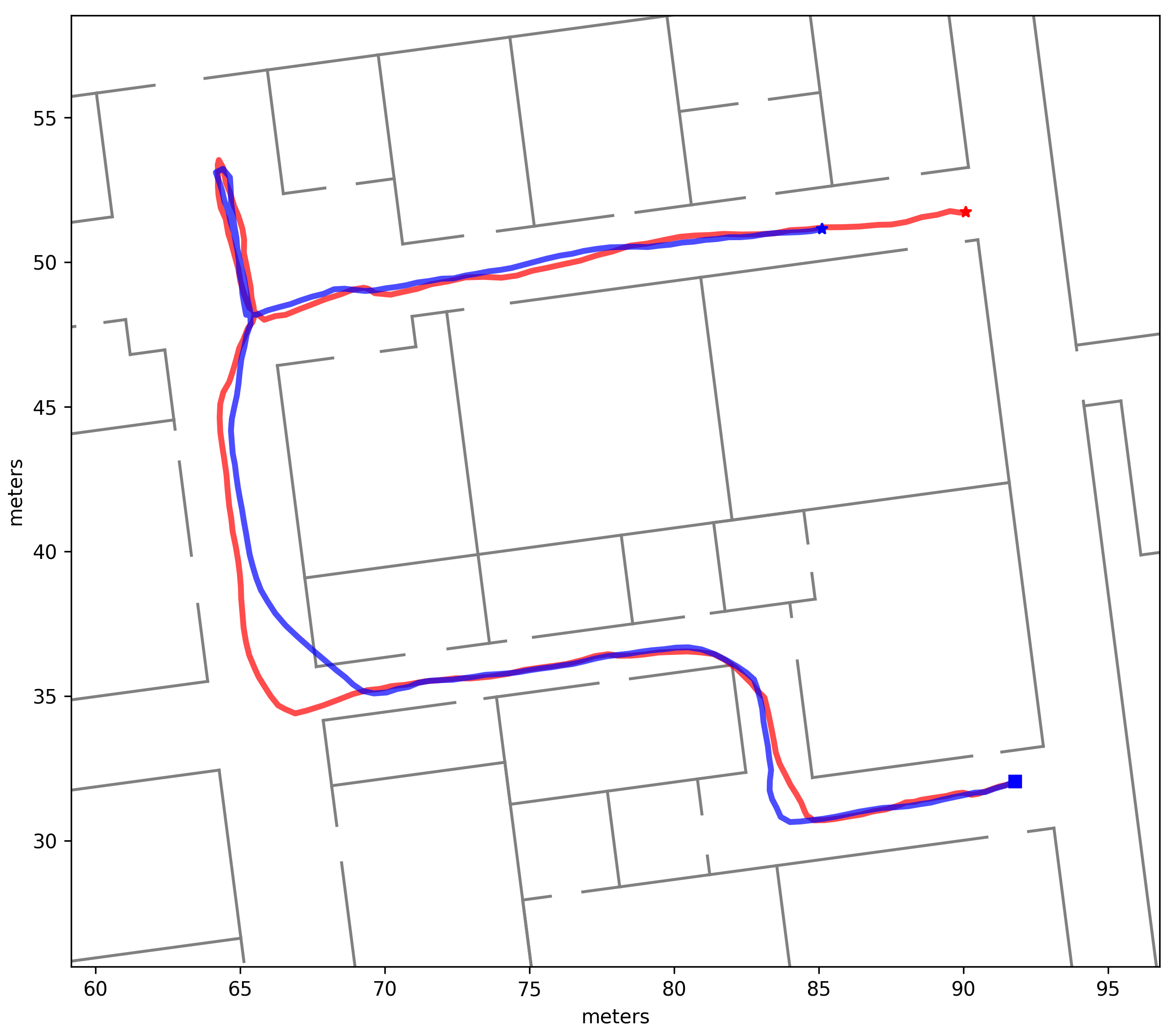}&
  \includegraphics[width=0.45\linewidth]{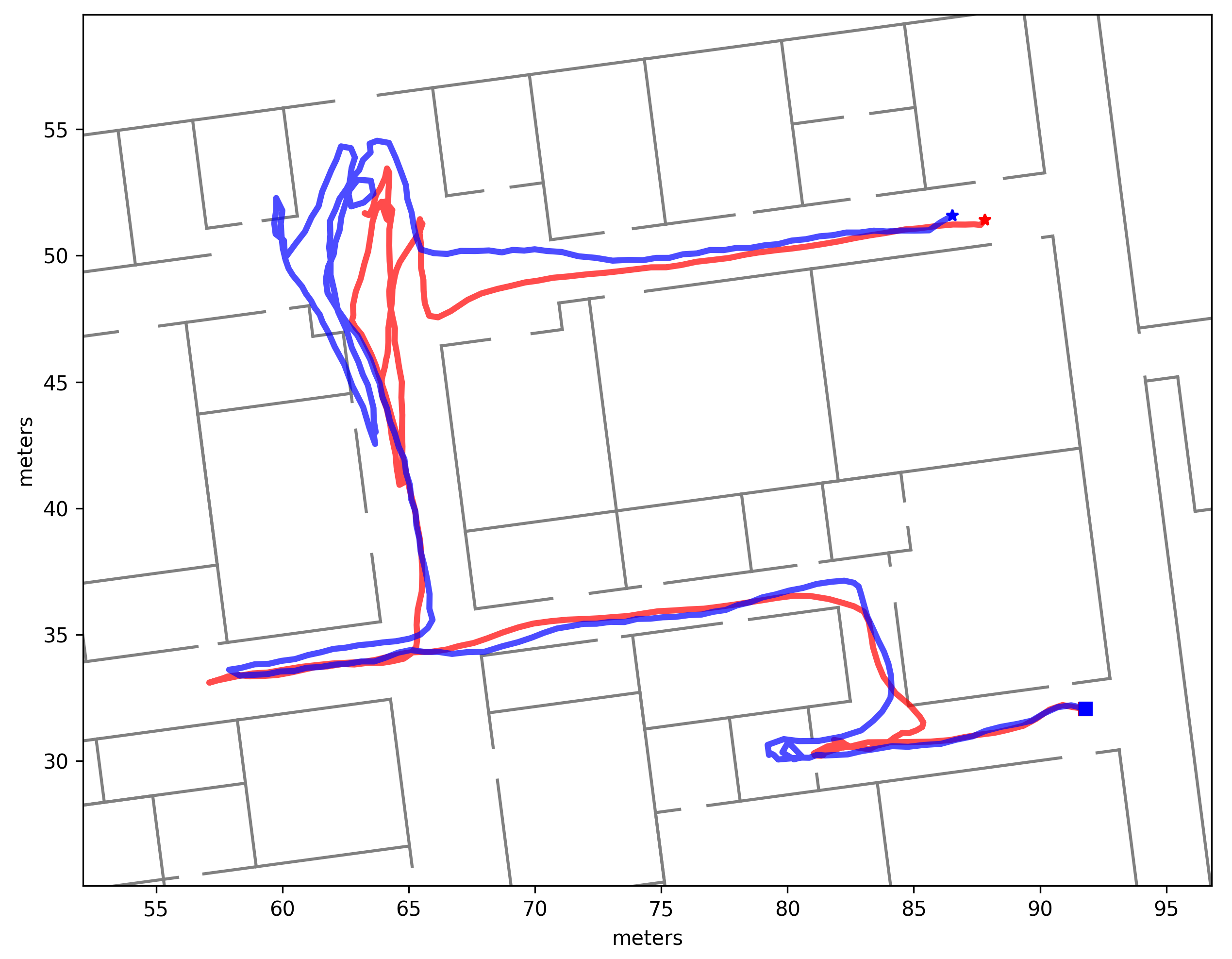}\\
  (c)&(d)  
  \end{tabular}
  \caption{Route R3W. (a): Floor plan of the considered building. Waypoints are shown in red, traversability graph edges are shown in gray. The start and end waypoints are marked with a square and a star, respectively. The shortest path is shown with a thick dark gray line. Landmarks are shown in blue and enumerated (see Tab.~\ref{tab:landmarks} for landmark listing.) (b)--(d): recorded paths using A/S (blue line) and RoNIN (red line). (b): P3. (c): P6. (d): P7. Note that P7 mistakenly entered a room (through an open door) soon after the beginning of the trial.}
  \Description{This figure has four panels. The top left panel shows a floor plan, with a number of small red circles at corridor junctions. Several pairs of these are joined by grey edges. Along the corridors, there are some numbered blue dots. A dashed grey line, overlapping a sequence of grey edges, joins a red circle framed by a square, with a red circle framed by a star.}
  \label{fig:R3W}
\end{figure}

The trials were conducted on the second floor of a campus building. Three routes were defined, as shown in Figs.~\ref{fig:R1W}--\ref{fig:R3W}~(a). When testing the Wayfinding app, participants first traversed route R1W (Fig.~\ref{fig:R1W}~(a); length: 123~m), then route R2W (Fig.~\ref{fig:R2W}~(a); length: 97~m), and finally route R3W (Fig.~\ref{fig:R3W}~(a); length: 72~m). Note that the endpoints of these routes were at the location of a closed exit door, and that 
the endpoint of each route coincided with the starting point of the next route. 
R1W traversed one L-junction, two T-junctions, and two X-junctions, and included 4 turns. R2W contained two L-junctions, two T-junctions, and two X-junctions, with 5 turns. R3W contained two L-junctions and two X-junctions, with 4 turns. The width of the corridors ranged from 1.9~m to 6.0~m.  The building was generally quiet, with a small number of students and researchers occasionally encountered in the corridors. 

As mentioned earlier, all junctions in this building were at $90^{\circ}$. A few potentially challenging situations are shown in Fig.~\ref{fig:pictures}, and include: a narrow door (always kept open) that had to be traversed (a); multiple alcoves that had to be negotiated (c); a side staircase in a narrow corridor that had to be avoided (e); a wide (10~m $\times$ 6~m) open space, partially visible in (d).

When testing the Backtracking app, the same routes were traversed in the opposite direction and in the reverse order (these routes are called R1B, R2B, and R3B). So in practice, each participant first walked R1W, R2W, and R3W, using the Wayfinding app, while the Backtracking app (in a different smartphone) collected way-in data for each route. Then, they tested the Backtracking app by traversing R3B, R2B, and R1B. We would like to emphasize that the Backtracking app did not have knowledge of the building layout.

A separate building was used for the practice trial when participants were shown how to use the two apps. The practice trial route was simpler, with only 2 turns, for a total length of 63 meters. The two buildings were at a short distance from each other.

\subsubsection{Modalities}
After being consented, each participant was explained the purpose of both apps and their functions and encouraged to ask questions if something was not clear. Particular care was taken to impress upon the participants that notifications about upcoming turns would be produced with advance notice. We explained that, upon hearing such a notification, they would need to search for the first place where they could take a turn. This place could be in their close proximity, or a few meters down the way. 

After this initial phase, participants underwent the simple calibration procedure described in Sec.~\ref{sec:localization}. Then, 
we walked with them to the location where the practice trial would start. Each participant carried two iPhones in their pants pocket (note that we asked the participants prior to the experiment to please wear pants with pockets on the experiment day). Both phones needed to be carried for the initial (Wayfinding) trials: while one of them (an iPhone 12) ran the Wayfinding app, the second one (an iPhone XR) recorded way-in data. Participants also wore a wireless bone conduction headset  (Shokz OpenRun), through which they could receive notifications from apps. They also wore an Apple Watch Series 8, used to interact with the apps. Before beginning the practice trial, we made sure that the VoiceOver speed and sound volume were set to the desired level.  We asked each participant whether they preferred directions to be given in units of meters, feet, or steps, and set the apps' parameters accordingly (see Tab.~\ref{tab:participants}). Note that for the first three participants (P1--P3), due to an implementation mistake, the  Backtracking app produced distances only in units of steps.

We asked the participants to practice making left and right swipes on the Watch. All participants eventually learned how to do this, though P2 had some difficulties at the beginning, as he interpreted ``right'' or ``left'' as referring to the axis of his left arm (where he wore the Watch), i.e. in a direction orthogonal to the forearm, rather than parallel to it.

At this point, the practice trial started: the practice route was traversed using the Wayfinding app; once at destination, it was traversed in the reverse direction using the Backtracking app. After the practice trial, participants were asked whether they wanted to turn off the sound of individual detected footsteps. All participants chose to leave the footstep sound on, with some commenting that hearing it made them feel reassured that the system was functioning.   We also asked them whether they wanted to turn off notification of nearby landmarks (a few landmarks were announced during the practice trial). They all chose to leave landmark notifications on.

Upon completion of the practice trials, the participant and the experimenters moved to the building where the actual experiment was to take place, and in particular to the starting point of the initial route (R1W). From there, the sequence of trials described in Sec.~\ref{sec:env} was started. At the beginning of each
trial, the participant was accompanied to the starting location of the route, and their body was oriented along the initial walking direction. Then, they were asked to select the next route from the list, using the  Watch. Once the route was selected, they were asked to rotate the Watch's crown to start the app. We also asked them to make a left swipe   on the Watch, to hear a description of the whole route. Then, the participant started walking the route. Upon arriving at destination, participants were asked to stop the app by rotating the Watch's crown. They were asked if they wanted to rest a bit before starting the next trial. Then, they were re-positioning to the starting point of the next route (same as the endpoint of the previous route), and oriented correctly, before starting the next route. During the trials, the experimenters followed the participants at a safe distance, taking care not to influence their routing decision.

At the transition between trials with the Wayfinding app and trials with the Backtracking app,  participants were handed a new headset and Watch (same models), and asked to substitute the old ones with these new ones. Both devices can be paired with one iPhone at a time; since the Backtracking app ran on a different iPhone than the Wayfinding app, this switch was necessary. 

At the end of the last trial (route R1B with the Backtracking app), participants and experimenters walked back to the first building, where participants were asked to participate in a questionnaire, involving the ten  System Usability Scale (SUS) questions, as well as a number of open ended questions.


\begin{table}[h]
\caption{Summary of the experiment for the Wayfinding and Backtracking routes. For successfully completed routes, we report the duration (in seconds). When displayed with grey background, the participant missed one or more turns, or took a wrong turn, but was able to walk back and complete the route with guidance from the app. $R$: required system reset. $E$: route was completed, but verbal input from experimenter was needed at some point. $\varprod$: trial had to be aborted due to app's inability to track the participant. In two cases, a second attempt was made after a trial had been aborted.}
\label{tab:performance}
\centering\begin{tabular}{c|c|c|c|c|c|c|c||c|}
 &\bf P1&\bf P2&\bf P3&\bf P4&\bf P5&\bf P6&\bf P7&\bf Length\\
  \hline
\bf  R1W&261 $(R,E)$&355 $(R)$&297 $(R)$& 216&\cellcolor{gray!20}271&223&\cellcolor{gray!20}206 &123~m\\
\hline
\bf R2W& \cellcolor{gray!20}304 $(E)$&\cellcolor{gray!20}209&134&163&211&\cellcolor{gray!20}262&\cellcolor{gray!20}171&97~m\\
  \hline\bf R3W& 125&\cellcolor{gray!20}170&98&127&144&\cellcolor{gray!20}139&\cellcolor{gray!20}330&72~m \\
  \hline
  \hline\bf R3B& 180&$\varprod$&115&134 $(E)$&136&$\varprod$&$\varprod$ &72~m\\
  \hline
  \bf R2B& 182 $(E)$&\cellcolor{gray!20}232&$\varprod$&\cellcolor{gray!20} 238&173&154&149 &97~m\\
  \hline\bf R1B& 187&206&149&167&$\varprod$, 163 $(E)$&184&$\varprod,\varprod$&123~m \\
  \hline  
\end{tabular}

\end{table}

\begin{table}[h!]
\caption{List of landmarks (see Figs.~\ref{fig:R1W}--\ref{fig:R3W}).}\label{tab:landmarks}
\centering\begin{tabular}{c|l}
     Landmark ID &Landmark type\\
     \hline
    1&  alcove\\
     \hline
2&benches\\
     \hline
3&staircase\\
     \hline
4&photocopiers\\
     \hline
5&alcove\\
     \hline
6&alcove\\
     \hline
7&pillar\\
     \hline
8&couches\\
     \hline
9&pillar\\
     \hline
10&pillar\\
     \hline
11&door\\
     \hline
12&cabinets\\
     \hline
\end{tabular}
\end{table}

\subsection{Observations}
\subsubsection{Tracking and Guidance Performance: Wayfinding}\label{sec:TGPW}
Tab.~\ref{tab:performance} reports the duration of successful route traversals. The average speed (route length divided by traversal time) was 0.50~m/s for all three routes.
The first route (R1W) required an intervention from the experimenters for participants P1, P2, and P3. For P1, the localization algorithm used to generate notifications (A/S by default) had to be switched to RoNIN after the first turn, because A/S was unable to correctly track the participant  (see Fig.~\ref{fig:R1W}~(b), blue line). For P2 and P3, we had to manually reset the user's location in the app. In all three cases, the remainder of the Wayfinding trials were completed without any issues (except for P1 in R2W, as discussed later). It is important to note that for these three participants, the step length update $s_i$ had yet to be incorporated in our Particle Filter implementation (Sec.~\ref{sec:localization}), hence A/S relied solely on the step length measured during calibration. The first route segment of R1W is a long (more than 40 meters) and narrow corridor, with several obstacles and a staircase (see Fig.~\ref{fig:pictures}~(e)) on its right side (participants were advised not to walk on the staircase). This caused the participants to walk with a substantially smaller stride length than during calibration. By the time they reached the end of the corridor, the accumulated length error was such that, even with Particle Filtering, tracking became exceedingly challenging. For P1, switching to RoNIN (which produced excellent results in this case; Fig.~\ref{fig:R1W}~(b), red line) did the trick. However, RoNIN also occasionally gave poor results in the same area (e.g., it was unable to track P2). After we upgraded the Particle Filtering to include step length values, this problem no longer occurred, and we were able to use A/S successfully for all other participants.

Tab.~\ref{tab:performance} also shows (cells filled in gray) situations in which participants successfully completed the trial, but occasionally had to trace back their path (as prompted by the app) because they had missed a turn. 
These situations are clearly visible in Fig.~\ref{fig:R1W}~(d), Fig.~\ref{fig:R2W}~(c) and (d), and Fig.~\ref{fig:R3W}~(c) and (d). Later analysis, using the recorded video and logged data from the app, brought to light the reasons for these missed turns. In some cases, participants were  distracted by passer-by (P5 in R1W), or they were speaking at the time notification was issued (P2 in R2W; Fig.~\ref{fig:R2W}~(c)). P6 missed a turn in R3W  while walking in a large open space (Fig.~\ref{fig:R3W}~(c)). Probably due to the lack of a nearby wall, she might have been unsure about where she should turn exactly, so she continued walking until she perceived a wall to the right. By that time, she was notified to turn around.
P7, who missed one turn in R1W and one in R2W, as well as multiple turns in R3W, walked very fast with a dog guide. In these situations, she had already walked past the junction by the time the notification was completed and she could process it. Of note, on route R3W, soon after the beginning of the trial, P7 walked straight into a small room whose door was open  (Fig.~\ref{fig:R3W}~(d)). She then missed other turns in the same route, as shown in the figure, one of which multiple times, until eventually she found the correct final route segment. Only one time (P6 in R2W) did the app issue a notification shortly after the participant had passed the junction, causing her to miss the turn. This was due to localization algorithm momentarily undershooting her position. However, after being  notified that she should turn around, she successfully negotiated the turn.


Participants using a long cane appeared to generally react proactively to early advance notice of upcoming turns. They typically would begin walking closer to the wall on the side where they were expected to turn, and tap the wall with their cane until they found an opening (Fig.~\ref{fig:pictures}~(f)). In some cases, an alcove (wall recess) was located right before the corridor junction, and some participants got ``stuck'' in this alcove, before finding their way out and proceeding to the junction (Fig.~\ref{fig:pictures}~(c)). A remarkable exception was represented by the mobility technique of P2. Rather than scanning the ground surface in front of himself, P2, a seasoned independent traveler, used the cane to periodically tap the floor surface, and made judgments about nearby surfaces based on the sound produced by this tapping. Upon hearing a notification of an upcoming junction, he would move somewhat closer to the corresponding side of the corridor (without tapping the wall) until he perceived the presence of an opening to his left or to his right, at which point he would make a $90^{\circ}$ turn into it. However, in one situation in the R3W route, he missed a turn and had to then turn back as guided by the system (he then commented ``I knew where that was exactly'', indicating that he may have indeed perceived the opening the first time around.)

The two participants with a dog guide displayed very different behaviors. Since dog guides are normally trained to understand ``left'' or ``right'' directions, the participants simply needed (in principle) to convey the direction from the app verbally to their dogs. 
For the case of P7, this mechanism worked well, except for the fact that, as mentioned earlier, she walked really fast with her dog, thereby often passing by a junction to then receive a ``turn around'' notification. P1, on the other hand, worked with a slower and more cautious dog. For example, in route R1W, right after the second turn, they got stuck in an alcove to the right, and needed help from the experimenter to get out of it (marked as $E$ in Tab.~\ref{tab:performance}). It is conceivable that if P1 had been using a cane, she could have searched left and right to find a way out of the alcove. In two other situations, her dog initially refused to turn into a specific corridor. In particular, at the beginning of R2W (Fig.~\ref{fig:R2W}~(d)), she missed the first turn because, by the time she gave her dog a ``right'' command after the notification, they had already walked past the intersection. Afterwards, they correctly turned around as advised by the app. However, as she issued a ``left'' command to return to the route, the dog refused to turn and instead walked straight towards the exit door. At that time, we intervened, lest they would leave the building. It took some time before P1 managed to convince her dog to walk again on that corridor. After the study was completed, P1 explained to us that her dog might have felt nervous due to the circumstances of the trials.

Both inertial sensing algorithms (A/S and RoNIN) were able to successfully track the participants through the routes, except for the initial part of R1W for the first three participants before the implementation of the adaptive step length mechanism, as discussed earlier. Occasionally, the reconstructed path ``cut'' some corners (see e.g. Fig.~\ref{fig:R1W}~(c), RoNIN) or momentarily overshot the location (Fig.~\ref{fig:R3W}~(b), A/S), but the drift tracking mechanism of our Particle Filtering implementation~\cite{ren2021smartphone} seemed to have worked well, at least for these trials. In particular, it is remarkable that both algorithms were able to track the rather chaotic path taken by P7 in R3W (Fig.~\ref{fig:R3W}~(d)).

\subsubsection{Tracking and Guidance Performance: Backtracking}\label{sec:TGPB}
As shown in Tab.~\ref{tab:performance}, six trials with the Backtracking app had to be aborted, because the app failed to track the walker during the return route. In one of these trials (P3 in R2B), the participant had taken a wrong turn (for a reason described later in this section). The app correctly issued a ``turn around'' notification. However (as he later told us), P3 decided to instead keep walking, as he mistakenly thought that he remembered the path he had taken. As he kept walking away from the route, the app soon became unable to track him.

Two trials had to be aborted for route R3B. One of them (with P7) was due to incorrect way-in path simplification, consequent to a ``meandering'' way-in path (Fig.~\ref{fig:optimization}~(b)). In the case of P2 and P6, the culprit was a large magnetic field discrepancy (visible for P6 in Fig.~\ref{fig:Backtracking-bad}; note the white horizontal line). 

The first two segments of R1B are located in very wide corridors (Fig.~\ref{fig:pictures}~(d)), and we noticed that the magnetic field changes dramatically from one side to the other of these corridors (as already remarked in Sec.~\ref{sec:localization}). The different locations of the walker in the corridors between way-in and return likely caused this large discrepancy.

Another common situation occurred in route R2B. As can be seen in Fig.~\ref{fig:R2W}~(a), the return route (starting from the star in that figure) goes through two L-junction, then crosses an X-junction. In our original implementation of the Backtracking app, the notification issued upon entering a new segment was similar to that of the Wayfinding app: ``Walk straight for about XX [meters/feet/steps]. Then, you will turn [left, right].'' This is the type of notification that was issued after the second L-turn in the return route R2B, indicating a final right turn after walking for approximately 15 meters. What happened with P2, P3, and P4 is that, rather than walking through the X-junction, they turned right on it, presumably because they knew they had to turn right at some point. This situation did not happen with the Wayfinding app (in the reverse route) because, as discussed in Sec.~\ref{sec:notifications}, in the proximity of the X-junction the Wayfinding app would notify the walker to ``Keep walking straight''. This option is not available to the Backtracking app. The reason is that the Backtracking app is only aware of the way-in route: it cannot possibly know that the user traversed an X junction at some point, and thus cannot remind the user, on the return route, to ``walk straight'' when in proximity of the junction. We thus decided to slightly change the notification format, by removing the last sentence (``Then, you will turn [left, right]'')  when entering a new segment, and used this new implementation from P5 on. This ostensibly minimal interface modification was sufficient to avoid incorrect turns at this junction for the remaining participants. 

Participant P5 was not able to complete R1B, a route that initially proceeds straight through an X-junction (Fig.~\ref{fig:R1W}~(a)). P5 soon started veering to the left, hit a wall, then started walking around trying to find his way, taking multiple turns that the app was unable to match against the way-in route. We decided to abort that trial and to try again from the starting point. This time, P5 turned right instead of going straight. He told us that he knew that he had to turn right at some point, based on the initial route description provided (through a left swipe on the Watch), which stated that he should walk for 27 meters, then turn right. Note that this is a similar situation as experienced in the X-junction for R2B, as described above. One experimenter made him notice that he had not walked for 27 meters yet. At that point, P5 changed course, and was able to successfully complete the trial. 

While traversing route R1B, participant P2, after being notified by the app of an upcoming left turn,  encountered difficulties when trying to locate the turn, and ended up getting trapped in an alcove before reaching the junction. After leaving the alcove, P2 faced another alcove at the other side of the corridor. Eventually, and following directions from the app, the participant was able to return to the correct route.

The trail for P7 on route R1B had to be aborted; a second trial was likely unsuccessful. P7, as mentioned earlier, walked with a fast and very confident dog, and zipped through the X-junction where she was supposed to turn right. By the time she processed the notification asking her to turn around, she was already a long way down the corridor. Unfortunately, the Backtracking app was not able to recover tracking afterwards.

\begin{figure}
  \centering  
  \begin{tabular}{cc}
    \includegraphics[width=0.4\linewidth]{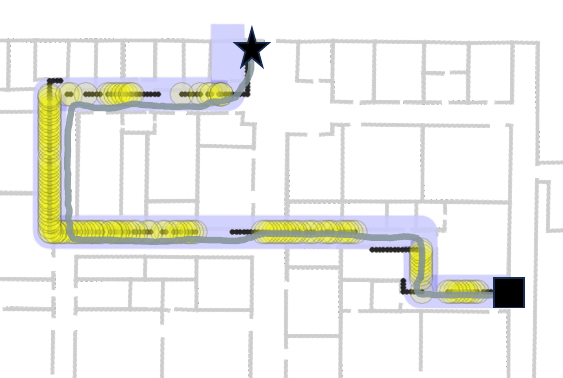} &
  \includegraphics[width=0.25\linewidth]{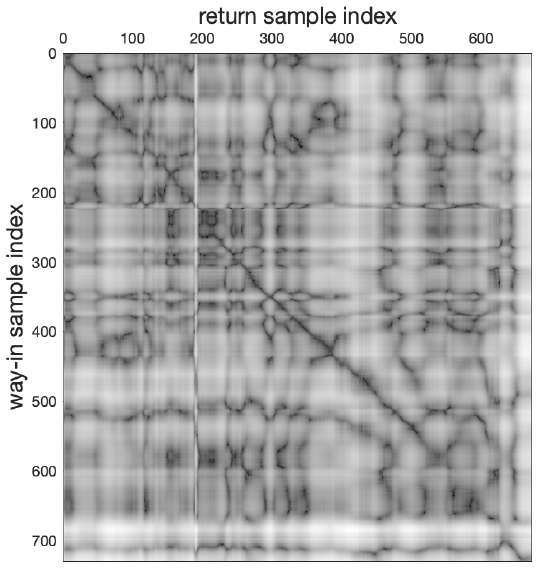}\\
    \includegraphics[width=0.4\linewidth]{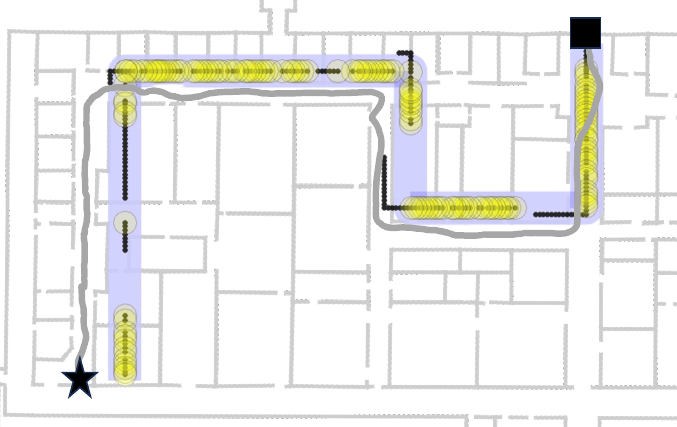} &
  \includegraphics[width=0.25\linewidth]{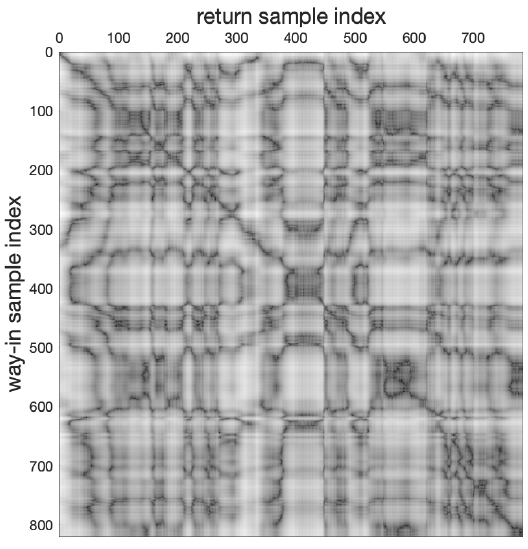} 
    \end{tabular}
 \caption{Examples of successful backtracking trials (hybrid matching). Top: Route R2B for participant P5. Bottom: R1B for P4. Left panel: The way-in path is shown with a thick purple line, ending at the black square. The length of each segment is given by the number of steps recorded, multiplied by the step length measured during calibration. The actual path of the participant during the return phase is shown by a gray line. Reliable matches are shown as yellow circles. Projected sequences are shown with black lines. Right panel: Magnetic discrepancy for all pairs $(i,j)$ of samples from way-in (vertical axis) and return (horizontal axis). Lighter gray indicates larger discrepancy.  } 
 \Description{This figure has two rows, each with two panels. The panels in the top row are similar to those in the bottom row. In each row, the left panel shows a partial floor plan of a building. There is a thick purple polyline, made by a chain of segments connecting a 90 degrees, starting from a black square and ending at a black star. There is a grey thin line, also starting at the black square, and approximately following the purple thick line. On the purple thick line, there are multiple yellow circles. In addition, there are short segments made by black dots, that are either straight or turn by 90 degrees. Each such short segment starts at a yellow circle. In the right panel, there is a two-dimensional graph. The axes labels read ``return sample index'' (horizontal axis) and ``way-in sample index'' (vertical axis). Each point in the graph is shown with a different brightness of gray. }
 \label{fig:Backtracking-good}
\end{figure}

\begin{figure}
  \centering  
  \begin{tabular}{cc}
    \includegraphics[width=0.4\linewidth]{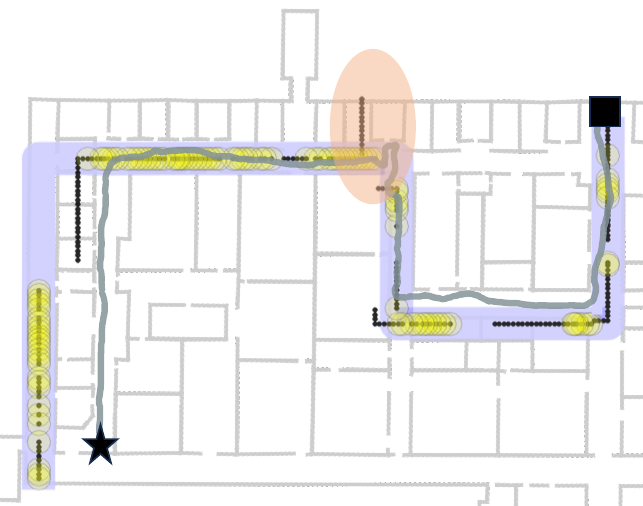} &
  \includegraphics[width=0.25\linewidth]{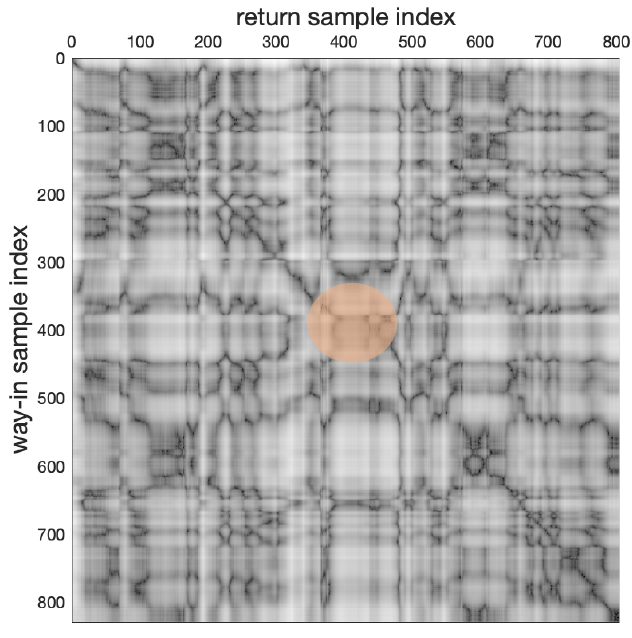} \\
    \includegraphics[width=0.4\linewidth]{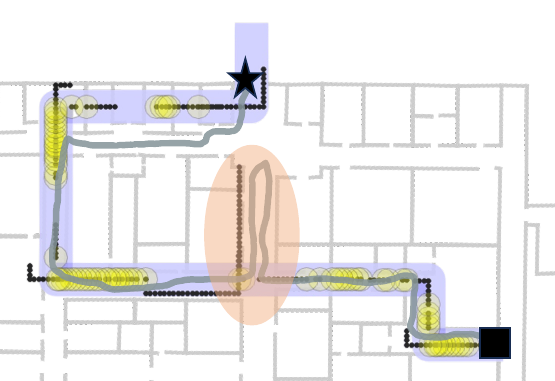} &
  \includegraphics[width=0.25\linewidth]{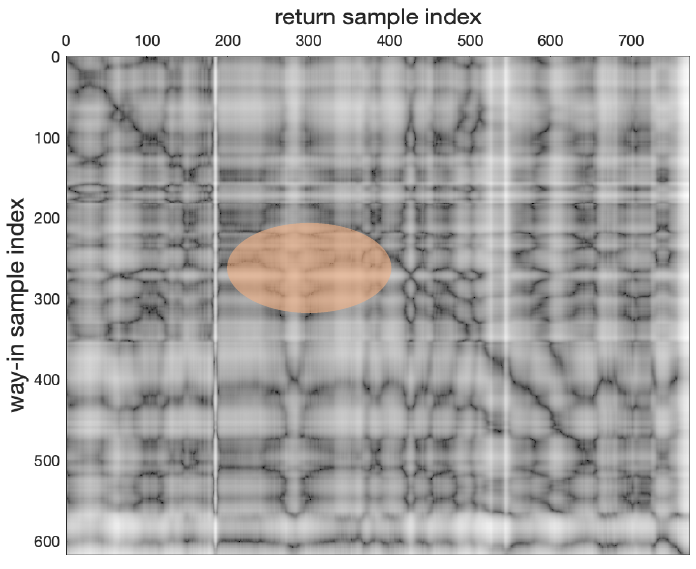} 
    \end{tabular}
\caption{See caption of Fig.~\ref{fig:Backtracking-good}. Top: route R1B, participant P2. Bottom: R2B, P2. Highlighted are situations in which the participant took a wrong path then walked back as directed by the app.  Turns by $180^{\circ}$ are shown with dash-dotted line in the right panel. }
 \Description{This figure has two rows, each with two panels. The panels in the top row are similar to those in the bottom row. In each row, the left panel shows a partial floor plan of a building. There is a thick purple polyline, made by a chain of segments connecting a 90 degrees, starting from a black square and ending at a black star. There is a grey thin line, also starting at the black square, and approximately following the purple thick line. On the purple thick line, there are multiple yellow circles. In addition, there are short segments made by black dots, that are either straight or turn by 90 degrees. Each such short segment starts at a yellow circle. In the right panel, there is a two-dimensional graph. The axes labels read ``return sample index'' (horizontal axis) and ``way-in sample index'' (vertical axis). Each point in the graph is shown with a different brightness of gray. } \label{fig:Backtracking-medium}
\end{figure}

\begin{figure}
  \centering  
  \begin{tabular}{cc}
    \includegraphics[width=0.3\linewidth]{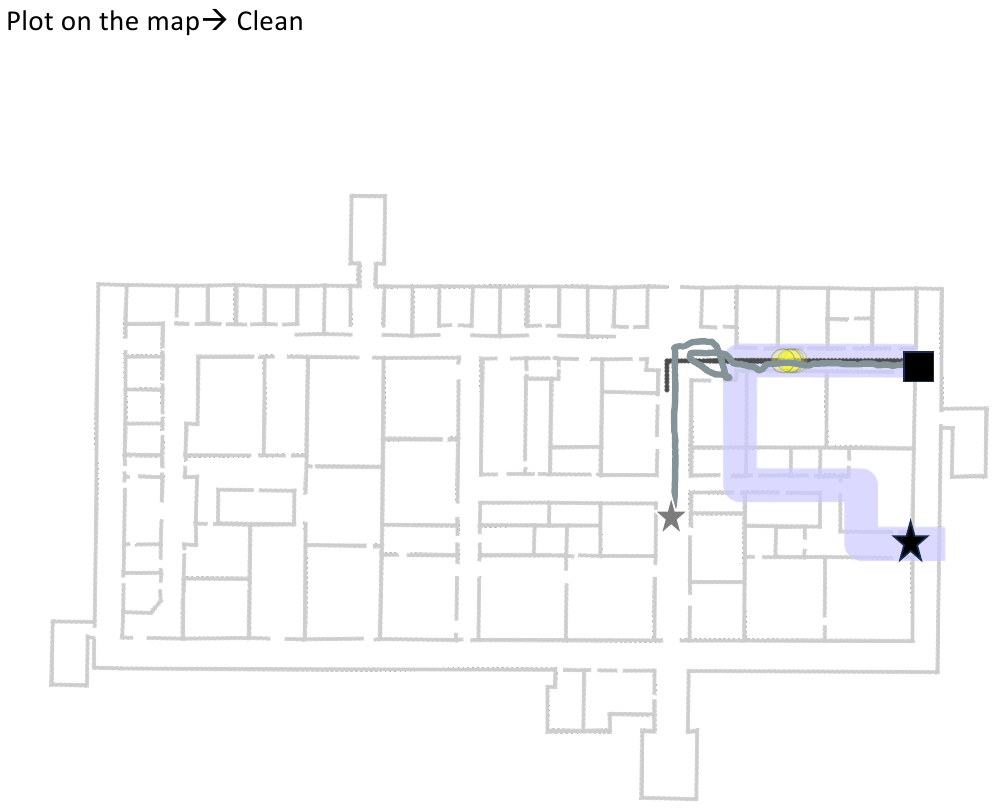} &
  \includegraphics[width=0.25\linewidth]{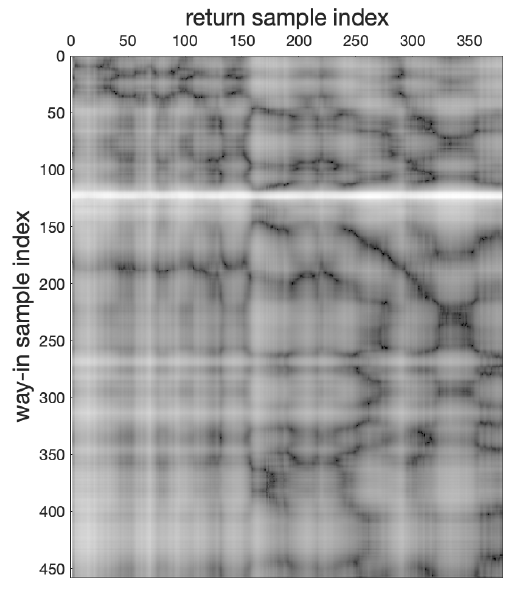}
    \end{tabular}
  \caption{See caption of Fig.~\ref{fig:Backtracking-good}. Route R3B, participant P6. In this case, the app failed to track the participant. The gray star represents the point at which the trial was aborted; the black star is the desired destination.}
 \Description{This figure has two rows, each with two panels. The panels in the top row are similar to those in the bottom row. In each row, the left panel shows a partial floor plan of a building. There is a thick purple polyline, made by a chain of segments connecting a 90 degrees, starting from a black square and ending at a black star. There is a grey thin line, also starting at the black square, and approximately following the purple thick line. On the purple thick line, there are multiple yellow circles. In addition, there are short segments made by black dots, that are either straight or turn by 90 degrees. Each such short segment starts at a yellow circle. In the right panel, there is a two-dimensional graph. The axes labels read ``return sample index'' (horizontal axis) and ``way-in sample index'' (vertical axis). Each point in the graph is shown with a different brightness of gray. }
\label{fig:Backtracking-bad}
\end{figure}

Examples of successful trials with the Backracking app are shown in Figs.~\ref{fig:Backtracking-good}--\ref{fig:Backtracking-medium}. In these plots, the way-in route (shown with a thick purple line) has segments whose length is obtained by multiplying the number of steps taken in the segments by the step length, computed through initial calibration. For this reason, these segments are not necessarily co-located with the corridors shown in the underlying floor plan. Note that this is not a problem in itself:  the goal of the Backtracking app is simply to match the walker's location during return with the same location during the way-in, in order to produce correct guidance notifications. Whether these reconstructed routes are metrically consistent with the floor plan is irrelevant for our purposes.

\subsubsection{Final Questionnaire}\label{sec:FQ}

The responses of the participants to the System Usability Scale questionnaire, administered at the end of the trials,  are reported in Tab.~\ref{tab:SUS}. The overall score~\cite{brooke1996sus} was 80.36. Though the interpretation of SUS scores is the object of debate~\cite{brooke2013sus}, this score converts to a percentile rank of 90\% based on the distribution of scores reported in~\cite{sauro2011practical}.

The open-ended questions, along with a summary of the responses, are listed below.

{\em 1. Do you think that the system always knew your location?} Most participants replied with variations of ``yes, most of the times'' (99\% of the times according to P7, 80\% according to P1). However, P2 and P3 replied with ``No''. According to P2, there were moments in which localization was correct. P3 felt that localization was sensitive to how one walks.

{\em 2. Do you think that the system gave you correct directions?} P4--P7 replied ``yes''. P2 thought that it gave correct directions most of the times (80\% of the times according to P1). P3 said: ``Yes, when it knew what it was doing.''

\begin{table} [h]
\caption{System Usability Scale (SUS) responses. The overall SUS score~\cite{brooke1996sus} was 80.36. } \label{tab:SUS}
\centering
\begin{tabular}{r|c|c|c|c|c|c|c||c|}
&P1&P2&P3&P4&P5&P6&P7&Mean  \\
\hline
{\em 1. I think that I would like to use this system frequently.}&3&4&1&5&5&5&4&3.86\\
\hline{\em 2. I found the system unnecessarily complex.}& 2&1&1&2&2&4&1&1.86\\
\hline {\em 3. I thought the system was easy to use.}&
4&4&5&5&5&5&5&4.71\\
\hline {\em 4. I think that I would need the support} &1&1&1&1&4&3&1&1.71\\
{\em of a technical person to be able to use this system.}&&&&&&&&\\
\hline {\em 5. I found the various functions in this system were well integrated.}&4&2&1&5&4&5&5&3.71\\
\hline{\em 6. I thought there was too much inconsistency in this system.}&3&3&1&1&2&4&1&2.14\\
\hline{\em 7. I would imagine that most people would learn}&3&3&5&4&5&5&5&4.29\\
{\em to use this system very quickly.}&&&&&&&&\\
\hline {\em 8. I found the system very cumbersome to use.}&2&1&1&1&2&1&1&1.29\\
\hline{\em 9. I felt very confident using the system.} &4&4&3&5&4&5&5&4.29\\
\hline{\em 10. I needed to learn a lot of things before }&2&1&1&2&1&4&1&1.71
\\ \hline
\end{tabular}
\end{table}


{\em 3. The system often gives turning directions (such as ``at the next junction, turn right'') with some advance notice, which means that you need to find the turn using your cane/dog. Was this a problem for you?} All participants stated that this was not a problem in general, with some adding specific comments. P1 said: ``Once  I got used to it, I sort of got it.'' P2 commented that it was not a problem, unless there are obstacles he could bump against when moving closer to the wall in preparation for the turn. P4 said that only one time this created a problem, and precisely when, in R2B, she took a right turn too soon (a situation discussed in Sec.~\ref{sec:TGPB}). She then added that she realized that she should have waited for a notification prompting her to take the turn. P6 
said that, while not a problem in itself, it would have been preferable if the notification times were more ``consistent'', in the sense of producing a prompt always at the same distance from the junction (whereas, due to localization errors, this distance was often variable). P7 commented that this was the only thing in the system that was not as accurate as she thought. 

{\em 4. Were the notifications understandable? Too many notifications? Too few?} All participants found that the notifications were ``fine'' or ``just right''. P2 elaborated on his reply, commenting that the system produced exactly the type of information he needed: it gave him a rough distance to the next turn, then it gave him a heads-up before the turn. P1 added that she would have liked to know if there is a door to open. (Note the  routes considered in the study did not require walkers to open any doors.) P4 and P6 commented very positively about landmark notifications. For example, P6  mentioned two announcements (``stairwell to the left'', ``bench to the right'') as examples of useful information. Likewise, P4 was enthusiastic about the landmark notifications, and thought they could be useful in practice. For example, she mentioned that sometimes she may feel thirsty, and it would be nice to know if she was passing by a drinking fountain. She also mentioned that landmark notification would be particularly useful when someone visits a place for the first time; this type of notification could be turned off once one is more familiar with the environment. Similarly, P2 thought that the amount of landmarks announced in the trials was just right. However, if he was really concentrated on the route, he probably would want fewer notifications.

{\em 5. Was it easy for you to use the Watch?}  All participants except for P2 said that they found it easy. P2 said that ``it was an adventure'' but also commented that with some more practice, he would have found it easier to use the Watch gestures.

{\em 6. What would you like to have in this app that is not already there?} Both P2 and P5 said that they would like to have more contextual information about the space they are visiting. P1 asked about how one would find the starting point, and especially the correct direction when starting a route. P3 mentioned that he would like to be able to scroll through a route description one step at the time. (In the current implementation, a left swipe produces a description of the remaining route, but does not let the user stop and repeat each step). P4 wanted the ability to manually add landmarks on the fly (e.g., when passing by a certain location of interest). P5 also mentioned that he would like to know which direction he was facing at each time, as this information would help him when he was confused.

{\em 7. Did you notice any difference between the Wayfinding system and the Backtracking system?} In general, the participants found that the two apps were very consistent with each other (with P5 mentioning that if we hadn't told him that they were different apps, he would not have noticed.) P2 and P3 lamented the fact that the two apps used different units. As mentioned earlier, this was due to an implementation mistake, which was later corrected.

{\em 8. Do you think that using this app would make you feel safer or more confident when traveling alone in a new place? [Asked for each app separately.]} All participants except for P3 replied with an enthusiastic ``yes''. P3 thought that he could use the Backtracking app for situations such as when at a conference with multiple tables in a large hall, if for example one wanted to go to the restroom then come back to the same table. P2 envisioned using it when visiting a new place (e.g., a large medical building). He would like to be able to have a menu from which he could narrow down the place he would need to go. He also mentioned that, even if the app were not able to take him precisely there, there would be a lot of value in at least helping him to get closer. P4 thought that the Wayfinding app would help even for buildings that she had visited before. She mentioned an example of a large office building she had started visiting recently. While walking there, she stopped to think about where to go next, and immediately several bystanders came to her, offering unrequested (and unwelcome) help. This app would help her in these situations of uncertainty. P4 also said that using this app would relieve her from having to constantly focus on keeping track of her position and deciding where to go - her brain would have more ``bandwidth'' for other things. This is particularly the case in noisy and distracting environments. She also mentioned a practical example of application of the Backtracking app: walking with a sighted guide in a medical building to reach a doctor's office, then having to walk back when there is no one around to help. P5 thought that the apps helped him create a mental map of the whole route, and enjoyed the fact that he didn't have to memorize the whole path because the apps would give him prompts when traversing it.

\section{Discussion}\label{sec:Discussion}
Inertial-based localization is attractive because it does not rely on external infrastructure, and because it doesn't require use of the smartphone's camera, which would force the user to hold the phone in one hand while walking, or attach it to their clothing or to a lanyard. Prior wayfinding systems for blind travelers using inertial sensors~\cite{fallah2012user,apostolopoulos2014integrated} relied on the ``user as a sensor'' modality to overcome localization errors due to accumulating drift. While the ``user as a sensor'' modality can be a powerful tool for mitigation of ambiguous situations (see e.g.~\cite{ren2023experiments}, where it was tested successfully in an indoor/outdoor wayfinding system), it can be demanding for the user, who is tasked with detecting and identifying the landmarks. With our work, we decided to ``push the envelop'' and see whether a Wayfinding or Backtracking system that only relies on inertial sensor data, without input from the user, could be feasible. Our experimental results suggest that, at least for buildings characterized by networks of corridors, this is indeed the case. Of course, confirmation from users when they reached some well perceivable landmarks could be integrated for improved robustness.

The environment in which we tested our systems contained narrow and wide corridors, as well as fairly large open spaces. It thus can be considered representative of many public spaces (office buildings, schools, health centers). However, larger buildings (e.g., airports, transit hubs, large shopping malls) may be challenging for inertial-only wayfinding systems: in very large opens spaces, particles disperse isotropically, and thus the Particle Filter may not be able to successfully correct for accumulating drift. The building considered in our study did not have junctions with corridors intersecting at other than 90 degrees. While arbitrary intersection angles may not represent a problem for Particle Filtering (which can adapt to any wall layout), a higher turning angle granularity may increase the error rate of the turn detector used for the Backtracking app (though 45 degrees turns were not shown to create problems in simulations with the WeAllWalk data set~\cite{tsai2021finding,flores2018weallwalk}).

In our study, interaction with the apps was only enabled through the Watch, a mechanism that was very well received by the participants. In practical situations, though, users may want to take the phone out of their pocket, e.g. to pick up a call or to send a text. These ``natural'' types of interactions will be allowed (and encouraged) in future studies. As mentioned in Sec.~\ref{sec:localization} the A/S algorithm assumes that the phone does not change orientation with respect to one's body, and therefore may produce incorrect values for the walking direction if one re-orients the phone with respect to their body. In these cases, RoNIN  would be a preferable localization algorithm, enabling  ``natural'' phone interactions. Qualitative analysis of the trajectories recorded from A/S and RoNIN (see e.g. Figs.~\ref{fig:R1W}-\ref{fig:R3W}) showed no clear winner, and it appears that both techniques can be used successfully, if followed by a carefully designed Particle Filter, whose purpose is to ensure that the reconstructed routes are consistent with the building layout.

For the Backtracking app, Particle Filtering was not an option, because the layout of the walls in the building is unknown. Our approach is to use structural geometric constraints (representing paths as sequences of straight segments and turns at discrete angles), graph-based sequence alignment, and magnetic field consistency, to match a return path with its way-in counterpart. The relatively large number of failed trials (6 out of 21) highlights the difficulty of the task, and the need for further research. In general, we found that our Backtracking app works well when users do not deviate too much from the way-in path. Large detours from the original route are difficult for the algorithm to handle, as it becomes liable to mismatches. Another problem is related to the spatial variability of the magnetic field within the width of a large corridor or hallway, which may cause mismatches when, during return, the user walks on a different trajectory within the same space. In future work, we will consider more adaptive statistical models for the magnetic anomalies in indoor environments~\cite{kuang2022consumer} (also relying on existing magnetic field data sets~\cite{torres2015ujiindoorloc}).

One of the goals of our investigation was to evaluate whether the  simple interface mechanism introduced in Sec.~\ref{sec:UI}, which notifies walkers of an incoming turn with large advance notice, would be feasible and acceptable. Prior work (e.g.~\cite{ahmetovic2023sonification}) studied optimal sonification techniques to ensure that walkers begin turning at the right time. These mechanisms are appropriate for localization systems (e.g. using computer vision or BLE beacons) that are substantially more accurate than inertial sensing~\cite{ohn2018variability}. In our case, the user interface must accommodate for localization errors, hence the early notification. We were not sure whether participants would have difficulty discovering the exact location of the junction, or whether they would find this modality burdensome or annoying.  We were please to find that, by and large, our participants were able to negotiate these situation successfully, even in the face of structural impediments (e.g., when they would get stuck in an alcove when looking for an opening). What's more, participants, when asked whether they thought that advance notification was a problem (Sec.~\ref{sec:FQ}, question 3.), they overwhelmingly asserted that this was not the case. The positive usability scores recorded from the SUS responses are another indication that our participants were generally satisfied with the interface design.  We are aware, however, that the advance notification approach could fail, for example, if there are two openings close to each other on the same side of the corridor (say, within two meters from each other). In this case, more contextual information could be produced by the system (e.g., notifying the user that they will encounter two junctions, and that they should take the second one).

Perhaps the most critical limitation of our approach is that users of both systems must start a route from a certain location, and begin walking in a certain direction. This initialization phase is necessary to geometrically calibrate the user's path with the floor plan frame. Note that the same requirement is found in prior work with inertial-based wayfinding~\cite{fallah2012user,apostolopoulos2014integrated}, and is a consequence of the dead-reckoning nature of this method. Note that in principle it is possible for the system to function even without knowledge of starting position and orientation. As shown in~\cite{crabb2023lightweight}, one may distribute a large number of particles uniformly across the environment, then after a certain period (several minutes) of walking time, the surviving particles would converge with good likelihood around the user's position. However, this would require the user to walk aimlessly and without guidance from the system for a while, something that may not be desirable in practice. We envision two possible solutions to the problem of starting location/orientation determination. One possibility is to use the ``user as a sensor'' paradigm. Users may start from a location that is well perceivable, such as the main entrance of building or other identifiable landmarks such as the end of a staircase, and be advised to start walking in a certain known direction (in our example, straight across building entrance or coasting a wall to the left or to the right of the staircase entrance.) A different solution, one that we are currently exploring, is to create a hybrid system that can use visual data (e.g. via automatic landmark recognition~\cite{chen2019multi}) for sporadic ``fixes'' using computer vision techniques when the location and orientation of the user needs to be ascertained. After that, the user may move the smartphone back to their pocket, and be tracked by the inertial system.  


Finally, we were heartened to hear from our participants that they found that the apps they tested could make them feel safer and more confident in their independent travel. While we are aware of the limitations of our systems, this type of feedback (along with the good scores from the SUS questionnaire) confirms that our proposed technology has serious potential for practical use as a navigational aid.

\section{Conclusions}\label{sec:Conclusions}
We have presented the results of an experiment with seven blind participants who used two custom-designed iPhone apps for wayfinding and backtracking. These apps use different inertial-based mechanisms to track the user in a known floor plan (Wayfinding) or to match the user's path with a previously taken path (Backtracking). Our experiments showed that inertial-based tracking, coupled with a carefully designed user interface, is a suitable technology for wayfinding in a building characterized by a network of corridors when the initial position/orientation of the user is known. Out study also showed that our approach to backtracking (which assumes no knowledge of the building's layout) has good potential, but improvements are needed to increase its robustness in practical scenarios.

\bibliographystyle{ACM-Reference-Format}
\bibliography{RouteMe2}


\begin{thebibliography}{63}


\ifx \showCODEN    \undefined \def \showCODEN     #1{\unskip}     \fi
\ifx \showDOI      \undefined \def \showDOI       #1{#1}\fi
\ifx \showISBNx    \undefined \def \showISBNx     #1{\unskip}     \fi
\ifx \showISBNxiii \undefined \def \showISBNxiii  #1{\unskip}     \fi
\ifx \showISSN     \undefined \def \showISSN      #1{\unskip}     \fi
\ifx \showLCCN     \undefined \def \showLCCN      #1{\unskip}     \fi
\ifx \shownote     \undefined \def \shownote      #1{#1}          \fi
\ifx \showarticletitle \undefined \def \showarticletitle #1{#1}   \fi
\ifx \showURL      \undefined \def \showURL       {\relax}        \fi
\providecommand\bibfield[2]{#2}
\providecommand\bibinfo[2]{#2}
\providecommand\natexlab[1]{#1}
\providecommand\showeprint[2][]{arXiv:#2}

\bibitem[Abu~Doush et~al\mbox{.}(2017)]%
        {abu2017isab}
\bibfield{author}{\bibinfo{person}{Iyad Abu~Doush}, \bibinfo{person}{Sawsan Alshatnawi}, \bibinfo{person}{Abdel-Karim Al-Tamimi}, \bibinfo{person}{Bushra Alhasan}, {and} \bibinfo{person}{Safaa Hamasha}.} \bibinfo{year}{2017}\natexlab{}.
\newblock \showarticletitle{ISAB: integrated indoor navigation system for the blind}.
\newblock \bibinfo{journal}{\emph{Interacting with Computers}} \bibinfo{volume}{29}, \bibinfo{number}{2} (\bibinfo{year}{2017}), \bibinfo{pages}{181--202}.
\newblock


\bibitem[Ahmetovic et~al\mbox{.}(2023)]%
        {ahmetovic2023sonification}
\bibfield{author}{\bibinfo{person}{Dragan Ahmetovic}, \bibinfo{person}{Federico Avanzini}, \bibinfo{person}{Adriano Barat{\`e}}, \bibinfo{person}{Cristian Bernareggi}, \bibinfo{person}{Marco Ciardullo}, \bibinfo{person}{Gabriele Galimberti}, \bibinfo{person}{Luca~A Ludovico}, \bibinfo{person}{Sergio Mascetti}, {and} \bibinfo{person}{Giorgio Presti}.} \bibinfo{year}{2023}\natexlab{}.
\newblock \showarticletitle{Sonification of navigation instructions for people with visual impairment}.
\newblock \bibinfo{journal}{\emph{International Journal of Human-Computer Studies}}  \bibinfo{volume}{177} (\bibinfo{year}{2023}), \bibinfo{pages}{103057}.
\newblock


\bibitem[Ahmetovic et~al\mbox{.}(2016)]%
        {ahmetovic2016navcog}
\bibfield{author}{\bibinfo{person}{Dragan Ahmetovic}, \bibinfo{person}{Cole Gleason}, \bibinfo{person}{Chengxiong Ruan}, \bibinfo{person}{Kris Kitani}, \bibinfo{person}{Hironobu Takagi}, {and} \bibinfo{person}{Chieko Asakawa}.} \bibinfo{year}{2016}\natexlab{}.
\newblock \showarticletitle{NavCog: a navigational cognitive assistant for the blind}. In \bibinfo{booktitle}{\emph{Proceedings of the 18th International Conference on Human-Computer Interaction with Mobile Devices and Services}}. \bibinfo{pages}{90--99}.
\newblock


\bibitem[Ahmetovic et~al\mbox{.}(2017)]%
        {ahmetovic2017achieving}
\bibfield{author}{\bibinfo{person}{Dragan Ahmetovic}, \bibinfo{person}{Masayuki Murata}, \bibinfo{person}{Cole Gleason}, \bibinfo{person}{Erin Brady}, \bibinfo{person}{Hironobu Takagi}, \bibinfo{person}{Kris Kitani}, {and} \bibinfo{person}{Chieko Asakawa}.} \bibinfo{year}{2017}\natexlab{}.
\newblock \showarticletitle{Achieving practical and accurate indoor navigation for people with visual impairments}. In \bibinfo{booktitle}{\emph{Proceedings of the 14th International Web for All Conference}}. \bibinfo{pages}{1--10}.
\newblock


\bibitem[Apostolopoulos et~al\mbox{.}(2014)]%
        {apostolopoulos2014integrated}
\bibfield{author}{\bibinfo{person}{Ilias Apostolopoulos}, \bibinfo{person}{Navid Fallah}, \bibinfo{person}{Eelke Folmer}, {and} \bibinfo{person}{Kostas~E Bekris}.} \bibinfo{year}{2014}\natexlab{}.
\newblock \showarticletitle{Integrated online localization and navigation for people with visual impairments using smart phones}.
\newblock \bibinfo{journal}{\emph{ACM Transactions on Interactive Intelligent Systems (TiiS)}} \bibinfo{volume}{3}, \bibinfo{number}{4} (\bibinfo{year}{2014}), \bibinfo{pages}{1--28}.
\newblock


\bibitem[Azenkot et~al\mbox{.}(2011)]%
        {azenkot2011smartphone}
\bibfield{author}{\bibinfo{person}{Shiri Azenkot}, \bibinfo{person}{Richard~E Ladner}, {and} \bibinfo{person}{Jacob~O Wobbrock}.} \bibinfo{year}{2011}\natexlab{}.
\newblock \showarticletitle{Smartphone haptic feedback for nonvisual wayfinding}. In \bibinfo{booktitle}{\emph{The proceedings of the 13th international ACM SIGACCESS conference on Computers and accessibility}}. \bibinfo{pages}{281--282}.
\newblock


\bibitem[Brooke(1996)]%
        {brooke1996sus}
\bibfield{author}{\bibinfo{person}{John Brooke}.} \bibinfo{year}{1996}\natexlab{}.
\newblock \showarticletitle{Sus: a “quick and dirty’usability}.
\newblock \bibinfo{journal}{\emph{Usability evaluation in industry}} \bibinfo{volume}{189}, \bibinfo{number}{3} (\bibinfo{year}{1996}), \bibinfo{pages}{189--194}.
\newblock


\bibitem[Brooke(2013)]%
        {brooke2013sus}
\bibfield{author}{\bibinfo{person}{John Brooke}.} \bibinfo{year}{2013}\natexlab{}.
\newblock \showarticletitle{SUS: a retrospective}.
\newblock \bibinfo{journal}{\emph{Journal of usability studies}} \bibinfo{volume}{8}, \bibinfo{number}{2} (\bibinfo{year}{2013}), \bibinfo{pages}{29--40}.
\newblock


\bibitem[Chen et~al\mbox{.}(2020)]%
        {chen2020meshmap}
\bibfield{author}{\bibinfo{person}{Lina Chen}, \bibinfo{person}{Jinbin Wu}, {and} \bibinfo{person}{Chen Yang}.} \bibinfo{year}{2020}\natexlab{}.
\newblock \showarticletitle{MeshMap: A magnetic field-based indoor navigation system with crowdsourcing support}.
\newblock \bibinfo{journal}{\emph{IEEE Access}}  \bibinfo{volume}{8} (\bibinfo{year}{2020}), \bibinfo{pages}{39959--39970}.
\newblock


\bibitem[Chen et~al\mbox{.}(2019)]%
        {chen2019multi}
\bibfield{author}{\bibinfo{person}{Lidong Chen}, \bibinfo{person}{Yin Zou}, \bibinfo{person}{Yaohua Chang}, \bibinfo{person}{Jinyun Liu}, \bibinfo{person}{Benjamin Lin}, {and} \bibinfo{person}{Zhigang Zhu}.} \bibinfo{year}{2019}\natexlab{}.
\newblock \showarticletitle{Multi-level scene modeling and matching for smartphone-based indoor localization}. In \bibinfo{booktitle}{\emph{2019 IEEE International Symposium on Mixed and Augmented Reality Adjunct (ISMAR-Adjunct)}}. IEEE, \bibinfo{pages}{311--316}.
\newblock


\bibitem[Chen and Liu(2000)]%
        {chen2000mixture}
\bibfield{author}{\bibinfo{person}{Rong Chen} {and} \bibinfo{person}{Jun~S Liu}.} \bibinfo{year}{2000}\natexlab{}.
\newblock \showarticletitle{Mixture Kalman Filters}.
\newblock \bibinfo{journal}{\emph{Journal of the Royal Statistical Society: Series B (Statistical Methodology)}} \bibinfo{volume}{62}, \bibinfo{number}{3} (\bibinfo{year}{2000}), \bibinfo{pages}{493--508}.
\newblock


\bibitem[Cheng et~al\mbox{.}(2005)]%
        {cheng2005accuracy}
\bibfield{author}{\bibinfo{person}{Yu-Chung Cheng}, \bibinfo{person}{Yatin Chawathe}, \bibinfo{person}{Anthony LaMarca}, {and} \bibinfo{person}{John Krumm}.} \bibinfo{year}{2005}\natexlab{}.
\newblock \showarticletitle{Accuracy characterization for metropolitan-scale Wi-Fi localization}. In \bibinfo{booktitle}{\emph{Proceedings of the 3rd international conference on Mobile systems, applications, and services}}. \bibinfo{pages}{233--245}.
\newblock


\bibitem[Cheraghi et~al\mbox{.}(2017)]%
        {cheraghi2017guidebeacon}
\bibfield{author}{\bibinfo{person}{Seyed~Ali Cheraghi}, \bibinfo{person}{Vinod Namboodiri}, {and} \bibinfo{person}{Laura Walker}.} \bibinfo{year}{2017}\natexlab{}.
\newblock \showarticletitle{GuideBeacon: Beacon-based indoor wayfinding for the blind, visually impaired, and disoriented}. In \bibinfo{booktitle}{\emph{2017 IEEE International Conference on Pervasive Computing and Communications (PerCom)}}. IEEE, \bibinfo{pages}{121--130}.
\newblock


\bibitem[Crabb et~al\mbox{.}(2023)]%
        {crabb2023lightweight}
\bibfield{author}{\bibinfo{person}{Ryan Crabb}, \bibinfo{person}{Seyed~Ali Cheraghi}, {and} \bibinfo{person}{James~M Coughlan}.} \bibinfo{year}{2023}\natexlab{}.
\newblock \showarticletitle{A Lightweight Approach to Localization for Blind and Visually Impaired Travelers}.
\newblock \bibinfo{journal}{\emph{Sensors}} \bibinfo{volume}{23}, \bibinfo{number}{5} (\bibinfo{year}{2023}), \bibinfo{pages}{2701}.
\newblock


\bibitem[Edel and K{\"o}ppe(2015)]%
        {edel2015advanced}
\bibfield{author}{\bibinfo{person}{Marcus Edel} {and} \bibinfo{person}{Enrico K{\"o}ppe}.} \bibinfo{year}{2015}\natexlab{}.
\newblock \showarticletitle{An advanced method for pedestrian dead reckoning using BLSTM-RNNs}. In \bibinfo{booktitle}{\emph{2015 International Conference on Indoor Positioning and Indoor Navigation (IPIN)}}. IEEE, \bibinfo{pages}{1--6}.
\newblock


\bibitem[Fallah et~al\mbox{.}(2012)]%
        {fallah2012user}
\bibfield{author}{\bibinfo{person}{Navid Fallah}, \bibinfo{person}{Ilias Apostolopoulos}, \bibinfo{person}{Kostas Bekris}, {and} \bibinfo{person}{Eelke Folmer}.} \bibinfo{year}{2012}\natexlab{}.
\newblock \showarticletitle{The user as a sensor: navigating users with visual impairments in indoor spaces using tactile landmarks}. In \bibinfo{booktitle}{\emph{Proceedings of the SIGCHI Conference on Human Factors in Computing Systems}}. \bibinfo{pages}{425--432}.
\newblock


\bibitem[Fan et~al\mbox{.}(2017)]%
        {fan2017accurate}
\bibfield{author}{\bibinfo{person}{Xirui Fan}, \bibinfo{person}{Jing Wu}, \bibinfo{person}{Chengnian Long}, {and} \bibinfo{person}{Yanmin Zhu}.} \bibinfo{year}{2017}\natexlab{}.
\newblock \showarticletitle{Accurate and low-cost mobile indoor localization with 2-D magnetic fingerprints}. In \bibinfo{booktitle}{\emph{Proceedings of the First ACM Workshop on Mobile Crowdsensing Systems and Applications}}. \bibinfo{pages}{13--18}.
\newblock


\bibitem[Fiannaca et~al\mbox{.}(2014)]%
        {fiannaca2014headlock}
\bibfield{author}{\bibinfo{person}{Alexander Fiannaca}, \bibinfo{person}{Ilias Apostolopoulous}, {and} \bibinfo{person}{Eelke Folmer}.} \bibinfo{year}{2014}\natexlab{}.
\newblock \showarticletitle{Headlock: a wearable navigation aid that helps blind cane users traverse large open spaces}. In \bibinfo{booktitle}{\emph{Proceedings of the 16th international ACM SIGACCESS conference on Computers \& accessibility}}. \bibinfo{pages}{19--26}.
\newblock


\bibitem[Flores et~al\mbox{.}(2015)]%
        {flores2015vibrotactile}
\bibfield{author}{\bibinfo{person}{German Flores}, \bibinfo{person}{Sri Kurniawan}, \bibinfo{person}{Roberto Manduchi}, \bibinfo{person}{Eric Martinson}, \bibinfo{person}{Lourdes~M Morales}, {and} \bibinfo{person}{Emrah~Akin Sisbot}.} \bibinfo{year}{2015}\natexlab{}.
\newblock \showarticletitle{Vibrotactile guidance for wayfinding of blind walkers}.
\newblock \bibinfo{journal}{\emph{IEEE transactions on haptics}} \bibinfo{volume}{8}, \bibinfo{number}{3} (\bibinfo{year}{2015}), \bibinfo{pages}{306--317}.
\newblock


\bibitem[Flores and Manduchi(2018a)]%
        {flores2018easy}
\bibfield{author}{\bibinfo{person}{German Flores} {and} \bibinfo{person}{Roberto Manduchi}.} \bibinfo{year}{2018}\natexlab{a}.
\newblock \showarticletitle{Easy return: an app for indoor backtracking assistance}. In \bibinfo{booktitle}{\emph{Proceedings of the 2018 CHI Conference on Human Factors in Computing Systems}}. \bibinfo{pages}{1--12}.
\newblock


\bibitem[Flores and Manduchi(2018b)]%
        {flores2018weallwalk}
\bibfield{author}{\bibinfo{person}{German~H Flores} {and} \bibinfo{person}{Roberto Manduchi}.} \bibinfo{year}{2018}\natexlab{b}.
\newblock \showarticletitle{Weallwalk: An annotated dataset of inertial sensor time series from blind walkers}.
\newblock \bibinfo{journal}{\emph{ACM Transactions on Accessible Computing (TACCESS)}} \bibinfo{volume}{11}, \bibinfo{number}{1} (\bibinfo{year}{2018}), \bibinfo{pages}{1--28}.
\newblock


\bibitem[Fox et~al\mbox{.}(2003)]%
        {fox2003bayesian}
\bibfield{author}{\bibinfo{person}{V Fox}, \bibinfo{person}{Jeffrey Hightower}, \bibinfo{person}{Lin Liao}, \bibinfo{person}{Dirk Schulz}, {and} \bibinfo{person}{Gaetano Borriello}.} \bibinfo{year}{2003}\natexlab{}.
\newblock \showarticletitle{Bayesian filtering for location estimation}.
\newblock \bibinfo{journal}{\emph{IEEE pervasive computing}} \bibinfo{volume}{2}, \bibinfo{number}{3} (\bibinfo{year}{2003}), \bibinfo{pages}{24--33}.
\newblock


\bibitem[Fusco and Coughlan(2020)]%
        {fusco2020indoor}
\bibfield{author}{\bibinfo{person}{Giovanni Fusco} {and} \bibinfo{person}{James~M Coughlan}.} \bibinfo{year}{2020}\natexlab{}.
\newblock \showarticletitle{Indoor localization for visually impaired travelers using computer vision on a smartphone}. In \bibinfo{booktitle}{\emph{Proceedings of the 17th International Web for All Conference}}. \bibinfo{pages}{1--11}.
\newblock


\bibitem[Ganz et~al\mbox{.}(2012)]%
        {ganz2012percept}
\bibfield{author}{\bibinfo{person}{Aura Ganz}, \bibinfo{person}{James Schafer}, \bibinfo{person}{Siddhesh Gandhi}, \bibinfo{person}{Elaine Puleo}, \bibinfo{person}{Carole Wilson}, {and} \bibinfo{person}{Meg Robertson}.} \bibinfo{year}{2012}\natexlab{}.
\newblock \showarticletitle{PERCEPT indoor navigation system for the blind and visually impaired: architecture and experimentation}.
\newblock \bibinfo{journal}{\emph{International journal of telemedicine and applications}}  \bibinfo{volume}{2012} (\bibinfo{year}{2012}).
\newblock


\bibitem[Giudice et~al\mbox{.}(2019)]%
        {giudice2019evaluation}
\bibfield{author}{\bibinfo{person}{Nicholas~A Giudice}, \bibinfo{person}{William~E Whalen}, \bibinfo{person}{Timothy~H Riehle}, \bibinfo{person}{Shane~M Anderson}, {and} \bibinfo{person}{Stacy~A Doore}.} \bibinfo{year}{2019}\natexlab{}.
\newblock \showarticletitle{Evaluation of an accessible, real-time, and infrastructure-free indoor navigation system by users who are blind in the mall of america}.
\newblock \bibinfo{journal}{\emph{Journal of Visual Impairment \& Blindness}} \bibinfo{volume}{113}, \bibinfo{number}{2} (\bibinfo{year}{2019}), \bibinfo{pages}{140--155}.
\newblock


\bibitem[Gleason et~al\mbox{.}(2018)]%
        {gleason2018crowdsourcing}
\bibfield{author}{\bibinfo{person}{Cole Gleason}, \bibinfo{person}{Dragan Ahmetovic}, \bibinfo{person}{Saiph Savage}, \bibinfo{person}{Carlos Toxtli}, \bibinfo{person}{Carl Posthuma}, \bibinfo{person}{Chieko Asakawa}, \bibinfo{person}{Kris~M Kitani}, {and} \bibinfo{person}{Jeffrey~P Bigham}.} \bibinfo{year}{2018}\natexlab{}.
\newblock \showarticletitle{Crowdsourcing the installation and maintenance of indoor localization infrastructure to support blind navigation}.
\newblock \bibinfo{journal}{\emph{Proceedings of the ACM on Interactive, Mobile, Wearable and Ubiquitous Technologies}} \bibinfo{volume}{2}, \bibinfo{number}{1} (\bibinfo{year}{2018}), \bibinfo{pages}{1--25}.
\newblock


\bibitem[Guerreiro et~al\mbox{.}(2020)]%
        {guerreiro2020virtual}
\bibfield{author}{\bibinfo{person}{Jo{\~a}o Guerreiro}, \bibinfo{person}{Daisuke Sato}, \bibinfo{person}{Dragan Ahmetovic}, \bibinfo{person}{Eshed Ohn-Bar}, \bibinfo{person}{Kris~M Kitani}, {and} \bibinfo{person}{Chieko Asakawa}.} \bibinfo{year}{2020}\natexlab{}.
\newblock \showarticletitle{Virtual navigation for blind people: Transferring route knowledge to the real-World}.
\newblock \bibinfo{journal}{\emph{International Journal of Human-Computer Studies}}  \bibinfo{volume}{135} (\bibinfo{year}{2020}), \bibinfo{pages}{102369}.
\newblock


\bibitem[Herath et~al\mbox{.}(2020)]%
        {herath2020ronin}
\bibfield{author}{\bibinfo{person}{Sachini Herath}, \bibinfo{person}{Hang Yan}, {and} \bibinfo{person}{Yasutaka Furukawa}.} \bibinfo{year}{2020}\natexlab{}.
\newblock \showarticletitle{RoNIN: Robust neural inertial navigation in the wild: Benchmark, evaluations, \& new methods}. In \bibinfo{booktitle}{\emph{2020 IEEE International Conference on Robotics and Automation (ICRA)}}. IEEE, \bibinfo{pages}{3146--3152}.
\newblock
\newblock
\shownote{Code available at https://github.com/Sachini/ronin}.


\bibitem[Hossain et~al\mbox{.}(2020)]%
        {hossain2020sightless}
\bibfield{author}{\bibinfo{person}{Md~Elias Hossain}, \bibinfo{person}{Khandker~M Qaiduzzaman}, {and} \bibinfo{person}{Mostafijur Rahman}.} \bibinfo{year}{2020}\natexlab{}.
\newblock \showarticletitle{Sightless helper: an interactive mobile application for blind assistance and safe navigation}. In \bibinfo{booktitle}{\emph{Cyber Security and Computer Science: Second EAI International Conference, ICONCS 2020, Dhaka, Bangladesh, February 15-16, 2020, Proceedings 2}}. Springer, \bibinfo{pages}{581--592}.
\newblock


\bibitem[Ishihara et~al\mbox{.}(2017)]%
        {ishihara2017beacon}
\bibfield{author}{\bibinfo{person}{Tatsuya Ishihara}, \bibinfo{person}{Jayakorn Vongkulbhisal}, \bibinfo{person}{Kris~M Kitani}, {and} \bibinfo{person}{Chieko Asakawa}.} \bibinfo{year}{2017}\natexlab{}.
\newblock \showarticletitle{Beacon-guided structure from motion for smartphone-based navigation}. In \bibinfo{booktitle}{\emph{2017 IEEE Winter Conference on Applications of Computer Vision (WACV)}}. IEEE, \bibinfo{pages}{769--777}.
\newblock


\bibitem[Krainz et~al\mbox{.}(2016)]%
        {krainz2016accessible}
\bibfield{author}{\bibinfo{person}{Elmar Krainz}, \bibinfo{person}{Viktoria Lind}, \bibinfo{person}{Werner Moser}, {and} \bibinfo{person}{Markus Dornhofer}.} \bibinfo{year}{2016}\natexlab{}.
\newblock \showarticletitle{Accessible way finding on mobile devices for different user groups}. In \bibinfo{booktitle}{\emph{Proceedings of the 18th International Conference on Human-Computer Interaction with Mobile Devices and Services Adjunct}}. \bibinfo{pages}{799--806}.
\newblock


\bibitem[Kuang et~al\mbox{.}(2022)]%
        {kuang2022consumer}
\bibfield{author}{\bibinfo{person}{Jian Kuang}, \bibinfo{person}{Taiyu Li}, \bibinfo{person}{Qijin Chen}, \bibinfo{person}{Baoding Zhou}, {and} \bibinfo{person}{Xiaoji Niu}.} \bibinfo{year}{2022}\natexlab{}.
\newblock \showarticletitle{Consumer-grade inertial measurement units enhanced indoor magnetic field matching positioning scheme}.
\newblock \bibinfo{journal}{\emph{IEEE Transactions on Instrumentation and Measurement}}  \bibinfo{volume}{72} (\bibinfo{year}{2022}), \bibinfo{pages}{1--14}.
\newblock


\bibitem[Kuang et~al\mbox{.}(2018)]%
        {Kuang2018}
\bibfield{author}{\bibinfo{person}{Jian Kuang}, \bibinfo{person}{Xiaoji Niu}, \bibinfo{person}{Peng Zhang}, {and} \bibinfo{person}{Xingeng Chen}.} \bibinfo{year}{2018}\natexlab{}.
\newblock \showarticletitle{Indoor positioning based on pedestrian dead reckoning and magnetic field matching for smartphones}.
\newblock \bibinfo{journal}{\emph{Sensors}}  \bibinfo{volume}{18} (\bibinfo{year}{2018}), \bibinfo{pages}{4142}.
\newblock
Issue 12.


\bibitem[Kuriakose et~al\mbox{.}(2023)]%
        {kuriakose2023turn}
\bibfield{author}{\bibinfo{person}{Bineeth Kuriakose}, \bibinfo{person}{Ida~Marie Ness}, \bibinfo{person}{Maja~{\AA} skov Tengstedt}, \bibinfo{person}{Jannicke~Merete Svendsen}, \bibinfo{person}{Terese Bj{\o}rseth}, \bibinfo{person}{Bijay~Lal Pradhan}, {and} \bibinfo{person}{Raju Shrestha}.} \bibinfo{year}{2023}\natexlab{}.
\newblock \showarticletitle{Turn Left Turn Right-Delving type and modality of instructions in navigation assistant systems for people with visual impairments}.
\newblock \bibinfo{journal}{\emph{International Journal of Human-Computer Studies}} (\bibinfo{year}{2023}), \bibinfo{pages}{103098}.
\newblock


\bibitem[Li et~al\mbox{.}(2012)]%
        {li2012feasible}
\bibfield{author}{\bibinfo{person}{Binghao Li}, \bibinfo{person}{Thomas Gallagher}, \bibinfo{person}{Andrew~G Dempster}, {and} \bibinfo{person}{Chris Rizos}.} \bibinfo{year}{2012}\natexlab{}.
\newblock \showarticletitle{How feasible is the use of magnetic field alone for indoor positioning?}. In \bibinfo{booktitle}{\emph{2012 International Conference on Indoor Positioning and Indoor Navigation (IPIN)}}. IEEE, \bibinfo{pages}{1--9}.
\newblock


\bibitem[Listgarten et~al\mbox{.}(2004)]%
        {listgarten2004multiple}
\bibfield{author}{\bibinfo{person}{Jennifer Listgarten}, \bibinfo{person}{Radford Neal}, \bibinfo{person}{Sam Roweis}, {and} \bibinfo{person}{Andrew Emili}.} \bibinfo{year}{2004}\natexlab{}.
\newblock \showarticletitle{Multiple alignment of continuous time series}.
\newblock \bibinfo{journal}{\emph{Advances in neural information processing systems}}  \bibinfo{volume}{17} (\bibinfo{year}{2004}).
\newblock


\bibitem[Luca and Alberto(2016)]%
        {luca2016towards}
\bibfield{author}{\bibinfo{person}{Dierna~Giovanni Luca} {and} \bibinfo{person}{Macha Alberto}.} \bibinfo{year}{2016}\natexlab{}.
\newblock \showarticletitle{Towards accurate indoor localization using iBeacons, fingerprinting and particle filtering}. In \bibinfo{booktitle}{\emph{2016 International Conference on Indoor Positioning and Indoor Navigation (IPIN)}}.
\newblock


\bibitem[Manduchi and Coughlan(2014)]%
        {manduchi2014last}
\bibfield{author}{\bibinfo{person}{Roberto Manduchi} {and} \bibinfo{person}{James~M Coughlan}.} \bibinfo{year}{2014}\natexlab{}.
\newblock \showarticletitle{The last meter: blind visual guidance to a target}. In \bibinfo{booktitle}{\emph{Proceedings of the SIGCHI Conference on Human Factors in Computing Systems}}. \bibinfo{pages}{3113--3122}.
\newblock


\bibitem[Manduchi et~al\mbox{.}(2010)]%
        {manduchi2010blind}
\bibfield{author}{\bibinfo{person}{Roberto Manduchi}, \bibinfo{person}{Sri Kurniawan}, {and} \bibinfo{person}{Homayoun Bagherinia}.} \bibinfo{year}{2010}\natexlab{}.
\newblock \showarticletitle{Blind guidance using mobile computer vision: A usability study}. In \bibinfo{booktitle}{\emph{Proceedings of the 12th international ACM SIGACCESS conference on Computers and accessibility}}. \bibinfo{pages}{241--242}.
\newblock


\bibitem[Martin et~al\mbox{.}(2020)]%
        {martin2020accessible}
\bibfield{author}{\bibinfo{person}{Elliot~W Martin}, \bibinfo{person}{Emily Farrar}, \bibinfo{person}{Evan Magsig}, \bibinfo{person}{Susan~A Shaheen}, \bibinfo{person}{Ashley Auer}, \bibinfo{person}{Sarah Hoban}, \bibinfo{person}{Booz~Allen Hamilton}, {et~al\mbox{.}}} \bibinfo{year}{2020}\natexlab{}.
\newblock \bibinfo{booktitle}{\emph{Accessible Transportation Technologies Research Initiative (ATTRI) Impact Assessment White Paper}}.
\newblock \bibinfo{type}{{T}echnical {R}eport}. \bibinfo{institution}{United States. Department of Transportation. Intelligent Transportation~…}.
\newblock


\bibitem[Morris(2021)]%
        {NYTimes21}
\bibfield{author}{\bibinfo{person}{Amanda Morris}.} \bibinfo{year}{2021}\natexlab{}.
\newblock \showarticletitle{Navigational Apps for the Blind Could Have a Broader Appeal}.
\newblock \bibinfo{journal}{\emph{The New York Times}} (\bibinfo{date}{20 December} \bibinfo{year}{2021}).
\newblock


\bibitem[Murata et~al\mbox{.}(2018)]%
        {murata2018smartphone}
\bibfield{author}{\bibinfo{person}{Masayuki Murata}, \bibinfo{person}{Dragan Ahmetovic}, \bibinfo{person}{Daisuke Sato}, \bibinfo{person}{Hironobu Takagi}, \bibinfo{person}{Kris~M Kitani}, {and} \bibinfo{person}{Chieko Asakawa}.} \bibinfo{year}{2018}\natexlab{}.
\newblock \showarticletitle{Smartphone-based indoor localization for blind navigation across building complexes}. In \bibinfo{booktitle}{\emph{2018 IEEE International Conference on Pervasive Computing and Communications (PerCom)}}. IEEE, \bibinfo{pages}{1--10}.
\newblock


\bibitem[Nair et~al\mbox{.}(2022)]%
        {nair2022assist}
\bibfield{author}{\bibinfo{person}{Vishnu Nair}, \bibinfo{person}{Greg Olmschenk}, \bibinfo{person}{William~H Seiple}, {and} \bibinfo{person}{Zhigang Zhu}.} \bibinfo{year}{2022}\natexlab{}.
\newblock \showarticletitle{ASSIST: Evaluating the usability and performance of an indoor navigation assistant for blind and visually impaired people}.
\newblock \bibinfo{journal}{\emph{Assistive Technology}} \bibinfo{volume}{34}, \bibinfo{number}{3} (\bibinfo{year}{2022}), \bibinfo{pages}{289--299}.
\newblock


\bibitem[Ohn-Bar et~al\mbox{.}(2018)]%
        {ohn2018variability}
\bibfield{author}{\bibinfo{person}{Eshed Ohn-Bar}, \bibinfo{person}{Jo{\~a}o Guerreiro}, \bibinfo{person}{Kris Kitani}, {and} \bibinfo{person}{Chieko Asakawa}.} \bibinfo{year}{2018}\natexlab{}.
\newblock \showarticletitle{Variability in reactions to instructional guidance during smartphone-based assisted navigation of blind users}.
\newblock \bibinfo{journal}{\emph{Proceedings of the ACM on interactive, mobile, wearable and ubiquitous technologies}} \bibinfo{volume}{2}, \bibinfo{number}{3} (\bibinfo{year}{2018}), \bibinfo{pages}{1--25}.
\newblock


\bibitem[Ren et~al\mbox{.}(2021)]%
        {ren2021smartphone}
\bibfield{author}{\bibinfo{person}{Peng Ren}, \bibinfo{person}{Fatemeh Elyasi}, {and} \bibinfo{person}{Roberto Manduchi}.} \bibinfo{year}{2021}\natexlab{}.
\newblock \showarticletitle{Smartphone-based inertial odometry for blind walkers}.
\newblock \bibinfo{journal}{\emph{Sensors}} \bibinfo{volume}{21}, \bibinfo{number}{12} (\bibinfo{year}{2021}), \bibinfo{pages}{4033}.
\newblock


\bibitem[Ren et~al\mbox{.}(2023)]%
        {ren2023experiments}
\bibfield{author}{\bibinfo{person}{Peng Ren}, \bibinfo{person}{Jonathan Lam}, \bibinfo{person}{Roberto Manduchi}, {and} \bibinfo{person}{Fatemeh Mirzaei}.} \bibinfo{year}{2023}\natexlab{}.
\newblock \showarticletitle{Experiments with RouteNav, A Wayfinding App for Blind Travelers in a Transit Hub}. In \bibinfo{booktitle}{\emph{Proceedings of the 25th International ACM SIGACCESS Conference on Computers and Accessibility}}. \bibinfo{pages}{1--15}.
\newblock


\bibitem[Riehle et~al\mbox{.}(2012)]%
        {riehle2012indoor}
\bibfield{author}{\bibinfo{person}{Timothy~H Riehle}, \bibinfo{person}{Shane~M Anderson}, \bibinfo{person}{Patrick~A Lichter}, \bibinfo{person}{Nicholas~A Giudice}, \bibinfo{person}{Suneel~I Sheikh}, \bibinfo{person}{Robert~J Knuesel}, \bibinfo{person}{Daniel~T Kollmann}, {and} \bibinfo{person}{Daniel~S Hedin}.} \bibinfo{year}{2012}\natexlab{}.
\newblock \showarticletitle{Indoor magnetic navigation for the blind}. In \bibinfo{booktitle}{\emph{2012 Annual International Conference of the IEEE Engineering in Medicine and Biology Society}}. IEEE, \bibinfo{pages}{1972--1975}.
\newblock


\bibitem[Riehle et~al\mbox{.}(2013)]%
        {riehle2013indoor}
\bibfield{author}{\bibinfo{person}{Timothy~H Riehle}, \bibinfo{person}{Shane~M Anderson}, \bibinfo{person}{Patrick~A Lichter}, \bibinfo{person}{William~E Whalen}, {and} \bibinfo{person}{Nicholas~A Giudice}.} \bibinfo{year}{2013}\natexlab{}.
\newblock \showarticletitle{Indoor inertial waypoint navigation for the blind}. In \bibinfo{booktitle}{\emph{2013 35th Annual International Conference of the IEEE Engineering in Medicine and Biology Society (EMBC)}}. IEEE, \bibinfo{pages}{5187--5190}.
\newblock


\bibitem[Ross and Blasch(2000)]%
        {ross2000wearable}
\bibfield{author}{\bibinfo{person}{David~A Ross} {and} \bibinfo{person}{Bruce~B Blasch}.} \bibinfo{year}{2000}\natexlab{}.
\newblock \showarticletitle{Wearable interfaces for orientation and wayfinding}. In \bibinfo{booktitle}{\emph{Proceedings of the fourth international ACM conference on Assistive technologies}}. \bibinfo{pages}{193--200}.
\newblock


\bibitem[Sato et~al\mbox{.}(2019)]%
        {sato2019navcog3}
\bibfield{author}{\bibinfo{person}{Daisuke Sato}, \bibinfo{person}{Uran Oh}, \bibinfo{person}{Jo{\~a}o Guerreiro}, \bibinfo{person}{Dragan Ahmetovic}, \bibinfo{person}{Kakuya Naito}, \bibinfo{person}{Hironobu Takagi}, \bibinfo{person}{Kris~M Kitani}, {and} \bibinfo{person}{Chieko Asakawa}.} \bibinfo{year}{2019}\natexlab{}.
\newblock \showarticletitle{NavCog3 in the wild: Large-scale blind indoor navigation assistant with semantic features}.
\newblock \bibinfo{journal}{\emph{ACM Transactions on Accessible Computing (TACCESS)}} \bibinfo{volume}{12}, \bibinfo{number}{3} (\bibinfo{year}{2019}), \bibinfo{pages}{1--30}.
\newblock


\bibitem[Sato et~al\mbox{.}(2017)]%
        {sato2017navcog3}
\bibfield{author}{\bibinfo{person}{Daisuke Sato}, \bibinfo{person}{Uran Oh}, \bibinfo{person}{Kakuya Naito}, \bibinfo{person}{Hironobu Takagi}, \bibinfo{person}{Kris Kitani}, {and} \bibinfo{person}{Chieko Asakawa}.} \bibinfo{year}{2017}\natexlab{}.
\newblock \showarticletitle{Navcog3: An evaluation of a smartphone-based blind indoor navigation assistant with semantic features in a large-scale environment}. In \bibinfo{booktitle}{\emph{Proceedings of the 19th International ACM SIGACCESS Conference on Computers and Accessibility}}. \bibinfo{pages}{270--279}.
\newblock


\bibitem[Sauro(2011)]%
        {sauro2011practical}
\bibfield{author}{\bibinfo{person}{Jeff Sauro}.} \bibinfo{year}{2011}\natexlab{}.
\newblock \bibinfo{booktitle}{\emph{A practical guide to the system usability scale: Background, benchmarks \& best practices}}.
\newblock \bibinfo{publisher}{Measuring Usability LLC}.
\newblock


\bibitem[Shahini et~al\mbox{.}(2022)]%
        {shahini2022friendly}
\bibfield{author}{\bibinfo{person}{Farzaneh Shahini}, \bibinfo{person}{Vanessa Nasr}, {and} \bibinfo{person}{Maryam Zahabi}.} \bibinfo{year}{2022}\natexlab{}.
\newblock \showarticletitle{A Friendly Indoor Navigation App for People with Disabilities (FIND)}. In \bibinfo{booktitle}{\emph{Proceedings of the Human Factors and Ergonomics Society Annual Meeting}}, Vol.~\bibinfo{volume}{66}. SAGE Publications Sage CA: Los Angeles, CA, \bibinfo{pages}{1922--1926}.
\newblock


\bibitem[Shu et~al\mbox{.}(2019)]%
        {Shu2019}
\bibfield{author}{\bibinfo{person}{Yuanchao Shu}, \bibinfo{person}{Zhuqi Li}, \bibinfo{person}{Börje Karlsson}, \bibinfo{person}{Yiyong Lin}, \bibinfo{person}{Thomas Moscibroda}, {and} \bibinfo{person}{Kang Shin}.} \bibinfo{year}{2019}\natexlab{}.
\newblock \showarticletitle{Incrementally-deployable indoor navigation with automatic trace generation}.
\newblock \bibinfo{journal}{\emph{IEEE INFOCOM 2019-IEEE Conference on Computer Communications}}, \bibinfo{pages}{2395--2403}.
\newblock
\showISBNx{1728105153}


\bibitem[{\v{S}}tancel et~al\mbox{.}(2021)]%
        {vstancel2021indoor}
\bibfield{author}{\bibinfo{person}{Martin {\v{S}}tancel}, \bibinfo{person}{Jan Hurtuk}, \bibinfo{person}{Michal Huli{\v{c}}}, {and} \bibinfo{person}{Jakub {\v{C}}erve{\v{n}}{\'a}k}.} \bibinfo{year}{2021}\natexlab{}.
\newblock \showarticletitle{Indoor atlas service as a tool for building an interior navigation system}.
\newblock \bibinfo{journal}{\emph{Acta Polytech. Hung}}  \bibinfo{volume}{18} (\bibinfo{year}{2021}), \bibinfo{pages}{87--110}.
\newblock


\bibitem[Storms et~al\mbox{.}(2010)]%
        {Storms2010}
\bibfield{author}{\bibinfo{person}{William Storms}, \bibinfo{person}{Jeremiah Shockley}, {and} \bibinfo{person}{John Raquet}.} \bibinfo{year}{2010}\natexlab{}.
\newblock \showarticletitle{Magnetic field navigation in an indoor environment}.
\newblock \bibinfo{journal}{\emph{2010 Ubiquitous Positioning Indoor Navigation and Location Based Service}}, \bibinfo{pages}{1--10}.
\newblock
\showISBNx{1424478790}


\bibitem[Subbu et~al\mbox{.}(2013)]%
        {subbu2013locateme}
\bibfield{author}{\bibinfo{person}{Kalyan~Pathapati Subbu}, \bibinfo{person}{Brandon Gozick}, {and} \bibinfo{person}{Ram Dantu}.} \bibinfo{year}{2013}\natexlab{}.
\newblock \showarticletitle{LocateMe: Magnetic-fields-based indoor localization using smartphones}.
\newblock \bibinfo{journal}{\emph{ACM Transactions on Intelligent Systems and Technology (TIST)}} \bibinfo{volume}{4}, \bibinfo{number}{4} (\bibinfo{year}{2013}), \bibinfo{pages}{1--27}.
\newblock


\bibitem[Torres-Sospedra et~al\mbox{.}(2015)]%
        {torres2015ujiindoorloc}
\bibfield{author}{\bibinfo{person}{Joaqu{\'\i}n Torres-Sospedra}, \bibinfo{person}{David Rambla}, \bibinfo{person}{Raul Montoliu}, \bibinfo{person}{Oscar Belmonte}, {and} \bibinfo{person}{Joaqu{\'\i}n Huerta}.} \bibinfo{year}{2015}\natexlab{}.
\newblock \showarticletitle{UJIIndoorLoc-Mag: A new database for magnetic field-based localization problems}. In \bibinfo{booktitle}{\emph{2015 International conference on indoor positioning and indoor navigation (IPIN)}}. IEEE, \bibinfo{pages}{1--10}.
\newblock


\bibitem[Tsai et~al\mbox{.}(2021)]%
        {tsai2021finding}
\bibfield{author}{\bibinfo{person}{Chia~Hsuan Tsai}, \bibinfo{person}{Peng Ren}, \bibinfo{person}{Fatemeh Elyasi}, {and} \bibinfo{person}{Roberto Manduchi}.} \bibinfo{year}{2021}\natexlab{}.
\newblock \showarticletitle{Finding Your Way Back: Comparing Path Odometry Algorithms for Assisted Return}. In \bibinfo{booktitle}{\emph{2021 IEEE International Conference on Pervasive Computing and Communications Workshops and other Affiliated Events (PerCom Workshops)}}. IEEE, \bibinfo{pages}{117--122}.
\newblock


\bibitem[Williams et~al\mbox{.}(2015)]%
        {williams2015not}
\bibfield{author}{\bibinfo{person}{Michele~A Williams}, \bibinfo{person}{Erin Buehler}, \bibinfo{person}{Amy Hurst}, {and} \bibinfo{person}{Shaun~K Kane}.} \bibinfo{year}{2015}\natexlab{}.
\newblock \showarticletitle{What not to wearable: using participatory workshops to explore wearable device form factors for blind users}. In \bibinfo{booktitle}{\emph{Proceedings of the 12th International Web for All Conference}}. \bibinfo{pages}{1--4}.
\newblock


\bibitem[Yoon et~al\mbox{.}(2019)]%
        {yoon2019leveraging}
\bibfield{author}{\bibinfo{person}{Chris Yoon}, \bibinfo{person}{Ryan Louie}, \bibinfo{person}{Jeremy Ryan}, \bibinfo{person}{MinhKhang Vu}, \bibinfo{person}{Hyegi Bang}, \bibinfo{person}{William Derksen}, {and} \bibinfo{person}{Paul Ruvolo}.} \bibinfo{year}{2019}\natexlab{}.
\newblock \showarticletitle{Leveraging augmented reality to create apps for people with visual disabilities: A case study in indoor navigation}. In \bibinfo{booktitle}{\emph{The 21st International ACM SIGACCESS Conference on Computers and Accessibility}}. \bibinfo{pages}{210--221}.
\newblock


\bibitem[Yu et~al\mbox{.}(2019)]%
        {yu2019shoesloc}
\bibfield{author}{\bibinfo{person}{Tuo Yu}, \bibinfo{person}{Haiming Jin}, {and} \bibinfo{person}{Klara Nahrstedt}.} \bibinfo{year}{2019}\natexlab{}.
\newblock \showarticletitle{Shoesloc: In-shoe force sensor-based indoor walking path tracking}.
\newblock \bibinfo{journal}{\emph{Proceedings of the ACM on Interactive, Mobile, Wearable and Ubiquitous Technologies}} \bibinfo{volume}{3}, \bibinfo{number}{1} (\bibinfo{year}{2019}), \bibinfo{pages}{1--23}.
\newblock


\bibitem[Zahabi et~al\mbox{.}(2022)]%
        {zahabi2022design}
\bibfield{author}{\bibinfo{person}{Maryam Zahabi}, \bibinfo{person}{Xi Zheng}, \bibinfo{person}{Azima Maredia}, {and} \bibinfo{person}{Farzaneh Shahini}.} \bibinfo{year}{2022}\natexlab{}.
\newblock \showarticletitle{Design of Navigation Applications for People with Disabilities: A Review of Literature and Guideline Formulation}.
\newblock \bibinfo{journal}{\emph{International Journal of Human--Computer Interaction}} (\bibinfo{year}{2022}), \bibinfo{pages}{1--23}.
\newblock


\end{thebibliography}

\end{document}